%% file: main.tex
\documentclass[
12pt, % The default document font size, options: 10pt, 11pt, 12pt
english, % ngerman for German
singlespacing, % Single line spacing, alternatives: onehalfspacing or doublespacing
headsepline, % Uncomment to get a line under the header
chapterinoneline, % Uncomment to place the chapter title next to the number on one line
]{MastersDoctoralThesis} % The class file specifying the document structure

\usepackage[utf8]{inputenc} % Required for inputting international characters
\usepackage[T1]{fontenc} % Output font encoding for international characters
\usepackage{mathpazo} % Use the Palatino font by default
\usepackage{amsthm,amsmath,latexsym,amssymb,amsfonts,amsbsy}
\usepackage{physics} 
\usepackage{tikz,esvect} %drawing
\usepackage{pgfplots}
\usepgfplotslibrary{patchplots}
\usetikzlibrary{patterns, positioning, arrows}
\usepackage{pgf}
\let\oldsqrt\sqrt
\usetikzlibrary{3d,calc,shapes}
\tikzset{%
  raindrop/.pic={
    code={\tikzset{scale=0.8}
 \clip [preaction={top color=black, bottom color=black!70}]
 (0,0)  .. controls ++(0,-1) and ++(0,1) .. (1,-2)
 arc (360:180:1)
 .. controls ++(0,1) and ++(0,-1) .. (0,0) -- cycle;
 \foreach \j in {1,3,...,20}
 \shade [top color=blue!50!cyan!50, shift=(270:0), xscale=1-\j/40,yscale=1-\j/80, white, opacity=1/15]
 [rotate=-\j] (0,0)  .. controls ++(0,-1) and ++(0,1) .. (1,-2)
 arc (360:180:1)
 .. controls ++(0,1) and ++(0,-1) .. (0,0) -- cycle;
  }}}
\definecolor{wvvxds}{rgb}{0.396078431372549,0.3411764705882353,0.8235294117647058}
\definecolor{rvwvcq}{rgb}{0.08235294117647059,0.396078431372549,0.7529411764705882}
\definecolor{wwccqq}{rgb}{0.4,0.8,0}
\definecolor{darkmagenta}{rgb}{0.55, 0.0, 0.55}
\definecolor{darkblue}{rgb}{0.0, 0.0, 0.55}
\definecolor{darkred}{rgb}{0.7, 0.0, 0.3}
\usepackage{empheq}
\usepackage{xcolor}
\usepackage{pgfplots}
\definecolor{lightgreen}{HTML}{90EE90}

   \def\sqrt{\mathpalette\DHLhksqrt}
   \def\DHLhksqrt#1#2{%
   \setbox0=\hbox{$#1\oldsqrt{#2\,}$}\dimen0=\ht0
   \advance\dimen0-0.2\ht0
   \setbox2=\hbox{\vrule height\ht0 depth -\dimen0}%
   {\box0\lower0.4pt\box2}} 
\usepackage[autostyle=true]{csquotes} % Required to generate language-dependent quotes in the bibliography
\newcommand{\HRulegrossa}{\rule{\linewidth}{1.2mm}}
\thesistitle{\sc de Sitter Entanglement and conformal description of the cosmological horizon} % Your thesis title, this is used in the title and abstract, print it elsewhere with \ttitle
\supervisor{Dr. Rodrigo \textsc{Olea}} % Your supervisor's name, this is used in the title page, print it elsewhere with \supname
\examiner{Dr. Per Sundell (UNAB) \\ Dr. Alberto Faraggi (UNAB)\\ Dr. Francisco Rojas (UAI)} % Your examiner's name, this is not currently used anywhere in the template, print it elsewhere with \examname
\degree{Master in science} % Your degree name, this is used in the title page and abstract, print it elsewhere with \degreename
\author{Felipe \textsc{Diaz}} % Your name, this is used in the title page and abstract, print it elsewhere with \authorname
\addresses{} % Your address, this is not currently used anywhere in the template, print it elsewhere with \addressname

\subject{Physics science} % Your subject area, this is not currently used anywhere in the template, print it elsewhere with \subjectname
 % Keywords for your thesis, this is not currently used anywhere in the template, print it elsewhere with \keywordnames
\university{\href{http://www.unab.cl}{Universidad Andres Bello}} % Your university's name and URL, this is used in the title page and abstract, print it elsewhere with \univname
\department{\href{http://fisica.unab.cl}{Physics Department}} % Your department's name and URL, this is used in the title page and abstract, print it elsewhere with \deptname
\group{Grupo de Altas Energias} % Your research group's name and URL, this is used in the title page, print it elsewhere with \groupname
\faculty{\href{http://facultades.unab.cl/cienciasexactas/}{Facultad de Ciencias Exactas}} % Your faculty's name and URL, this is used in the title page and abstract, print it elsewhere with \facname

\AtBeginDocument{
\hypersetup{pdftitle=\ttitle} % Set the PDF's title to your title
\hypersetup{pdfauthor=\authorname} % Set the PDF's author to your name
\hypersetup{pdfkeywords=\keywordnames} % Set the PDF's keywords to your keywords
\hypersetup{colorlinks=false}
\hypersetup{urlcolor=red}
}

\begin{document}

\frontmatter % Use roman page numbering style (i, ii, iii, iv...) for the pre-content pages

\pagestyle{plain} % Default to the plain heading style until the thesis style is called for the body content

%----------------------------------------------------------------------------------------
%	TITLE PAGE
%----------------------------------------------------------------------------------------

\begin{titlepage}
\begin{center}
\includegraphics[width=0.3\textwidth]{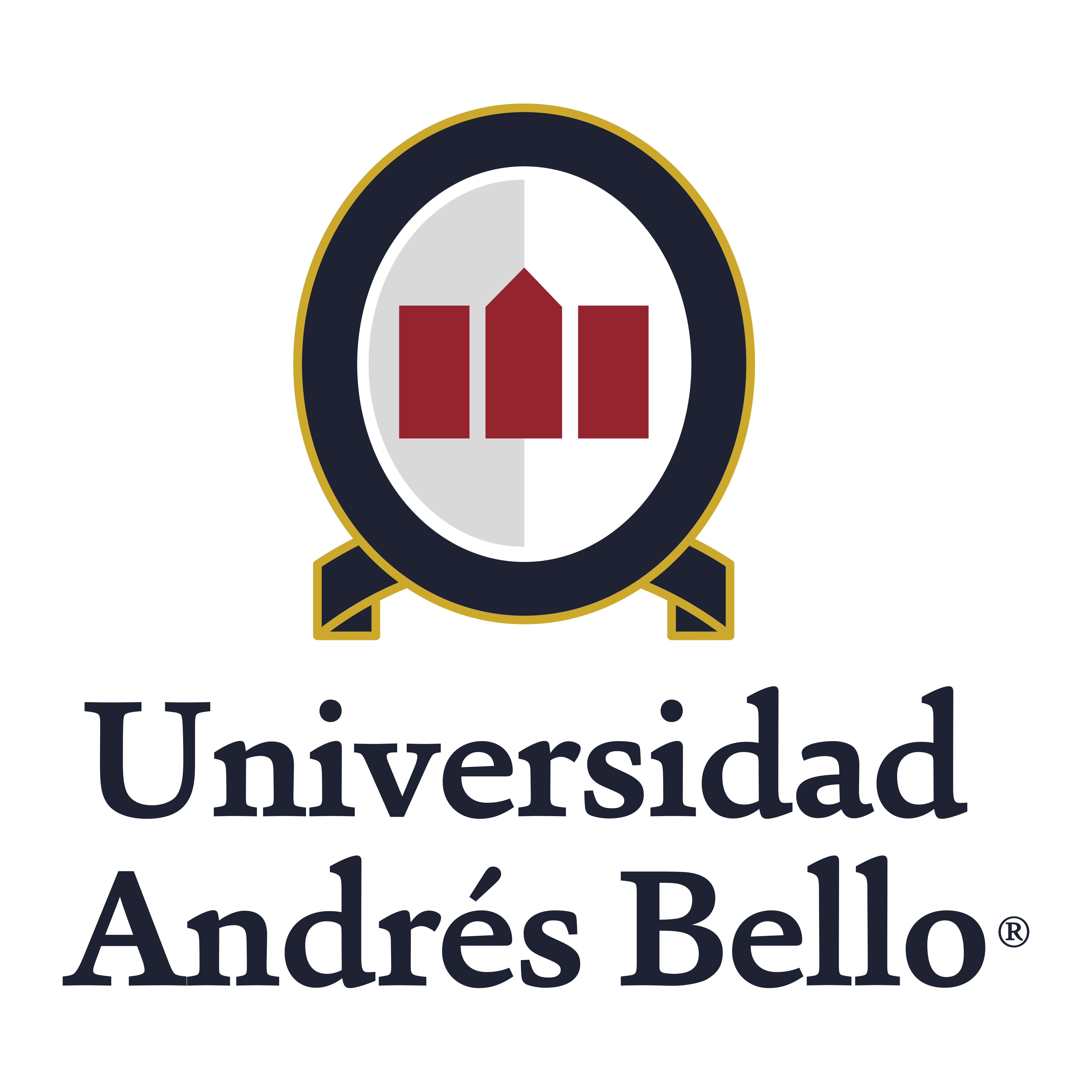}

%\vspace{1mm}
%\vspace*{.06\textheight}
{\scshape\LARGE \univname\par}\vspace{1.5cm} % University name
\textsc{\Large Master Thesis}\\[0.5cm] % Thesis type
\HRulegrossa \\
\HRule \\[0.4cm] % Horizontal line
{\huge \bfseries \ttitle\par}\vspace{0.4cm} % Thesis title
\HRule \\ % Horizontal line
 \HRulegrossa \\[1.5 cm]
 
\begin{minipage}[t]{0.4\textwidth}
\begin{flushleft} \large
\emph{Author:}\\
\href{http://www.Hologravity.com}{\authorname} % Author name - remove the \href bracket to remove the link
\end{flushleft}
\end{minipage}
\begin{minipage}[t]{0.4\textwidth}
\begin{flushright} \large
\emph{Supervisor:} \\
\href{http://www.hologravity.com}{\supname} % Supervisor name - remove the \href bracket to remove the link  
\end{flushright}
\end{minipage}\\[1cm]
 
\emph{Jury Members:}\\
\examname
\vspace{4mm}

\large \textit{A thesis submitted in fulfillment of the requirements\\ for the degree of \degreename}\\[0.3cm] % University requirement text
\textit{in the}\\[0.4cm]
\deptname\\[2mm] % Research group name and department name
 
{\large \today} % Date

\end{center}
\end{titlepage}

%----------------------------------------------------------------------------------------
%	DECLARATION PAGE
%----------------------------------------------------------------------------------------

\begin{declaration}
\addchaptertocentry{\authorshipname} % Add the declaration to the table of contents
\noindent I, \authorname, declare that this thesis titled, \enquote{\ttitle} and the work presented in it are my own. I confirm that:

\begin{itemize} 
\item This work was done wholly or mainly while in candidature for a research degree at this University.
\item Where any part of this thesis has previously been submitted for a degree or any other qualification at this University or any other institution, this has been clearly stated.
\item Where I have consulted the published work of others, this is always clearly attributed.
\item Where I have quoted from the work of others, the source is always given. With the exception of such quotations, this thesis is entirely my own work.
\item I have acknowledged all main sources of help.
\item Where the thesis is based on work done by myself jointly with others, I have made clear exactly what was done by others and what I have contributed myself.\\
\item This thesis is based on \cite{Arias:2019aa, Arias:2019zug, Arias:2019lzk}.
\end{itemize}
 \vspace{2cm}
\noindent Signed:\\
\rule[0.5em]{25em}{0.5pt} % This prints a line for the signature
 
\noindent Date:\\
\rule[0.5em]{25em}{0.5pt} % This prints a line to write the date
\end{declaration}

\cleardoublepage

%----------------------------------------------------------------------------------------
%	QUOTATION PAGE
%----------------------------------------------------------------------------------------

\vspace*{0.2\textheight}

\noindent\enquote{\itshape In this house, young lady, we obey the laws of thermodynamics!}\bigbreak

\hfill Homer Simpson

%----------------------------------------------------------------------------------------
%	ABSTRACT PAGE
%----------------------------------------------------------------------------------------

\begin{abstract}
\addchaptertocentry{\abstractname} % Add the abstract to the table of contents
In this thesis, we aim to understand the microscopic details and origin of the Cosmological Horizon, produced by a static observer in four-dimensional de Sitter (dS$_4$) spacetime. We consider a deformed extension of dS spacetime by means of a single $\mathbb Z_q$ quotient, which resembles an Orbifold geometry. 

The Orbifold parameter induces a pair of codimension two minimal surfaces given by 2-spheres in the Euclidean geometry. 
Using dimensional reduction on the two-dimensional plane where the minimal surfaces have support, we use the Liouville field theory and the Kerr/CFT mechanism in order to describe the underlying degrees of freedom of the Cosmological Horizon. 
 We then show, that in the large $q$-limit, this pair of codimensions two surfaces can be realized as the conformal boundaries of dS$_3$. We notice that the central charge obtained using Liouville theory, in the latter limit, corresponds to the Strominger central charge obtained in the context of the dS/CFT correspondence. This is a realization of the duality on lower dimension as a limit of the orbifold geometry. 
 
In addition, a formulation of entanglement entropy for de Sitter spacetimes is given in terms of dS holography and also a different approach in which the entanglement between two disconnected bulk observers is described in terms of the topology of the spacetime. Therefore, a quarter of the area formula is proposed, in which the area corresponds to the are of the set of fixed points of an $S^2/\mathbb Z_q$ orbifold. 
\end{abstract}

%----------------------------------------------------------------------------------------
%	ACKNOWLEDGEMENTS
%----------------------------------------------------------------------------------------

\begin{acknowledgements}
\addchaptertocentry{\acknowledgementname} % Add the acknowledgements to the table of contents
I would like to thanks to my whole family, for the support and love that they have been giving me through my whole life. There are not enough words or ways to express how grateful I am to you. 

Many thanks to the "Husares'' for all the funny and emotional moments through the last 15 years. You have stuck by me through so many ups and downs and I want you to know that I will always be there for you. Tons of Thanks.

I am deeply grateful to the guys on the high energy physics group; Ignacio, Giorgos, Gabriel, David, Juan Pablo, Ivan, Gustavo, Pablo, Javier, Yihao and Israel. This guys have shown to me what it means to have passion for physics. 

It is a pleasure to thanks to Per Sundell. The door to Prof. Sundell office was always open whenever I ran into a trouble spot or had a question about my research or writing. He consistently allowed this paper to be my own work, but steered me in the right the direction whenever he thought I needed it.

This thesis would have not been possible without all the help of Cesar \textit{Super}-Arias. I must express my very profound gratitude to Cesar for being such a good friend,  forencouraging me to learn new methods in physics, and for always be there when I have been in trouble with understanding of new concepts, performing certain calculations, or simply struggling with life.

A special acknowledge to my advisor Rodrigo Olea, not only for the infnite patience at the moment to help me to understand such a beautiful theory as gravity and all that
this implies, but also for all the long talks about music, life, and physics. More than an advisor, a friend.

I would also thanks to all the professors of the physics department at UNAB; A. Faraggi, B. Valillo, M. Cambiasso and R. Aros. I have learn all the foundations and concepts on physics from them. 
\\

Finally, I sincerely appreciate to Daniela for all the confidence and support, you have captured my heart. 
\\

This work is supported by the MSc Scholarship provided by the Dirección General de Investigación of Universidad Andres Bello 
(DGI-UNAB). I acknowledge to the financial support as a M.Sc Thesis student given by FONDECYT Regular grant No1170765 
\end{acknowledgements}

%----------------------------------------------------------------------------------------
%	LIST OF CONTENTS/FIGURES/TABLES PAGES
%----------------------------------------------------------------------------------------

\tableofcontents % Prints the main table of contents

\dedicatory{Dedicated to my mother} 

%----------------------------------------------------------------------------------------
%	THESIS CONTENT - CHAPTERS
%----------------------------------------------------------------------------------------

\mainmatter % Begin numeric (1,2,3...) page numbering

\pagestyle{thesis} % Return the page headers back to the "thesis" style

% Include the chapters of the thesis as separate files from the Chapters folder
% Uncomment the lines as you write the chapters

\include{Chapters/Introduction}
\include{Chapters/dS}
\include{Chapters/CFT}

\include{Chapters/Liouville}

\include{Chapters/HEE} 
\include{Chapters/KerrCFT}

\include{Chapters/Entanglement} 
\include{Chapters/Microscopic}

\include{Chapters/Conclusions}

%----------------------------------------------------------------------------------------
%	THESIS CONTENT - APPENDICES
%----------------------------------------------------------------------------------------

\appendix % Cue to tell LaTeX that the following "chapters" are Appendices

% Include the appendices of the thesis as separate files from the Appendices folder
% Uncomment the lines as you write the Appendices

%\include{Appendices/AppendixA}
\include{Appendices/AppendixB}
\include{Appendices/AppendixC}

%----------------------------------------------------------------------------------------
%	BIBLIOGRAPHY
%----------------------------------------------------------------------------------------

%\printbibliography[heading=bibintoc]
\bibliographystyle{JHEP}
\bibliography{thesis.bib}

%----------------------------------------------------------------------------------------

\end{document}

%% file: Chapters/Introduction.tex
% Chapter 1

\chapter{Introduction} % Main chapter title

\label{Intro} % For referencing the chapter elsewhere, use \ref{Chapter1} 

%----------------------------------------------------------------------------------------

% Define some commands to keep the formatting separated from the content 
\newcommand{\keyword}[1]{\textbf{#1}}
\newcommand{\tabhead}[1]{\textbf{#1}}
\newcommand{\code}[1]{\texttt{#1}}
\newcommand{\file}[1]{\texttt{\bfseries#1}}
\newcommand{\option}[1]{\texttt{\itshape#1}}

%----------------------------------------------------------------------------------------

The $d$-dimensional de Sitter spacetime (dS$_d$) corresponds to a maximally symmetric positively curved exact solution to Einstein equations with positive cosmological constant $\Lambda$, and its symmetry group corresponds to $SO(1,d)$. Particularly in $d=4$, the dS$_4$ geometry can be cast into a Freidman-Robertson-Walker metric with exponential inflation in time. Recent astrophysical observations \cite{Perlmutter_1999, 1538-3881-116-3-1009} indicates that our universe may end up in this scenario. Moreover dS$_4$ is used to model the cosmological inflation \cite{Guth:1980zm}. The fact that we live on an expanding universe implies that an inertial observer would have access to only one portion of the whole spacetime, therefore will be causally connected only with a subregion of $dS$ and will be surrounded by the so-called \emph{cosmological horizon} ($\mathcal H$), that provides the concept of \emph{observable universe} \cite{III:2005aa}. This because the expansion of the space is so fast that the light-rays do not manage to propagate to the entire universe.  
\\

In \cite{Gibbons:1977mu} Gibbons and Hawking have realized, using Euclidean quantum gravity methods, that the Cosmological Horizon shares thermodynamic properties with the ones of black holes. An inertial observer inside the $dS$ spacetime would detect thermal radiation at the horizon with a corresponding thermodynamic entropy conjugated that satisfy the Bekenstein-Hawking area law  \cite{Bekenstein:1974ax, Bekenstein:1973ur, Bekenstein:1972tm, Hawking:1976de}, that is reffered as the Gibbons--Hawking entropy. In \cite{Witten:aa, Banks:2000fe} it has been pointed out that the Hilbert space may have finite dimension which implies that standard Einstein theory with positive cosmological constant may not be able to be quantized and this may become for a more fundamental theory that determines the possible values of $\Lambda$. Then a universal microscopic description of this thermal features of $dS$ remains unknown. 
\\

Some attempts to this consist on try to describe the near horizon symmetries \cite{Carlip:1999cy}. The constrained algebra of general relativity acquires a central extension under suitable set of boundary conditions, which in terms of Cardy's formula \eqref{MicroCardy} resembles the quarter of the area law for entropy. In three dimensions, the Chern-Simons formulation of gravity can be obtained for a $dS$ background by using $SL(2,\mathbb C)$ as a gauge group, which in a spatial disc boundary, reduces to a Wess-Zumino-Witten conformal model \cite{Witten:1983ar}. Then, the $dS_3$ entropy arises as the entropy of highly excited thermal states \cite{Maldacena:1998ab} and its in agreement with the semiclassical formula. 
\\

Furthermore for three-dimensional gravity, inspired in Brown-Henneaux construction of asymptotic symmetries of three dimensional anti-de Sitter (AdS) gravity \cite{Brown:1986nw}, Strominger has proposed to describe the microscopic underlying degrees of freedom at spacelike infinity of $dS$ spacetime that can be described in terms of a Euclidean (probably non-unitary) conformal field theory (CFT) on a sphere. Therefore a dS/CFT correspondence was proposed \cite{Strominger:2001pn} and some concrete realizations of this duality can be found in \cite{Strominger:2001aa, Anninos:2011aa, Klemm:2002ab, Kabat:2002aa, Narayan:2017aa, Park:1998qk, Park:2008aa, Bernardo:2018cow, Bernardo:2019bbi}. In three dimensions, the Gibbons-Hawking formula has been recovered from this duality \cite{Klemm:2002ab}, by using the Brown-York stress tensor \cite{Brown:1993aa}, whose infinitesimal variation under conformal transformation leads to a given central charge which is exactly equal to the one found it for its AdS relative. Also this correspondence has provide to reconstruct the same holographic conformal anomaly as from the AdS/CFT dictionary \cite{Maldacena:1998ac, Gubser:1998aa, Witten:1998aa} in four dimensions \cite{NOJIRI2001145}.  For the three-dimensional case it is also worth to notice that the $dS/$CFT correspondence has been provided before the Strominger formulation, in which a generalization for Chern-Simon theory with boundary its dual to a certain CFT\footnote{We thank professor Mu-In Park for point this out.} \cite{Park:1998qk, Park:2008aa, Park:1998yw, Banados:1998tb, Banados:1998ta}, which by using $SL(2,\mathbb C)$ as a gauge group corresponds to $dS_3$ and using $SL(2,\mathbb R)$ corresponds to $AdS_3$.
\\

Moreover, find stable vacua solutions in string theory has always been a challenging problem \cite{Kachru:2003aw, Maloney:2002rr}, conjecturing that the dS solutions are included in the so-called swampland of string theory \cite{Vafa:2005ui, Ooguri:2006in,Garg:2018reu}.  Therefore, understanding dS spacetime is a fundamental key if we aim to someday describe a complete theory of quantum gravity. The above mentioned AdS/CFT correspondence is founded in string theory and attempts to extend the correspondence into dS using different foliations of AdS space can be found in \cite{Alishahiha:2005ab, Alishahiha:2005aa}  relating $dS_d$ gravity theory with the low energy limit of a conformal field theory coupled to gravity on a one-lower dimensional $dS$ spacetime, has been used in \cite{Dong:2018aa, Geng:2019bnn} to describe the entanglement entropy interpretation of $dS$. Some realizations of this latter proposal can be found in \cite{Geng:2019zsx, Geng:2019ruz, Lewkowycz:2019xse, Gorbenko:2018oov, Geng:2019ruz}.
\\

Here, we shall perform a conformal description of the horizon by using the two-dimensional defects that arises after orbifolding the horizon. These two-dimensional antipodal defects are coupled by a tension on top of the gravity action. This motivates the conformal description of the horizon via the trace anomaly formula for two-dimensional CFTs \cite{Alves_2004, Polchinski_1998}. In order to describe this microscopically we make use of the Liouville conformal field theory as an effective action for the defects, where we have identify the couplings of the theory with the ones of the gravity side, and use this in order to obtain the central charge what resembles the dS entanglement entropy by using Cardy's formula. 
As a different approach, it is possible to mimic the near-horizon geometry of Kerr black holes and use the Kerr/CFT correspondence \cite{Guica:2008mu}, which realize the dual description of the near-horizon region of extreme black holes with a given chiral conformal field theory, this has allow us to reproduce the Virasoro algebra where the central extension has also been used in the Cardy formula leading to the same entropy obtained via Liouville and via conformal trace anomaly. Both entropies are identified with the relative entropy that exist between the branes and the horizon, and its in agreement with the $dS_4$ entropy.
\\

A different interpretation of the $dS$ entropy corresponds to the entanglement entropy on quantum field theory. This has been inspired by the calculation of entanglement via holographic techniques \cite{Ryu:2006ef, Ryu:2006bv, Nishioka:2009un} in the context of AdS/CFT, which relates this quantum property with geometry of the bulk gravity theory.  The $dS$ entropy has been attempted to be interpreted as entanglement entropy from certain CFT, this has been done using the dS/CFT correspondence by seeking for extremal co-dimension two surfaces $dS$ \cite{Narayan:2017aa} which has been proposed to corresponds to a thermofield double state for a ghost-CFT \cite{Jatkar:aa}, and also relation of the analytically continued extremal surfaces via Eculidean AdS space has been pointed out in \cite{Sato:2015aa}. A  different holographic proposal called $dS/dS$ correspondence \cite{Alishahiha:2005ab, Alishahiha:2005aa}, which relates $dS_d$ gravity theory with the low energy limit of a conformal field theory coupled to gravity on a one-lower dimensional $dS$ spacetime, has been used in \cite{Dong:2018aa, Geng:2019bnn} to describe the entanglement entropy interpretation of $dS$.
\\

In this thesis we manage to propose how to understand the $dS_4$ entropy as entanglement entropy, first between two disconnected conformal field theories living at the spacelike boundary of the spacetime, and also for two disconnected bulk geometries at the interior. For the first we have assumed the existence of an holographic duality whereby quantum gravity on $dS$ is dual to two copies of certain Euclidean conformal field theory, one copy per boundary. In turn, the bulk observer can be described in terms of a thermofield double state in the tensor product of the field theory Hilbert space, therefore the density matrix turns to be thermal. 

Noticing that both boundaries conformal field theory have the topology of a 3-sphere and the boundary replica symmetry, defined by azimuthal identification, is guaranteed and the construction of branched cover boundary manifold is trivial, and we made use of the replica trick \cite{0305-4608-5-5-017, Calabrese:2004eu} allow us to compute entanglement between this two. 

For the second approach we have made use of the fact that the topology of the bulk corresponds to the $S^2\times_w S^2$ which describes two disconnected Rindler observers, and the entanglement entropy recovers the quarter of the area law but in terms of the area of the set of fixed points. In this case the deformed $S^2/\mathbb Z_q$ orbifolded horizon posses a manifest bulk replica symmetry, that can be thought of as the observers back-reaction whose set of fixed points defines a pair of minimal surfaces. 

In both approaches the antipodal symmetry, that maps every point into its antipodal opposite, plays a fundamental role. It has allows us to write the quantum gravity partition function of a single CFT or Rindler wedge respectively.
\\

We close the thesis some discussions and conclusions. Some potentially interesting directions of future work are given.

%% file: Chapters/dS.tex
% Chapter 1

\chapter{De Sitter spacetime} % Main chapter title
\label{dS} % For referencing the chapter elsewhere, use \ref{Chapter1} 

As it is motivated in the Introduction, we will now study the four-dimensional $dS$ spacetime, which in presence of an observer enhances a cosmological horizon that surround this one. In this chapter, we will made a short review the main features of $dS$ and how it can be seen it as an embedding of a higher-dimensional flat spacetime in Lorentzian signature. This is carried out by choosing different embedding coordinates we can describe what is called the global patch or the static patch for an observer. It is also showed that the cosmological horizon shares properties that resembles the ones of the black holes.

 Later on the chapter, we introduce a different embedding that corresponds to an extended static patch that covers both Rindler wedges of the interior and describes the cosmological horizon as an sphere. At the end of this chapter we proceed to made use of this extended coordinates in order to perform an Orbifold that will be used during the computations to describe the $dS$ entropy. 
\\

For this chapter we use great textbooks and notes to understand the dS spacetime \cite{Hawking_1973, Carroll:2004st, HartmanLectures}.
%----------------------------------------------------------------------------------------

% Define some commands to keep the formatting separated from the content 
%----------------------------------------------------------------------------------------

\section{Coordinates and embbedings}
De Sitter spacetime can be constructed and understood as an embedding on a higher dimensional space. It corresponds simply to a timelike hyperbola on a higher-dimensional Minkowski spacetime characterized by the radius $\ell$, i.e., $dS_d \hookrightarrow \mathcal M_{d+1}$. The embedding metric is just the Lorentzian flat metric of $d+1$-dimensions
\begin{align}
dS_d \hookrightarrow\mathcal M^{d+1}~,\quad ds_{\mathcal M^{d+1}}^2 = -dX_0^2 + \sum_{1\leq i \leq d} dX_i^2 ~,
\end{align}
and $dS$ is defined as the hypersurface defined by the equation
\begin{equation}
\label{hyperboloid}
X^2 =\ell^2~.
\end{equation}
The hyperboloid \eqref{hyperboloid} has the topology of $\mathbb R \times S^{d-1}$ with manifest $O(d,1)$ symmetrys. 
\subsection{Global coordinates}
Even though the construction of de Sitter spacetime can be done in an arbitrary dimension, the analysis performed during this thesis corresponds to the four-dimensional case.
We can define a global set of coordinates for the hyperbola by just defining
\begin{equation}
\label{global}
X_0=\ell \sinh (T/\ell)~, \quad 
X_i= \ell \cosh(T/\ell) z_i ~,~~  1\leq i \leq 4~,
\end{equation}
where $z_i$ coordinatize the unit $3$-sphere, such that $\vec{z}^{\,2} = 1$.
Then, the resulting metric for the hyperbola hat satisfy the hypersurface equation \eqref{hyperboloid} is
\begin{equation}\label{globalmetric}
ds^2 = -dT^2 + \ell^2 \cosh^2(T/\ell)\,d\Omega^2_{3}~,
\end{equation}
where $-\infty < T < \infty$ and $0\leq\Theta\leq\pi$, also $d\Omega^2_{3} = d\Theta^2 + \sin^2 \Theta d\Omega_2^2$ is the unit $3$-sphere line element, which is foliated by $2-$spheres $d\Omega^2_2$. This are called global coordinates, because they cover the whole hyperbola and $0 \leq \Theta \leq \pi$. The points $\Theta = \{0 ,\pi\}$ are usually referred to as the north and south poles of the global $3-$sphere.  
This is an exact vacuum solution to the Einstein equation with positive cosmological constant, relating the $dS$ radius by
\begin{align}
\Lambda = \frac{(d-1)(d-2)}{2\ell^2} = \frac{3}{\ell^2}~,
\end{align}
which, in four dimensions, can be related with the Hubble parameter as
\begin{align}
H = \sqrt{\frac{\Lambda}{3}}~.
\end{align}
Using conformal time $\tau$ such that
\begin{align}
\tan \frac{\tau}{2} \equiv \tanh \frac{T}{2\ell}~,
\end{align} 
the line element yields
\begin{align}
ds^2 = \frac{-d\tau^2 + d\Omega_3^2}{\cos^2\tau}~,
\end{align}
and $-\pi/2 < \tau < \pi/2$. The past and future infinities $\mathcal I^{\pm}$ define the conformal boundaries at
\begin{align}
\mathcal I^{\pm} \equiv \tau^{-1}(\pm\pi/2)~,
\end{align}
where both have the topology of an $S^3$.
\vspace{0.5cm}
\begin{center}
\begin{tikzpicture}
\node at (0.7,3.3) {$\mathcal I^+$};
\node at (0.7,-0.3) {$\mathcal I^-$};
\node[right, rotate=90] at (-1.3,0.5) {\footnotesize North pole};
\node[right, rotate=-90] at (2.3,2.3) {\footnotesize South pole}; 
\draw [thick] (-1,0)--(-1,3)--(2,3)--(2,0)--(-1,0);
\draw[thick, dashed] (-1,0)--(2,3);
\draw[thick, dashed] (-1,3)--(2,0);
%
%Caption~~~~~~~~~~~~~~~~~~~~
\node[text width=16cm, text justified] at (1.5,-1.7){
\small {\hypertarget{Fig:1} \bf Fig.~1}: 
\sffamily{Penrose diagram of $dS_4$. 
The horizontal axis is the polar coordinate $\Theta$ 
of\\  the global $S^3$, with north and south poles defined by
the points $\Theta=0, \pi$, respectively. \\
The dashed lines denote the future and past cosmological horizons $\mathcal H^{\pm}$.
}};
\end{tikzpicture}
\end{center}

\subsubsection{Static patch and Rindler wedges}\label{Rwedge}
It is possible to fulfill the hyperboloid condition \eqref{hyperboloid} by a different choice of embedding coordinates
\begin{equation}\label{static}
X^0= \sqrt{\ell^2- \hat r^2} \sinh(\hat t/\ell)~, \quad
X^1=\sqrt{\ell^2- \hat r^2} \cosh(\hat t/\ell)~,\quad 
X^i = \hat r \hat y^i ~,
\end{equation}
where 
\begin{align}
-\infty<\hat t<\infty,\qquad 0\leqslant \hat r <\ell, \qquad \sum_{i} (\hat y^i)^2=1~, \qquad i=2,3,4~.
\end{align}
The above parametrization 
yields the line element
\begin{equation}
\label{staticmetric}
ds^2|_{\mathfrak R_S}= 
- \left(1-\frac{\hat r^2}{\ell^2}\right) d\hat t^2 
+ \frac{d\hat r^2}{1 -  \frac{\hat r^2}{\ell^2}}    
+ \hat r^2 d\hat\Omega^2_2~.
\end{equation}
This describes the worldline of an observed $\mathcal O$ located at $\hat r = 0$. It can be seen that this metric has a manifestly timelike Killing vector and that this observer is surrounded by a cosmological horizon $\mathcal H$, placed at $\hat r = \ell$.
This souther inertial observer is causally connected to its southern Rindler wedge 
\begin{align}
\mathfrak R_S \equiv \mathcal O^+ \cap \mathcal O^- ~,
\end{align}
where $\mathcal O^\pm$ corresponds to the causal future or past of the southern observer $\mathcal O_S$. This Rindler wedge represents the set of all points in $dS_4$ where he can send signals to and receive signals from $\mathcal O$. Its boundary defines the cosmological horizon that surrounds the observer 
\begin{align}
\partial\mathfrak R_S \equiv \mathcal H~.
\end{align}
The worldline corresponds, in the embedding coordinates, to 
\begin{align}
X^1 (\mathcal O) = \sqrt{\ell^2 + (X^0 (\mathcal O))^2}~, \qquad X^i (\mathcal O) = 0~.
\end{align}
The  Killing  vector $\partial_{\hat{t}}$ is  timelike  inside $\mathfrak R$ and null  along  the (Killing) horizon $\mathcal H$~.
\begin{center}
\begin{tikzpicture}
\fill[fill=blue!10] (3,0)--(1.5,1.5)--(3,3);
\node at (2.4,1.5) {$\mathfrak R_S$};
\fill[fill=yellow!10] (0,0)--(1.5,1.5)--(0,3);
\node at (0.7,1.5) {$\mathfrak R_N$};
%Iplus
\draw [thick] (0,3)--node[above]{$\mathbf{\mathcal I^+}$}(3,3);
\draw [thick] (0,0) --node[below] {$\mathbf{\mathcal I^-}$}(3,0); 
%South                                
\draw [thick] (3,0)--(3,3);
%\node[right, rotate=-90] at (4.3,2.8) {\footnotesize South pole}; 
\node [red!50] at (3,1.5) {\textbullet};
\node at (3.5,1.5) {$\mathcal O_S$};
%\node[right, rotate=-90] at (3.75,2.5) {$\hat r=0$};
%North
\draw [thick] (0,0) --(0,3);
\node [red!50] at (0,1.5) {\textbullet};
\node at (-0.4,1.5) {$\mathcal O_N$};
%\node[right, rotate=90] at (-0.3,1) {\footnotesize North pole}; 
%
\draw [thick, dashed] (0,0) --(3,3); 
\node at (1.1,2.4) {$\mathcal H$};
\node at (2,2.4) {$\mathcal H$};
\draw [thick, dashed] (0,3)--(3,0);
%Caption~~~~~~~~~~~~~~~~~~~~
\node[text width=8cm, text justified] at (9,1.5){
\small {\hypertarget{Fig:2} \bf Fig.~2}: 
\sffamily{
Rindler wedge $\mathfrak R_S=\mathcal O^- \cap \mathcal O^+$
of an observer $\mathcal O_S$ at the south pole.
The cosmological horizon $\mathcal H=\partial\,\mathfrak R_S$ 
is a bifurcated Killing horizon, with bifurcation point the 
intersection $\mathcal H^+\cap\mathcal H^-$ at center of the 
diagram.
}};
\end{tikzpicture}
\end{center}
It has been shown in \cite{Gibbons:1977mu} that the cosmological horizon shares thermodynamical properties with black holes \cite{Bekenstein:1974ax, Bekenstein:1973ur, Bekenstein:1972tm, Bardeen:1973gs, Hawking:1974sw, Hawking:1976de}. This can be seen by zooming in the horizon region by using a small dimensionless parameter $\varepsilon << 1$, 
\begin{align}
\hat r = \ell\left( 1 - \frac{\varepsilon^2}{2} \right)~,
\end{align}
such that $\hat r \rightarrow \ell$ when $\varepsilon \rightarrow 0$. The line element is approximated by the Rindler space as
\begin{align}
ds^2 \approx \ell^2\left(- \frac{\varepsilon^2}{\ell^2}d\hat t^2 + d\varepsilon^2\right) + \dots~.
\end{align}
Demanding the absence of conical defects (or excess) in the Euclidean sector we obtain
\begin{align}\label{TdS}
T_{dS} = \frac{1}{2\pi \ell}~,
\end{align}
with a conjugated Gibbons-Hawking entropy
\begin{align}\label{dSS}
\boxed{\mathcal S_{dS} = \frac{\mathcal A_{\mathcal H}}{4\hbar G_4} = \frac{\pi \ell^2}{\hbar G_4}}~, 
\end{align}
which follows the Bekenstein area law of black hole horizons. 

\section{Antipodal defects and de Sitter space}\label{SubSec:Deformation} 
We present a different set of embedding coordinates that also fulfill the hyperboloid equation \eqref{hyperboloid} but cover both northen and southern Rindler wedges and describe a pair of causally disconnected antipodal observers. 

\subsection{Maximally extended coordinates}\label{Sub:maxext}
Parametrizing the embedding coordinates as 
\begin{align}
\label{embedding}
X_0 = &\sqrt{\ell^2 - \xi^2}\,\cos\theta\,\sinh(t/\ell)~,\quad
X_1 =\sqrt{\ell^2 - \xi^2}\,\cos\theta\,\cosh(t/\ell)~,  \\ \nonumber
X_2 &= \xi\,\cos\theta ~,\quad
X_3 =\ell \sin\theta \cos\phi~,\quad
X_4 = \ell \sin\theta \sin\phi ~,
\end{align}
with the coordinates running as 
\begin{equation}
-\infty<t<\infty~,\quad 
-\ell<\xi<\ell~,\quad 
0\leq \theta\leq\pi~,\quad  
0\leq\phi<2\pi~.
\end{equation}
This choice leads to the following line element
\begin{equation}
\label{metric}
ds^2=\ell^2 ( d\theta^2 + \sin^2\theta \, d\phi^2 )
+\cos^2\theta \left[- \left(  1 -  \frac{\xi^2}{\ell^2}\right) dt^2 
+ \frac{d\xi ^2}{1 -  \frac{\xi^2}{\ell^2}} \, \right]~,
\end{equation}
what describes the union of the Rindler wedges of two inertial observers. It follows that the Rindler wedges are causally disconnected at $\theta=\pi/2$, and the observers are then defined at
\begin{align}\label{NSconventions}
\theta\big\rvert_{\mathfrak R_N} \in \bigg[ 0 , \frac{\pi}{2}\bigg)~, \qquad \theta\big\rvert_{\mathfrak R_S} \in \bigg( \frac{\pi}{2}, \pi \bigg]~.
\end{align}
We can also check that the timelike isometry, defined by the Killing vector
\begin{align}\label{tkilling}
\partial_t^2 = \frac{\cos^2\theta}{\ell^2}\left(\xi^2 - \ell^2\right)~,
\end{align}
flips the sign when passing from one Rindler wedge to other, and vanishes at $\xi = \pm \ell$. This defines the bifurcated Killing horizon $\mathcal H$ which has a fixed time topology of a $2-$sphere of radius $\ell$. 
\\ 

This topology corresponds of a fibration of $dS_2$ over $S^2$, with manifest $SL(2,\mathbb R)$ isometry of $dS_2$, which contains the timelike generator \eqref{tkilling}, and breaks the $SO(3)$ symmetry of the $S^2$ down to $U(1)$. Then the isometries in the equator $\theta = \pi/2$ corresponds to $U(1)$, for the north and south poles its $SL(2,\mathbb R)$ and for every other point $SL(2,\mathbb R)\times U(1)$.
\\

Comparing the parametrizations of the embedding coordinates 
$X^0$ and $X^1$ in equations~\eqref{global} and~\eqref{embedding}, 
we find 
\begin{equation}\label{transf}
\cosh^2(T/\ell)\sin^2\Theta
=\frac{\xi^2}{\ell^2}\cos^2\theta + \sin^2\theta ~, \qquad
\frac{\sinh^2(T/\ell)}{1-\cosh^2(T/\ell)\sin^2\Theta}
= \sinh^2(t/\ell)~.
\end{equation}
From the first equation (and by the choice \eqref{NSconventions}), 
it follows that the worldlines of $\mathcal O_N$ and ${\cal O}_S$ 
are mapped to the north and south poles of the global $S^3$, 
located at $\Theta=0, \pi$, respectively.
From the second equation and by comparing the signs of $X^0$, 
it also follows that
\begin{equation}
t|_{\Theta=0}=T~,\qquad 
t|_{\Theta=\pi}=-T~,
\end{equation}
that is, the local time runs forwards in 
$\mathfrak R_N$ and backwards in $\mathfrak R_S$.

By combining \eqref{transf} with \eqref{tkilling} we can compare the cosmological horizon $\mathcal H$ with the global coordinates
\begin{equation}
\sin^2\Theta=\cos^2\tau~,
\end{equation}
and check that form two diagonals on the Penrose diagram, \emph{viz.}
\begin{equation}
\mathcal H = \partial (\mathfrak R_N \cup \mathfrak R_S)~.
\end{equation}
\\

By using the parametrization of the (transversal section) $2$-sphere 
\begin{equation}\label{sphereparam}
d\hat\Omega^2_2=d\hat\theta^2 + \sin^2\hat\theta d\hat\phi^2~.
\end{equation}
Comparing the maximally-extended embedding coordinates \eqref{embedding} and the static patch \eqref{static}, 
\begin{equation}\label{map1}
\hat t= {\rm sign}(\cos\theta)\,t~,\qquad
\hat \phi=\phi~,
\end{equation}
and 
\begin{equation}\label{map2}
\tan \hat \theta=\frac{\ell\tan\theta}{\xi}~,\qquad 
\hat r= \sqrt{\ell^2\sin^2\theta+\xi^2\cos^2\theta}~.
\end{equation}
The last two relations can be inverted as follows
\begin{equation}\label{map3}
\sin \theta =\frac{\hat r\sin\hat\theta}{\ell}~,\qquad 
\xi=\frac{\ell\hat r \cos \hat \theta}{\sqrt{\ell^2
-\hat r^2\sin^2\hat \theta}}~.
\end{equation}

Transformations~\eqref{map1}--\eqref{map3} make explicit 
the map from the regions of the double Rindler wedge with 
$0\leq \theta<\pi/2$ and $\pi/2 <\theta\leqslant \pi$, to
$\mathfrak R_N$ and $\mathfrak R_S$, respectively.

%%%%%%%%%%%%%%%%%%
\vspace{0.5cm}
\begin{center}
\begin{tikzpicture}[scale=0.9]
%North Hemisphere
\fill[fill=yellow!10] (-5,2)-- (-3,2)    
arc (0:180:20mm);                        
%South Hemisphere
\fill[fill=blue!10] (-5,2)-- (-3,2)    
arc (0:-180:20mm); 
\fill[fill=yellow!10] (-3,2) 
arc[start angle=0,end angle=-180, 
x radius=2, y radius=0.5];
\fill[fill=yellow!10] (-5,4)--(-7,2)--(-3,2);
\draw [thick] (-5,2) circle (2cm);
\node [red!60] at (-5,4) {\textbullet};
\node at (-5,4.4) {$\mathcal O_N$};
\node [red!60] at (-5,0) {\textbullet};
\node at (-5,-0.4) {$\mathcal O_S$};
%\node at (-7,4) {$\mathcal H\cong S^2$};
\draw[thick, dashed] (-3,2) 
arc[start angle=0,end angle=180, 
x radius=2, y radius=0.5];
\draw[thick] (-3,2) 
arc[start angle=0,end angle=-180, 
x radius=2, y radius=0.5];
%Penrose
\draw [thick, ->] (-2.5,2) -- (-1.2,2);
\fill[fill=yellow!10] (0,0)--(2,2)--(0,4);
\fill[fill=blue!10] (4,0)--(2,2)--(4,4);
%Iplus
\node [red!60] at (0,2) {\textbullet};
\node [red!60] at (4,2) {\textbullet};
%\node [darkblue] at (2,2) {\textbullet};
%\node [darkred] at (4,2) {\textbullet};
%\draw [thick,darkblue](0,2)--(4,2);
\draw [thick] (0,4) --node[above] {$\mathbf{\mathcal I^+}$}(4,4);
\draw [thick] (0,0) --node[below] {$\mathbf{\mathcal I^-}$}(4,0);                        
\draw [thick] (4,0)--(4,4); 
\node at (4.5,2) {$\mathcal O_S$};
\draw [thick] (0,0) --(0,4);
\node at (-0.5,2) {$\mathcal O_N$};
\node at (1,2) {$\mathfrak R_N$};
\node at (3,2) {$\mathfrak R_S$};
%\node[right, rotate=90] at (-0.3,1) {\footnotesize North pole}; 
%
\draw [thick, dashed] (0,0) --(4,4); 
%\draw [thick] (2,2)--node[midway, above, sloped] {$\mathcal H$}(4,0);
\draw [thick, dashed] (0,4)--(4,0);
%CAPTION
\node[text width=16cm, text justified] at (-0.6,-3){
\small {\hypertarget{Fig:3} \bf Fig.~3}: 
\sffamily{
Global depiction of the double Rindler wedge coordinates. 
The worldlines \\ of the observers $\mathcal O_N$ and $\mathcal O_S$ 
are sent to the global north and south poles, 
respectively. 
\\ The north hemisphere of the 2-sphere is mapped to the northern 
Rindler wedge $\mathfrak R_N$. \\ Likewise, the south hemisphere 
is sent to the southern Rindler wedge $\mathfrak R_S$.
}}; 
\end{tikzpicture}
\end{center}
Henceforth we will refer to this geometry as the de Sitter spacetime and will be deformed in the following section.
%%%%%%%%%%%%%%
%%%%%%%%%%%%%%
\section{Orbifolding the Horizon}\label{Sec:conical}
By considering the maximally-extended static patch \eqref{metric} it is possible to generate what is known as the spindle geometry \cite{thurston2014three}.  
\\

A two dimensional orbifold $\Gamma_n = \Gamma (q_1, q_2)$ corresponds to a Riemann surface $\Gamma$ that can be endowed with a metric structure for particular values of the anisotropy parameters $q_i$, that has $n$-marked points, referred as orbifold points. Near an orbifold point it is possible to identify cyclically by $z_i \in \mathbb C /\mathbb Z_{q_i}$ with $q_i$ non-zero positive integers
\begin{align}
z_i \sim \exp\left\{\frac{2\pi i}{q_i}\right\}z_i~.
\end{align}
A possible definition for a Riemmanian metric $g(\Gamma_n)$ can be obtain by imposing compatibility with the orbifold structure such that, in polar coordinates $z_i = r\exp\{i\phi\}$, near an orbifold point  one has
\begin{align}
g = dr^2 + \frac{r^2}{q_i^2}d\phi^2~.
\end{align}
When the anisotropy parameters are just $q_i = 1$ we recover the metric of the cone. Therefore, we can see that the anisotropy parameter induces conical singularities on $g$ at every orbifold point, with deficit angles
\begin{align}\label{deficit}
\Delta \phi_i = 2\pi\left(1 - \frac{1}{q_i}\right)~.
\end{align}
Then, assuming analytic continuation of the anisotropy parameter $q$ to the real numbers, in the limit $q_i \rightarrow 1$, the orbifold points $x_i$ becomes non-singular marked points. 
\\

It is possible to achieve an orbifold as the quotient $\Gamma = \hat\Gamma /\Sigma$, with $\hat\Gamma$ an smooth surface and $\Sigma$ a discrete group. The particular case of an $\hat\Gamma = S^2$ will be used to define the orbifolded cosmological horizon, this particular two-dimensional orbifold recovers the called \textit{spindle geometry} $S^2(q_1,q_2)$.
\\
 
 An orbifold is called \textit{good} orbifold, if it has some covering orbifold which is a manifold and \textit{bad} orbifold if it does not. An example of a \textit{bad} orbifold is teardrop

\begin{center}
\begin{tikzpicture}
node [red] at (1,17) {\textbullet};
node at (2,18) {$\mathbb Z_q$}
\path (-0.63,0.17) pic {raindrop};
 %CAPTION
\node[text width=16cm] at (1,-3.5){
\small {\hypertarget{Fig:4} \bf Fig.~4}: 
\sffamily{The underlying space for a teardrop is $S^2$. The set of marked points \\ consists of a single point, whose neighborhood is modeled on $R^2 / \mathbb Z_q$, where \\ $\mathbb Z_q$ acts by rotations.}}; 
\end{tikzpicture}
\end{center}
Comparing the upper and lower halfs of the teardrop it is no possible to have connected covering of both hemispheres and therefore it is not possible to generate a covering manifold orbifold. 
It can be proof that if $M$ is a manifold and $\Sigma$ is a group acting properly discontinuously on $M$, then $M/\Sigma$ has the structure of an orbifold.\\

For the $S^2$ with two singular marked points, each one associated with groupd $\mathbb Z_{q_1}$ and $\mathbb Z_{q_2}$ is a \textit{bad} orbifold unless $q_1 = q_2 = q$. The case of more than two marked points on $S^2$, this always corresponds to a \textit{good} orbifold.
\\

We are interested in the case when we can endow a metric structure onto this case. The case when $q_1 = q_2 = q$ corresponds to a \textit{good} orbifold and can be expressed as
\begin{align}
S^2(q,q) \cong S^2 /\mathbb Z_q~. 
\end{align}
By using the parametrization \eqref{sphereparam} of the unit two-sphere, the spindle metric can be taken to be
\begin{align}
ds^2_{spindle} = d\theta^2 + \frac{\sin^2\theta}{Q(\theta)^2}d\phi^2~,
\end{align}
with $\theta= 0,\pi$ marked points and $Q(\theta)$ is an arbitrary smooth positive function that satisfy the asymptotic behaviour \cite{Geroch:1987aa}
\begin{equation}\label{asymp}
Q(\theta)=\left\{ 
\begin{array}{cc} 
q + \mathcal O(\theta^2)~, 
&\theta\rightarrow 0~, \\[1mm]
q + \mathcal O((\theta-\pi)^2)~,
&\theta\rightarrow \pi~.
\end{array} \right.
\end{equation}
As this is the exact geometry presented by the maximally extended coordinates, we follow to generate the spindle geometry onto the cosmological horizon $\mathcal H \cong S^2$ by choosing two $\mathbb Z_q$ orbifold point with same anisotropy parameter $q$. As a result the $dS_4$ geometry \eqref{metric} is deformed with conical singularities as
\begin{align}
ds^2 = \ell^2 ds_{spindle}^2 + w^2 ds^2_{dS_2}~,
\end{align}
where the warp factor $w = \cos\theta$ that satisfy the boundary conditions
\begin{align}\label{bc}
w^2 \big\rvert_{\theta = 0, \pi} = 1~, \qquad (w^2)' \big\rvert_{\theta = 0, \pi} =0~.
\end{align}
The resulting geometry will be denoted by 
\begin{align}
\widehat{dS}_{4} \equiv dS_4 / \mathbb Z_q~.
\end{align}
The induced two dimensional geometry at the orbifold points, is given by the boundary conditions \eqref{bc}
\begin{equation}
\label{SigmaNS}
(\Sigma_N, h)
=\big(\widehat{dS}_4, g\big)\big|_{\theta=0}~,\quad
(\Sigma_S, h)
=\big(\widehat{dS}_4, g\big)\big|_{\theta=\pi}~,
\end{equation}
with induced metric structure 
\begin{equation}
\label{h}
h =- \left(  1 -  \frac{\xi^2}{\ell^2}\right) dt^2 
+ \frac{d\xi^2}{   1 -  \frac{\xi^2}{\ell^2} } ~.
\end{equation}
We will refer to the codimension two submanifolds $(\Sigma_N, h)$ 
and $(\Sigma_S, h)$ defined in~\eqref{SigmaNS} as \emph{defects}, 
and to the space filling up the region between defects as the \emph{bulk}\footnote{The metric on the defects corresponds to dS$_2$ what has an isometry group the Lorentz group in three dimensions what is locally isomorphic to the special linear group $SO(1,2)\cong SL(2,\mathbb R)/Z_2$ exactly as in the AdS$_2$ case. }.
\\

The conical deficit induced by the anisotropy parameter $q$ contributes with a $\delta-$function distribution to the Riemann tensor on each orbifold point \cite{Fursaev:1995ef}. The resulting Ricci scalar reads
\begin{align}
R^q = R - \sum_{i = N,S} 4\pi\left( 1 - \frac{1}{q}\right)\delta \Sigma_i~.
\end{align}
with $\delta_{\Sigma_i}$ a projector on the defects. By integrating the Ricci scalar in the singular manifold $\widehat{dS}_4$ we generate the Einstein-Hilbert action coupled to a pair of Nambu-Goto action (one for each defect), \textit{viz}
\begin{align}\label{Itotal}
I[\widehat{dS}_4] &
\equiv \frac{1}{16\pi G_4}\int_{\widehat{dS}_4}d^4 x \sqrt{-g}
\Big( R^{(q)}-\frac{6}{\ell^2}\Big)\nonumber\\
&~=\frac{1}{16\pi G_4} 
\int_{\widehat{dS}_4\setminus(\Sigma_N\cup\Sigma_S)}
d^4 x \sqrt{-g}\Big( R- \frac{6}{\ell^2}\Big) -\sum_{i=N, S} 
\mathcal T_q\int_{\Sigma_i} d^2 y\sqrt{-h}~,
\end{align}
which is simply a bulk integral excluding the defects, plus two Nambu-Goto actions that has been added in order to solve the Einstein equations. Actually these two defect actions corresponds only to a fix metric action. Therefore there is no dynamics on the codimension-2 integrals and corresponds to a fixed Nambu-Goto string. These both are coupled to the tension 
\begin{align}\label{braneT}
\mathcal T_q = \frac{1}{4G_4}\left(1- \frac{1}{q} \right)~.
\end{align}
The tension contribution to the stress-tensor which is localized on the defects
\begin{align}\label{Tij}
T_{ij}^N =T_{ij}^S = \mathcal T_q h_{ij}~,
\end{align}
As said above, these stress tensor are supported on each defects. The orbifolded horizon is schematically illustrated in Figure~\hyperlink{Fig:5}5

\begin{center}
\begin{tikzpicture}[scale=0.9]
%Sphere
\draw [thick] (-4,0) circle (2cm);
\node [darkred] at (-4,2) {\textbullet};
\node at (-4,2.3) {$x_N$};
\node [darkred] at (-4,-2) {\textbullet};
\node at (-4,-2.3) {$x_S$};
\node at (-6,2) {$\mathcal H\cong S^2$};
\draw[thick, dashed] (-2,0) 
arc[start angle=0,end angle=180, 
x radius=2, y radius=0.5];
\draw[thick] (-2,0) 
arc[start angle=0,end angle=-180, 
x radius=2, y radius=0.5];
%Orbifold arrow
\node at (-0.5,0.4) {$S^2/\mathbb Z_q$};
\draw [thick, ->] (-1.3,0) -- (0.3,0);
%Spindle
\fill[fill=yellow!10] (1,2.3)--(0,1.7)--(5,1.7)--(6,2.3);
\fill[fill=blue!10] (1,-1.7)--(0,-2.3)--(5,-2.3)--(6,-1.7);
\draw [thick] (1,2.3) --(6,2.3);
\draw [thick] (0,1.7) --(5,1.7);
\draw [thick] (0,1.7) --(1,2.3);
\draw [thick] (6,2.3) --(5,1.7);
\coordinate (N) at (3,2);\coordinate (S) at (3,-2);
\node [darkred] at (3,2) {\textbullet};
\node [darkred] at (3,-2) {\textbullet};
\draw[thick] (N)to[out=-20,in=20](S);
\draw[thick] (N)to[out=-150,in=150](S);
%\draw [thick](3.05,0) ellipse (1.05cm and 0.3cm);
\draw[thick, dashed] (4.1,0) 
arc[start angle=0,end angle=180, 
x radius=1.05, y radius=0.3];
\draw[thick] (4.1,0) 
arc[start angle=0,end angle=-180, 
x radius=1.05, y radius=0.3];
\node at (6,1.7) {$(\Sigma_N, h)$};
\node at (0.1,-1.6) {$(\Sigma_S, h)$};
\draw [thick] (1,-1.7) --(6,-1.7);
\draw [thick] (0,-2.3) --(5,-2.3);
\draw [thick] (0,-2.3) --(1,-1.7);
\draw [thick] (6,-1.7) --(5,-2.3);
\draw [thick, ->] (3.05,0) -- (4.05,0);
\node at (5.1,0) {$\ell_q=\ell/q $};
%CAPTION
\node[text width=16cm, text justified] at (0,-5) 
{\small {\hypertarget{Fig:5}\bf Fig.~5}: 
\sffamily{
dS horizon with two marked points $x_N$ and $x_S$.
The orbifolding of $\mathcal H$ deforms \\ the spherical geometry to $S^2/\mathbb Z_q$ spindle, with radius $\ell_q\equiv\ell/q$. }
};
\end{tikzpicture}
\end{center}
The geometry of de Sitter space is recovered in the limit $q\rightarrow 1$. We will refer to it as \textit{tensionless limit} henceforth.
\begin{align}\label{tensionless}
\widehat{dS}_4 \big\rvert_{q=1} = dS_4~.
\end{align}
Another feature of the metric \eqref{metric} is that by the transformation
\begin{align}
t &\rightarrow i \ell \varphi ~, \qquad\, 0\leq\varphi<2\pi, \\
\xi &\rightarrow \ell\cos\vartheta~,\quad 0\leq\vartheta\leq\pi~,
\end{align}
this cast the metric into 
\begin{equation}\label{gE}
\frac{ds^2}{\ell^2}=  d\theta^2 + \sin^2\theta \, d\phi^2 
+ \cos^2\theta (d\vartheta^2 + \sin^2\vartheta \, d\varphi^2)~,
\end{equation}
which corresponds to the geometry of a direct product
\begin{equation}\label{S2wS2}
g = S_L^2 \times_w S_R^2~, 
\end{equation}
where the left and right 2-spheres (labeled by $L$ and $R$) are identical, 
and they both admit a $\mathbb Z_q$ action (discrete $\phi$ and $\varphi$ 
identifications). Then, when actually implementing the $\mathbb Z_q$ action, 
there are two possible choices. This will be used to describe the entropy of $\mathcal H$ as entanglement entropy in \autoref{Chp:dSEntanglement}.
%

%% file: Chapters/CFT.tex
% Chapter 1

\chapter{Conformal Field Theories} % Main chapter title
\label{CFT} % For referencing the chapter elsewhere, use \ref{Chapter1} 

%----------------------------------------------------------------------------------------

%----------------------------------------------------------------------------------------

A short review on conformal field theories in two- and higher-dimensions is given. This review is based on \cite{Di_Francesco_1997, Polchinski_1998, Ginsparg:aa, Ginsparg:ab, Qualls:2015aa, Schellekens_1996}.

\section{Conformal Group in $d\geq 3$}
For a metric tensor $g_{\mu\nu}$ describing a $d$-dimensional spacetime, we can define the conformal transformations of the coordinates as the invertible map $x\rightarrow\bf{x'}$ that would leave the metric invarian up to a scale factor, this can be read as
\begin{equation}\label{conftransf}
g'_{\mu\nu}({x}') = \Omega({x})g_{\mu\nu}({x}).
\end{equation}
This transformation is locally equivalent to a (pseudo) rotation and a dilation. The set of conformal transformations forms a group, which includes Poincar\'e as a sub-group, since this last is achieved if the scale $\Omega = 1$. 

Now, let us continue by the study the change of the conformal transformation by transforming the coordinates by $x^\mu \rightarrow x'^\mu = x^\mu + \varepsilon^\mu({\bf {x}})$, with $\varepsilon({x})$ as the local infinitesimal parameter. The metric transform as
\begin{align}
g_{\mu\nu}({x}) \rightarrow g'_{\mu\nu}({x}) = \frac{\partial x^\rho}{\partial x'^\mu}\frac{\partial x^\sigma}{\partial x'^\nu}g_{\rho\sigma}
\end{align}
Usually, if a transformation of the metric is such that maps a point on a manifold $\mathcal M$ into a different one, and leaves the metric invariant, it is called an \textit{isometry}
\begin{align}\label{isometry}
g'_{\mu\nu}({x}') = g_{\mu\nu}({x}'),
\end{align}
which, for infinitesimal transformations $x^\mu \rightarrow x'^\mu = x^\mu + \varepsilon^\mu(x)$ reads
\begin{align}\label{metricunderinf}
g'_{\mu\nu} \equiv{}& \frac{\partial x^\rho}{\partial x'^\mu}\frac{\partial x^\sigma}{\partial x'^\nu}g_{\rho\sigma} \nonumber \\ ={}& \left( \delta^\rho_\mu - \partial_\mu \varepsilon^\rho\right)\left( \delta^\sigma_\nu - \partial_\nu \varepsilon^\sigma\right) \nonumber \\ ={}& g_{\mu\nu} - \left( \partial_\mu\varepsilon_\nu + \partial_\nu\varepsilon_\mu\right) + \mathcal O(\varepsilon^2).
\end{align}
As is mentioned above, a isometry satisfy the condition \eqref{isometry}. Using \eqref{metricunderinf}, and dropping the second order contribution on the infinitesimal parameter, we obtain the \textit{Killing equation}
\begin{align}
\nabla_\mu \varepsilon_\nu + \nabla_\nu \varepsilon_\mu = 0,
\end{align}
and the exact solutions to this equation are called \textit{Killing vectors}. This equation corresponds to the vanishing of the Lie derivative of the metric along the Killing vector $\xi_\mu$. 
For the case of conformal symmetry, this acquires an extra term that comes from \eqref{conftransf} of the form
\begin{align}\label{f1}
\partial_\mu\varepsilon_\nu + \partial_\nu\varepsilon_\mu = f({x})g_{\mu\nu},
\end{align}
where $f({x})$ is a function of the coordinates that remains to be determined. In order to do this, we follow to take trace of Eq.\eqref{f1} and use $g_{\mu\nu}g^{\mu\nu} = d$
\begin{align}\label{f2}
f({x}) = \frac{2}{d}\partial_\alpha \varepsilon^\alpha.
\end{align}
For sake of simplicity, we will continue using the Minkowski metric, i.e. $g_{\mu\nu} = \eta_{\mu\nu} = \text{diag}(-1,1,\dots,1)$. Applying an extra derivative on \eqref{f1}, making a cyclic permutation of the indices and combine the resulting equations, we can write
\begin{align}
\partial_\mu \partial_\nu \varepsilon_\alpha = \frac{1}{2}\left(\eta_{\mu\alpha}\partial_\nu f + \eta_{\nu\alpha}\partial_\mu f - \eta_{\mu\nu}\partial_\alpha f \right),
\end{align}
and contracting the $\mu,\nu$-indices
\begin{align}
\partial^2\varepsilon_\mu = \frac{2-d}{2}\partial_\mu f~.
\end{align}
The derivative of the $f$-function can be achieved by just by applying a derivative on \eqref{f1}, which leads to
\begin{align}\label{f3}
(2-d)\partial_\mu \partial_\nu f = \eta_{\mu\nu}\partial^2 f~,
\end{align}
here we can clearly see that two-dimensional CFTs are quite different. Contracting the indices with a metric tensor we get a final equation for $f$
\begin{align}\label{ffinal}
(d-1)\partial^2 f = 0.
\end{align}
We can immediately see that in one-dimensional spacetime any function $f$ is solution of the above equation. Then any smooth transformation is conformal in one dimension, what was expected by the fact that in one dimension is there no notion of angles. 

Another important point is manifested by \eqref{f3}, that conformal transformations in $d=2$ are also a particular case and will be revisited later. For $d\geq 3$, \eqref{ffinal} implies that $\partial_\mu\partial_\nu f = 0$. Then, $f$ is at most linear in the coordinates, and we can write the generic form
\begin{align}\label{fin3d}
f({x}) = A + B_\mu x^\mu,
\end{align}
with $A$ and $B_\mu$ constant. Applying this Ansatz to \eqref{f1}, we get that $\partial_\mu\partial_\nu\varepsilon_\alpha = \text{Constant}$ and therefore we can rewrite 
\begin{align}
\varepsilon_\mu = a_\mu + b_{\mu\nu}x^\nu + c_{\mu\nu\rho}x^\nu x^\rho,
\end{align}
with $c_{\mu\nu\rho}$ totally symmetric tensor.  

As these equations hold for all $x^\mu$, we may treat each power of the coordinate separately. It follows immediately that $a_\mu$ is an arbitrary free constant, and is related to infinitesimal translations. The second term, which is linear in the coordinates, plugged in \eqref{f2} implies that $b_{\mu\nu}$ must correspond to a sum of an antisymmetric part and a pure trace. This corresponds to an infinitesimal rigid rotation and an infinitesimal scale transformation, respectively. 
For $c_{\mu\nu\rho}$, the same analysis leads to  
\begin{align}
c_{\mu\nu\rho} = \frac{1}{d}\left(\eta_{\mu\rho}c^\alpha_{~\alpha\nu} + \eta_{\mu\nu}c^\alpha_{~\alpha\rho} - \eta_{\nu\rho}c^\alpha_{~\alpha\mu}\right).
\end{align} 
Defining $b_\mu \equiv \frac{1}{d}c^\alpha_{~\alpha \mu}$, we can now write the whole infinitesimal transformation $x^\mu \rightarrow x'^\mu = x^\mu + \varepsilon^\mu({x})$ completely
\begin{align}
x'^\mu = x^\mu + 2({x\cdot b})x^\mu - b^\mu {x}^2.
\end{align}
what is often called \textit{special conformal transformations} (SCT).
Also we have found the most general conformal transformation
\begin{align}
\varepsilon^\mu ({x}) = a^\mu + m^\mu_{~\nu} x^\nu + \alpha x^\mu + 2({\bf b\cdot x})x^\mu - b^\mu{x}^2,
\end{align}
with $m_{\mu\nu} = -m_{\nu\mu}$ and $\alpha$ a free constant. 

demanding that metric (restoring back to an arbitrary curved metric) is preserved under this infinitesimal transformation, up to an overall scale factor, namely
\begin{align}
g'_{\mu\nu}({x}') = \Omega^{-2}({x})g_{\mu\nu}({x}')
\end{align}
such that
\begin{align}
\Omega^2({x})g'_{\mu\nu}({x}) = \frac{\partial x'^\rho}{\partial x^\mu}\frac{\partial x'^\sigma}{\partial x^\nu}g_{\rho\sigma},
\end{align}
and under infinitesimal conformal transformations leads to the so-called \textit{Conformal Killing equation}
\begin{align}\label{confKilling}
\nabla_\mu \varepsilon_\nu + \nabla_\nu \varepsilon_\mu = \frac{2}{d}\nabla_\gamma \varepsilon^\gamma g_{\mu\nu}({x})~,
\end{align}
where the conformal factor in this case, was also fixed by taking the trace of the above equation.
We now can fin the corresponding generators. As it is usual, let us defined them by
\begin{align}
\delta \phi ({x}) \equiv \phi'({x}) - \phi({x}) = -i\left(a^\mu P_\mu + \frac{1}{2}m_{\mu\nu}L_{\mu\nu} + \alpha D + b^\mu K_\mu \right)\phi(x)~,
\end{align}
where $\phi'({x})$ is a field obtained from $\phi'({x})$, by acting with an infinitesimal transformation. To obtain the generators, we will consider the simplest case, which corresponds to the scalar field, which satisfy 
\begin{align}
\phi'({x}') = \phi({x})~.
\end{align}
For the infinitesimal transformation on the coordinates follows
\begin{align}
\delta\phi({x}) = -\varepsilon^\mu({x})\partial_\mu \phi({x}).
\end{align}
Plugging the explicit form of $\varepsilon^\mu({x})$ we obtain the generator
\begin{align}\label{generators}
P_\mu = -i\partial_\mu, \quad L_{\mu\nu} = i\left(x_\mu\partial_\nu - x_\nu\partial_\mu \right),\quad D=-ix^\mu\partial_\mu,\quad K_\mu = i\left(x^2\partial_\mu - 2x_\mu x^\nu \partial_\nu \right). \nonumber \\
\end{align}
We can recognize the first three, as the operators that generate translations, rotations and dilatations respectively. The last one is the aforementioned SCT generator.
The finite version of the generators is given by the exponentiation of this last
\begin{align}
x'^\mu ={}& \exp\{i a^\nu P_\nu\}x^\mu &&= x^\mu + a^\mu, &&&\text{translations}\nonumber \\
x'^\mu ={}& \exp\{\frac{i}{2}m^{\rho\nu}L_{\rho\nu}\}x^\mu &&= M^\mu_{~\nu}x^\nu  &&&\text{rotations}\nonumber \\
x'^\mu ={}& \exp\{i\alpha D\}x^\mu &&= \alpha x^\mu, &&&\text{dilations} \nonumber \\
x'^\mu ={}& \exp\{ib^\nu K_\nu\}x^\mu &&= \frac{x^\mu - b^\mu {x}^2}{1 - 2({\bf b\cdot x}) + {\bf b}^2{x}^2} &&&\text{SCT},
\end{align}
with $\det(M^\mu_{~\nu}) = 1$. The finite version of the SCT can be rewritten in the following form
\begin{align}
\frac{x'^\mu}{{x'}^2} = \frac{x^\mu}{{x'}^2} - b^\mu~,
\end{align}
This transformation is pictorially represented in Fig.~1
%SCTDRAW
\begin{figure}[h]
\begin{center}
\begin{tikzpicture}[scale=3]

\draw[<->, very thick] (-1.5,0) -- (1.5,0) coordinate (x axis);

\draw[<->, very thick] (0,-0.3) -- (0,1.3) coordinate (y axis);

\draw (-1,0) .. controls (-1,0.555) and (-0.555,1) .. (0,1)
               .. controls (0.555,1) and (1,0.555) .. (1,0);

\foreach \x/\xtext in {-1, 1}
\draw (\x cm,1pt) -- (\x cm,-1pt) node[anchor=north] {$\xtext$};

\draw[->]  (-0.3,0.2) -- (-0.8,0.9) 
	node (a) at (-0.92,1.05) {$\frac{x^\mu}{x\cdot x}$}
	node at (-0.22,0.15) {$x^\mu$};
\filldraw [red] (-0.3,0.2) circle (0.6pt);

\draw[->]  (-0.81,0.93) -- (0.72,0.93) 
	node[sloped,above] (b) {$\frac{x'^\mu}{x'\cdot x'}$}
	node at (-0.4,1) {$-b^\mu$};
\filldraw [blue] (-0.81,0.93) circle (0.6pt);

\draw[->]  (0.75,0.93) -- (0.35,0.38) 
	node at (-0.4,1) {$-b^\mu$}
	node at (0.35,0.27) {$x'^\mu$};
\filldraw [orange] (0.75,0.93) circle (0.6pt);
\filldraw [red!30] (0.33,0.35) circle (0.6pt);
\label{fig:6}
\end{tikzpicture}
\end{center}
\end{figure}
%Caption~~~~~
\begin{center}
\begin{tikzpicture}
\node[text width=16cm, text justified] at (0,0){
\small {\bf Fig.~6\label{Fig6}}: 
\sffamily{A pictorially finite Special Conformal Transformation. Here it is shown that the \\ finite SCT corresponds to an inversion, which maps a point in the circle for one \\ outside of it, a translation and finally an inversion again. Leading the point $x^\mu$ to \\ a different point $x'^\mu$. Original figure from \cite{Blumenhagen_2009}}};
\end{tikzpicture}
\end{center}
Due to the particular point $x^\mu = b^\mu/b^2$ which maps the point $x^\mu$ to infinity, we observe that the SCT are not globally defined.

Going back to the generators \eqref{generators}, this form the Conformal Algebra by obeying the following commutation relations
\begin{align}\label{CAlgebra}
[D, P_\mu] ={}& iP_\mu~, \nonumber \\
[K_\mu , D] ={}& iK_\mu~, \nonumber \\
[K_\mu, P_\mu] ={}& i2\left( \eta_{\mu\nu}D - L_{\mu\nu} \right)~, \nonumber \\
[K_\mu ,L_{\alpha\beta}] ={}& i\left( \eta_{\mu\alpha}K_\beta - \eta_{\mu \beta}P_{\alpha} \right)~, \nonumber \\
[L_{\mu\nu}, L_{\alpha\beta}] ={}& i\left(\eta_{\nu\alpha}L_{\mu\beta} - \eta_{\mu\alpha}L_{\nu\beta}  + \eta_{\mu\beta}L_{\nu\alpha} -  \eta_{\nu\beta}L_{\mu\alpha}  \right)~,
\end{align}
while the other ones commutes
\begin{align}
[P_\mu, P_\nu] = [L_{\mu\nu}, D] = [K_\mu , K_\nu] = 0~.
\end{align}
Then the Conformal Algebra contains as a sub-algebra the Poincar\'e algebra formed by the generators $P_\mu$ and $L_{\mu\nu}$ . 

It is possible to rewrite this commutation relations by defining a higher-dimensional generator $J_{AB}$ which is anti-symmetric $J_{AB} = -J_{BA}$, and $A,B \in \{-1,\mu,d\}$, with $d$ the dimension of the spacetime. This generator have the following properties
\begin{align}
J_{\mu\nu} ={}& L_{\mu\nu}, && J_{-1,\mu} = \frac{1}{2}(P_\mu - K_{\mu}) \nonumber \\
J_{-1,d} ={}& D, && J_{d,\mu} = \frac{1}{2}(P_\mu + K_\mu)~.
\end{align}  
This new generators satisfy the algebra of rotations in $d+1$ dimensions $so(1,d+1)$. The explicit algebra is off course
\begin{align}\label{LorentzAlgebra}
[J_{AB}, J_{CD}] = i\left(\eta_{AD}J_{CD} - \eta_{AC}J_{BD} + \eta_{BC}J_{AD} - \eta_{BD}J_{AC} \right)~.
\end{align}
This explicitly shows the isomorphism between the Lorentz group in $(d+1)-$dimensions, i.e. $SO(1,d+1)$, and the Conformal Group in $d$ dimensions.
\subsection{Representations of the Conformal Group}
An infinitesimal transformation
\begin{align}\label{infinitesimal}
x'^\mu ={}& x^\mu + \omega_a\frac{\delta x^\mu}{\delta \omega_a}~, \nonumber \\
\phi'(x') ={}& \phi(x) + \omega_a\frac{\delta \phi'(x')}{\delta \omega_a}~,
\end{align}
where $\{\omega_a\}$ is a set of infinitesimal parameters. Now we can define the \textit{generators} $G_a$ of a symmetry transformation by 
\begin{align}
\delta_\omega \phi(x) \equiv \delta'(x) - \delta(x) \equiv -i\omega_a G_a\phi(x)~.
\end{align}
And comparing with \eqref{infinitesimal} we obtain the following expression for the generators
\begin{align}
iG_a \phi(x) = \frac{\delta x^\mu}{\delta \omega_a}\partial_\mu \phi(x) - \frac{\delta \phi'(x')}{\delta \omega_a}~.
\end{align}
Now we can seek for a matrix representation $T_g$ of a conformal transformation parametrized by $\omega_g$, such that a field transforms as
\begin{align}
\phi'(x') = \left( 1- i\omega_g T_g\right)\phi(x)~.
\end{align}
To obtain the $T_g$ generators we will start by studying the subgroup of Poincare group that leaves the point $x=0$ invariant, which corresponds to the Lorentz group. Suppose that we have a field at the origin, that transform in some matrix representation of the Lorentz group in the following way
\begin{align}
L_{\mu\nu}\phi(0) = S_{\mu\nu}\phi(0)~.
\end{align}
Here $S_{\mu\nu}$ is some matrix. Now, it follows to translate the Lorentz generator away from the origin. This by using the translation operator $P_\mu$
\begin{align}
L_{\mu\nu} \phi(x) ={}& \exp\{ix^\mu P_\mu\}S_{\mu\nu} \, \phi(0)~ \nonumber \\
                             ={}& \exp\{ix^\mu P_\mu\}S_{\mu\nu} \exp\{-ix^\nu P_\nu\}\exp\{ix^\rho P_\rho \}\phi (0)~, \nonumber \\
                             ={}& \exp\{ix^\mu P_\mu\}S_{\mu\nu} \exp\{-ix^\nu P_\nu\}\phi(x)~.
\end{align}
It follows to apply the Baker-Campbell-Hausdorff formula, which for $A$ and $B$ some operators reads
\begin{align}\label{hausdorff}
\exp\{-A\}B \exp\{A\} = B + [B,A] + \frac{1}{2}[[B,A],A] + \dots~, 
\end{align}
which leads to
\begin{align}
\exp\{i x^\alpha P_\alpha\}L_{\mu\nu}\exp\{-ix^\alpha P_\alpha \} = S_{\mu\nu} -x_\mu P_\nu + x_\nu P_\mu~.
\end{align}
And using this, we can see that the generators act as
\begin{align}
P_\mu \phi(x) ={}& -i\partial_\mu \phi(x)~, \nonumber \\
L_{\mu\nu} \phi(x) ={}& \left[i\left(x_\mu\partial_\nu - x_\nu\partial_\mu \right) + S_{\mu\nu}\right]\phi(x)~.
\end{align}
By using the Hausdorff formula \eqref{hausdorff} we can use the commutation relations \eqref{CAlgebra} to express the generators as
\begin{align}
\exp\{ix^\nu P_\nu \}K_\mu \exp\{-ix_\nu P_\nu\} ={}& K_\mu + 2x_\mu D - 2x^\nu L_{\mu\nu} + 2x_\mu x^\nu P_\nu - x^2 P_\mu~, \nonumber \\
\exp\{ix^\nu P_\nu \}D\exp\{-ix_\mu P_\nu\} ={}& D + x^\alpha P_\alpha~.
\end{align}
In the full conformal algebra, only the subgroup of rotations, dilations and special conformal transformations leaves the origin invariant. We denote $S_{\mu\nu}, -i\Delta$ and $k_\mu$ the respective values of those generators at $x= 0$, \textit{viz}
\begin{align}\label{gentwo}
L_{\mu\nu}\phi(0) ={}& S_{\mu\nu}\phi(0)~, \nonumber \\
D\phi(0) ={}& -i\Delta\phi(0)~, \nonumber \\
K_\mu \phi(0) ={}& k_\mu \phi(0)~.
\end{align}
These values must form a matrix representation for the reduced algebra, the commutation relations are
\begin{align}
[k_\mu, k_\nu] ={}& 0~, \nonumber \\
[\Delta, S_{\mu\nu}] ={}& 0~, \nonumber \\
[\Delta, k_\mu] ={}& k_\mu~,\nonumber \\
[S_{\mu\nu}, k_\alpha ] ={}& -i\{\eta_{\alpha\mu}k_\nu - \eta_{\alpha\nu}k_{\mu}\}~, \nonumber \\
[S_{\mu\nu}, S_{\alpha\beta}] ={}& -i\{\eta_{\mu\alpha}S_{\nu\beta} - \eta_{\nu\alpha}S_{\mu\beta} + \eta_{\mu\beta}S_{\alpha\nu} - \eta_{\nu\beta}S_{\alpha\mu}\}~.
\end{align}
With this we can see from Eq.\eqref{gentwo} that we can obtain other transformation rules
\begin{align}
K_\mu \phi(x) ={}& \left( k_\mu - i2x_\mu\Delta - x^\nu S_{\mu\nu} - 2ix_\mu x^\nu\partial_\nu + ix^2 \partial_\mu \right)\phi(x)~, \nonumber \\
D \phi(x) ={}& -i\left(x^\mu\partial_\mu + \Delta \right)\phi(x)~.
\end{align}
If we now imply that the field belongs to an irreducible representation of the Lorentz group, then by the Schur's lemma, any quantity that is commuting with the generators $S_{\mu\nu}$, then that quantity must be a multipleof the identity operator. This forces to all the $k_\mu$ to vanish. As a consequence, $\Delta$ must be some number known as the \textit{scaling dimensions}. Thus,
\begin{align}
L_{\mu\nu}\phi(0) ={}& S_{\mu\nu}\phi(0)~,\nonumber \\
D\phi(0) ={}& -i\Delta \phi(0)~, \nonumber \\
K_\mu \phi(0) ={}& 0~.
\end{align}
A field that satisfies this last is called a \textit{primary}. Then, the lowering operator $K_\mu$ are annihilates the primary state at $x=0$, and we can obtain an infinite number of states called $descendants$ by cting repeatedly with the translation operator $P_\mu$ which corresponds to just derivatives of the primary. 
\begin{align}
DP_\mu \phi(0) ={}& \left([D,P_\mu] + P_\mu D \right)\phi(0)~, \nonumber \\
={}& \left(iP_\mu + P_\mu D \right)\phi(0)~, \nonumber \\
={}& -i(\Delta -1)P_\mu (\Delta -1)\phi(0)~. 
\end{align}
With this we can form the descendants. The field $\phi(x) = \exp\{ix^\mu P_\mu \}\phi(0)$ is not a descendant either, but an infinite linear combination of descendants. 

The scaling dimensions $\Delta$ of a field is defined by the action of a scale transformation 
\begin{align}
\phi(\lambda x) = \lambda^{-\Delta}\phi(x)~.
\end{align}
For the scale factor of \eqref{conftransf}, the Jacobian of the conformal transformation $x\rightarrow x'$ is
\begin{align}
\bigg\rvert\frac{\partial x'}{\partial x}\bigg\rvert = \frac{1}{\sqrt{det g'_{\mu\nu}}} = \Omega^{-d/2}~,
\end{align}
then it can be shown that a spinless field ($S_{\mu\nu} = 0$) transforms as 
\begin{align}\label{quasi}
\phi(x) \rightarrow \phi'(x') = \bigg\rvert\frac{\partial x'}{\partial x}\bigg\rvert^{-\Delta/d}\phi(x)~,
\end{align}
any field that transform like \eqref{quasi} is called a \textit{quasi-primary}. More generally, the transformation properties of a field are determined by $\Delta$ and its spin.
%%%%%%%%%%%%%%%%%%%%

\subsection{Symmetries and Stress-Energy Tensor}
%%%%%%%%%%%%%%%%%%%%
Considering an action functional 
\begin{align}
S[\phi] = \int d^dx \mathcal L(\phi, \partial\phi)~,
\end{align}
if we have a transformation $\phi(x)\rightarrow \phi'(x)$, we can call this transformation a symmetry if 
\begin{align}
S[\phi'] = S[\phi]~.
\end{align}
The symmetries play a fundamental role in physics. One of the most beautiful theorems in physics was proposed by the mathematician Emmy N\"{o}ther \cite{Noether:1918zz} which says that \emph{every differentiable symmetry of the action of a physical system has a corresponding conservation law}.
For a generic infinitesimal variation of the field 
\begin{align}
\phi'(x) = \phi(x) + \delta \phi(x)~,
\end{align}
acts on the action in the following form
\begin{align}
\delta S[\phi,\delta\phi] ={}& S[\phi'] - S[\phi] \nonumber \\ ={}& \int d^d x \left[ \frac{\partial\mathcal L}{\partial\phi} - \partial_\mu \left(\frac{\partial\mathcal L}{\partial_\mu \phi}\right) \right]\delta\phi + \int d^d x \partial_\mu \left(\frac{\partial\mathcal L}{\partial_\mu \phi}\delta\phi\right)~,
\end{align}
where the first term corresponds to the equations of motion. Going on-shell we finally get 
\begin{align}
\delta S[\phi,\delta\phi] \approx \int d^d x \partial_\mu \left(\frac{\partial\mathcal L}{\partial_\mu \phi}\delta\phi\right)~.
\end{align}
Therefore, the vanishing of this boundary term ensures that the transformation is a symmetry of the theory. In general we could shift the current by some vector $k^\mu$ that its proportional to the infinitesimal transformation parameter exactly as $\delta\phi$, let us called $\lambda^a$. Then the total-derivative term $\partial_\mu j^\mu$ reads
\begin{align}
j^\mu = k^\mu - \frac{\partial\mathcal L}{\partial_\mu \phi}\delta\phi~,
\end{align}
where the fields satisfy the field equations. The N\"other current is proportional to the infinitesimal transformation $j^\mu = \lambda^a J_a^\mu$, and all this parameters are independent from each other and each $J^\mu_a$ must be conserved. The $J^\mu_a$ are called  \emph{N\"other currents}.
\\
Also $Q_a$ are called \emph{N\"other charges} which are associated to each continuous symmetry
\begin{align}
Q_a = \oint d^{d-1}x~n_\mu j^\mu~.
\end{align}
 They satisfy, under Poisson brackets
\begin{align}
\delta_a \phi(x) = \{\phi(x), Q_a\}
\end{align}
if $\delta\phi$ is a symmetry of the action.
\\

One of the most important quantities in field theory is the Stress-Energy Tensor (or Energy-Momentum tensor), this will be particularly relevant in two-dimensional CFT's that are going to be described in \autoref{CFT2}. The Energy-Momentum Tensor, is the conserved current associated to translation invariance, whose components give the density and flux density of energy and momentum. For a Lagrangian of the form $\mathcal L = \mathcal L(\phi({x}), \partial_\mu \phi({x}))$, an infinitesimal transformation on the coordinates $x^\mu \rightarrow x^\mu + \varepsilon^\mu$ leads to the following canonical conserved current (See Appendix B).
\begin{align}\label{deltaSinf}
\delta S = \int d^d x~T^{\mu\nu}\delta g_{\mu\nu},
\end{align}
with
\begin{align}\label{canonicalTmunu}
T^{\mu\nu} = -g^{\mu\nu} \mathcal L + \frac{\partial\mathcal L}{\partial(\partial_\mu \phi)}\partial^\nu \phi~,
\end{align}
with a conservation law $\partial_\mu T^{\mu\nu} = 0$ satisfied. Now we will see how the conformal symmetry will be expressed in terms of Ward identities into the Stress tensor. 
Considering conformal transformation in \eqref{deltaSinf} 
\begin{align}
\delta S = \int d^d x~T^{\mu\nu}\partial_\mu\varepsilon_\nu = \frac{1}{2}\int d^d x~T^{\mu\nu}(\partial_\mu\varepsilon_\nu + \partial_\nu \varepsilon_\mu).
\end{align}
using \eqref{f1} and \eqref{f2} we obtain
\begin{align}
\delta S =  \frac{1}{d}\int d^d x~T^{\mu}_{~~\mu}\partial_\alpha\varepsilon^\alpha~.
\end{align}
And we can see that if the energy-momentum is traceless, this ensures that the theory is conformal invariant. The inverse of the last statement is not always true, due to the fact that $\partial_\alpha \varepsilon^\alpha$ is not an arbitrary function.
\\

For Lorentz rotations of the scalar field the conserved current corresponding to conformal transformation is obtained by the N\"other theorem, and depends on the stress-tensor
\begin{align}
j^{\mu\nu\rho} = T^{\mu\nu}x^\rho - T^{\mu\rho}x^\nu~,
\end{align}
which follows from the conservation of $T^{\mu\nu}$ and using the symmetry of this. Similarly for dilatations 
\begin{align}
j^\mu = T^{\mu\nu}x_\nu~,
\end{align}
which is conserved only if $T^{\mu\nu}$ is traceless. Then the conserved charges associated to conformal transformation depends on the stress-tensor.
%%%%
\section{Two-dimensional Conformal Field Theories}\label{CFT2}
\subsection{Conformal Group}
Consider the coordinates $(z^0, z^1)$ on the plane. Now applying a coordinate system change $z^\mu \rightarrow w^\mu (x)$, then the metric transform as
\begin{align}
g^{\mu\nu} \rightarrow \left( \frac{\partial w^\mu}{\partial z^\alpha}\frac{\partial w^\nu}{\partial z^\beta}\right)g^{\alpha\beta}~.
\end{align}
Now, the condition that defines a conformal transformation \eqref{conftransf} imposes that the metric transformation behaves as $g(w)\propto g(z)$, explicitly
\begin{align}
\left(\frac{\partial w^0}{\partial z^0}\right)^2 + \left(\frac{\partial w^0}{\partial z^1}\right)^2 = \left(\frac{\partial w^1}{\partial z^0}\right)^2 + \left(\frac{\partial w^1}{\partial z^1}\right)^2~,
\end{align}
 expanding 
 \begin{align}
 \frac{\partial w^0}{\partial z^0}\frac{\partial w^1}{\partial z^0} +  \frac{\partial w^0}{\partial z^1}\frac{\partial w^1}{\partial z^1} = 0~. 
 \end{align}
 These conditions are equivalent either to
 \begin{align}\label{CauchyHolo}
 \frac{\partial w^1}{\partial z^0} = \frac{\partial w^0}{\partial z^1}~, \quad \text{and}\quad  \frac{\partial w^0}{\partial z^0} = -\frac{\partial w^1}{\partial z^1}~,
 \end{align}
or to
 \begin{align}\label{CauchyAntiHolo}
 \frac{\partial w^1}{\partial z^0} = -\frac{\partial w^0}{\partial z^1}~, \quad \text{and}\quad  \frac{\partial w^0}{\partial z^0} = \frac{\partial w^1}{\partial z^1}~.
 \end{align}
 The first equations \eqref{CauchyHolo} are recognized as the Cauchy-Riemann equations for holomorphic functions, whereas \eqref{CauchyAntiHolo} defines antiholomorphic functions. 
 We can use this last to motivate the use of complex coordinates $z$ and $\bar z$, with the following translation rules
 \begin{align}
 z = z^0 + iz^1~, \quad \bar z = z^0 - iz^1~,
 \end{align}
 the inverse transformations
 \begin{align}\label{euctransf}
 z^0 = \frac{1}{2}(z + \bar z)~, \quad z^1 = -\frac{i}{2}(z - \bar z)~.
 \end{align}
 For the derivatives 
 \begin{align}
 \partial_z  = \frac{1}{2}(\partial_0 - i\partial_1) \equiv \partial~, \quad \partial_{\bar z} = \frac{1}{2}(\partial_0 + i \partial_1) \equiv \bar\partial~,
 \end{align}
 or
  \begin{align}
\partial_0 = \partial + \bar\partial~,\quad \partial_1 = i(\partial - \bar\partial)~.
 \end{align}
 In terms of the complex coordinates, the metric tensor then reads
 \begin{align}
 g_{\mu\nu} = \frac{1}{2}\begin{pmatrix}
0 & 1 \\
1 & 0 
\end{pmatrix} ~, \quad g^{\mu\nu} = 2\begin{pmatrix}
0 & 1 \\
1 & 0 
\end{pmatrix}~.
 \end{align}
 where $\mu = (z,\bar z)$. This metric tensor allow to rise and lower (anti)holomorphic-indices. In this lenguage, the holomorphic Cauchy-Riemann equations \eqref{CauchyHolo} can be rewritten
 \begin{align}
 \bar\partial w(z,\bar z) = 0~,
 \end{align}
 whose solution is any holomorphic mapping (no $\bar z$ deppendence)
 \begin{align}
 z \rightarrow w(z).
 \end{align}
 We can now analyze the Stress-Tensor in two-dimensional Euclidean CFT, it has some important properties. Using the coordinate change $T_{\mu\nu} = \frac{\partial x^\alpha}{\partial x^\mu}\frac{\partial x^\beta}{\partial x^\nu}T_{\alpha\beta}$
 for \eqref{euctransf} we obtain
 \begin{align}
 T_{zz} ={}& \frac{1}{4}(T_{00} -2i T_{10} -T_{11})~, \nonumber \\
 T_{\bar z \bar z} ={}& \frac{1}{4}(T_{00} + 2i T_{10} -T_{11})~, \nonumber \\
 T_{z \bar z} ={}& \frac{1}{4}(T_{00} + T_{11})\frac{1}{4}T^\mu_{\,\,\mu} = 0~.
 \end{align}
 Where we have used $T^\mu_{\,\,\mu} = 0$. This implies that
 \begin{align}
 T_{zz} = \frac{1}{2}(T_{00} - iT_{10})~, \qquad T_{\bar z \bar z} = \frac{1}{2}(T_{00} + iT_{10})~, 
 \end{align}
 now, it is known from the Noether theorem that the stress tensor corresponds to a conserved quantity $\partial_{\mu}T^{\mu\nu} = 0$, then two condition arise 
 \begin{align}
 \partial_{0}T_{00} + \partial_1 T_{10} = 0~, \qquad \partial_{0}T_{01} + \partial_1 T_{11} = 0~,
 \end{align}
from which follows that
\begin{align}
\bar \partial T_{zz} ={}& \frac{1}{4}\left(\partial_0 +i\partial_1 \right)(T_{00} -iT_{11})~, \nonumber \\
={}& \frac{1}{4}(\partial_0 T_{00} + \partial_{1}T_{10} + i(\partial_1 T_{00} - \partial_0 T_{10})~, \nonumber \\
={}& 0~.
\end{align}
The same calculation for the complex counterpart leads to the same result $\partial T_{\bar z\bar z} = 0$. This means that the two non-vanishing components of the Stress-Tensor are a \textit{chiral} and \textit{anti-chiral} field
\begin{align}
T_{zz}(z,\bar z) = T(z)~, \qquad T_{\bar z\bar z}(z,\bar z) = T(\bar z)~.
\end{align}
 \subsection{Global Conformal Transformations}
 The past section was a local analysis of the transformation for the metric tensor. Now let us impose that the conformal transformation can be defined evrywhere and be invertible in order to form a group. 
 The invertible map that whole plane into itself form a set of global conformal transformation the we call the special conformal group. It turns out that the complete set of such mapping is 
 \begin{align}\label{mobius}
 f(z) = \frac{az + b}{cz + d}~, \quad \text{with}\quad ad-bc = 1~,
 \end{align}
 and $a,b,c,d \in \mathbb C$. These mappings are called projective transformations or \textit{Mobious transformations}, and can be associated to the matrix 
 \begin{align}
 A = \begin{pmatrix}
a & b \\
c & d 
\end{pmatrix}~. 
 \end{align}
 Also, is easy to check that the conformal group in two dimensions is isomorphic to $SL(2,\mathbb C)$. 
 
 \subsection{Generators}\label{SubSec:Generators2D}
 We can write any holomorphic infinitesimal local transformation as
 \begin{align}\label{inftran}
 z' = z + \varepsilon(z)~.
 \end{align}
 Now, it follows to assume that the infinitesimal parameter admits a Laurent expansion around $z = 0$
 \begin{align}
 \varepsilon(z) = \sum^\infty_{n = -\infty}c_n z^{n+1}~.
 \end{align}
For a scalar field, the infinitesimal transformation \eqref{inftran} implies
\begin{align}
\delta \phi = \sum_n \left(c_n \ell_n + \bar c_n \bar\ell_n\right) \phi(z, \bar z)~.
\end{align}
 where the generators $\ell$ and $\bar \ell$ have the form
 \begin{align}\label{laugen}
 \ell_n = -z^{n+1}\partial_z~,\quad \bar\ell_n = -\bar z^{n+1}\bar\partial~.
 \end{align}
 The commutation relations for the generators are
 \begin{align}\label{Witt}
 [\ell_m , \ell_n ] ={}& z^{m+1}\partial \left(z^{n+1}\partial\right) - z^{n+1}\partial \left(z^{m+1}\partial\right)~,  \nonumber \\
 ={}& (n+1)z^{m+n+1}\partial - (m+1)z^{m+n+1}\partial~, \nonumber \\
 ={}& -(m-n)z^{m+n+1}\partial~,\nonumber \\ 
 ={}& (m-n)\ell_{m+n}~, \nonumber \\
 [\bar\ell_m , \bar\ell_n ] ={}& (m-n)\bar\ell_{m+n}~, \nonumber \\
 [\ell_m,\bar\ell_n] ={}& 0~.
 \end{align}
This corresponds to the so-called \textit{Witt algebra}, which is the direct sum of two isomorphic algebras. 
If we restrict $n = -1, 0 ,1$, this is also a subalgebra $SL(2,\mathbb C)$, which is contained in the full Witt algebra. Actually these subalgebra, is the one associated to the global conformal transformations, from \label{laugen} reads
\begin{align}
n ={}& -1~, &&\ell_{-1} = -\partial~,&&&\text{translations}~, \nonumber \\
n ={}& 0~, &&\ell_0\,\,\,\, = -z\partial~, &&&\text{scale transformations}~, \nonumber \\
n  ={}& 1~, &&\ell_1\,\,\,\, = -z^2\partial~, &&&\text{special conformal transformations}~.
\end{align}
Analoge for $\bar \ell_n$.
In polar coordinates $z = r\exp\{i\phi \}$ the generators reads
\begin{align}
\ell_0 = -\frac{1}{2}\left(-r\partial_r + i\partial_\phi\right)~, \qquad \bar\ell_0 = \frac{1}{2}\left(r\partial_r + i\partial_\phi \right)~,
\end{align}
and is easier to see that 
\begin{align}\label{cngen}
\ell_0 + \bar\ell_0 = -r\partial_r~,\qquad i\left( \ell_0 - \bar\ell_0\right) = -\partial_\phi~,
\end{align}
are the generators of dilatation and rotations, respectively. 
\subsubsection{Primary Fields in Two-Dimensions}
In two dimensions, the rotation group $SO(2)$ is labeled simply by just one number, and there the effect of the spin is easy to conclude. \\
Applying a rotation $z = \exp\{i\theta\}z$ on a field of spin $s$, this transform as
\begin{align}
\phi' (z',\bar z') = \exp\{-i s\theta\}\phi(z, \bar z)~.
\end{align}
If a quasi-primary field has scaling dimensions $\Delta$, the transformation $z' = \lambda z$ with $\lambda \in \mathbb C$, which corresponds to a rotation with angle $\exp\{i\theta\} = (\lambda\bar\lambda^{-1})^\frac{1}{2}$. Such a dilatation renders to
\begin{align}
\phi'(z',\bar z') ={}& (\lambda\bar\lambda)^{-\frac{\Delta}{2}}(\lambda\bar\lambda^{-1})^{-\frac{\Delta}{2}}\phi(z,\bar z)~, \nonumber\\ ={}& \lambda^{-h}\lambda^{-\bar h}\phi(z,\bar z)~,
\end{align}
where $h$ is known as the \emph{conformal dimension} of the field. It is defined (and the antiholomorphic counterpart $\bar h$) as
\begin{align}
h = \frac{1}{2}(\Delta + s)~,\qquad \bar h = \frac{1}{2}(\Delta - s)~.
\end{align}
It can be checked that this definition holds for any globally defined conformal transformation
\begin{align}\label{primtrans}
\phi'(z',\bar z') = \left(\frac{dz'}{dz}\right)^{-h}\left(\frac{d\bar z'}{d\bar z}\right)^{-\bar h}\phi(z,\bar z)~.
\end{align}
A \emph{primary field} in two-dimensions is the field that transform in the above fashion \eqref{primtrans}, even for two dimension that we have an infinite number of locally defined conformal transformations. Under infinitesimal transformations $z' = z + \epsilon(z)$, these transform as
\begin{align}\label{deltaphi}
-\delta \phi(z,\bar z) = \left(\epsilon\partial\phi + h\phi\partial\epsilon\right) + \left(\bar\epsilon\bar\partial\phi + \bar h \phi\bar\partial\bar\epsilon\right)~.
\end{align}
%%%%%%%%%
%%%%%%%%%
\subsection{Radial quantization}
Instead following the standard canonical quantization procedure, we will follow the method called \emph{radial quantization}, in which the time coordinate corresponds to the radial direction on the plane. \\
\vspace{0.5cm}
\begin{center}
\begin{tikzpicture}[scale=0.9]
\node [draw, cylinder, shape aspect=6, rotate=90, minimum height=4.2cm, minimum width=3cm, thick] at (1,0) {};
\draw [line width=1.2mm, <-] (-3.5,0.5) -- (-1.5,0.5);
\node[scale=0.7] at (-2.4,-0.2) {$z = \exp\{-iw\}$};
%%%%%%LINES ON THE CYLINDER%%%%%%%%
\draw[thick] (2.65,0) arc[start angle=0,end angle=-180, x radius=1.65, y radius=0.7];\draw[thick] (2.65,0.3) arc[start angle=0,end angle=-180, x radius=1.65, y radius=0.7];\draw[thick] (2.65,0.6) arc[start angle=0,end angle=-180, x radius=1.65, y radius=0.7];\draw[thick] (2.65,0.9) arc[start angle=0,end angle=-180, x radius=1.65, y radius=0.7];
\draw[thick,dashed] (2.65,0) arc[start angle=0,end angle=180, x radius=1.65, y radius=0.7];\draw[thick,dashed] (2.65,0.3) arc[start angle=0,end angle=180, x radius=1.65, y radius=0.7];\draw[thick,dashed] (2.65,0.6) arc[start angle=0,end angle=180, x radius=1.65, y radius=0.7];\draw[thick,dashed] (2.65,0.9) arc[start angle=0,end angle=180, x radius=1.65, y radius=0.7];\draw[thick] (2.65,1.2) arc[start angle=0,end angle=-180, x radius=1.65, y radius=0.7];;\draw[thick] (2.65,1.5) arc[start angle=0,end angle=-180, x radius=1.65, y radius=0.7];
\draw[thick,dashed] (2.65,0) arc[start angle=0,end angle=180, x radius=1.65, y radius=0.7];\draw[thick,dashed] (2.65,0.3) arc[start angle=0,end angle=180, x radius=1.65, y radius=0.7];\draw[thick,dashed] (2.65,0.6) arc[start angle=0,end angle=180, x radius=1.65, y radius=0.7];\draw[thick,dashed] (2.65,0.9) arc[start angle=0,end angle=180, x radius=1.65, y radius=0.7];\draw[thick,dashed] (2.65,1.2) arc[start angle=0,end angle=180, x radius=1.65, y radius=0.7];\draw[thick,dashed] (2.65,1.5) arc[start angle=0,end angle=180, x radius=1.65, y radius=0.7];
\draw[thick] (2.65,-0.3) arc[start angle=0,end angle=-180, x radius=1.65, y radius=0.7];
\draw[thick,dashed] (2.65,-0.3) arc[start angle=0,end angle=180, x radius=1.65, y radius=0.7];

\draw[line width=0.4mm,->] (3,-1)--(3,0);
\draw[line width=0.4mm,<-] (2.7,-1.5) arc[start angle=0,end angle=-70, x radius=1.65, y radius=0.7];
\node at (3.4,0.2) {$x^0$};
\node at (3.2,-1.5) {$x^1$};
%%%%%%%%%%%%%%%%%%%%%
%%%%%%%%%%%%%%%%%%%%%
\draw[thick, opacity=0.7] (-6.4,0.5) circle (2.1cm);\draw[thick, opacity=0.7] (-6.4,0.5) circle (1.4cm);\draw[thick, opacity=0.7] (-6.4,0.5) circle (1.1cm);\draw[thick, opacity=0.7] (-6.4,0.5) circle (0.8cm);\draw[thick, opacity=0.7] (-6.4,0.5) circle (0.4cm);
\filldraw [black] (-6.4,0.5) circle (2pt);
%%%%%%%%%%%%%%%%%%%%%
%%%%%%%%%%%%%%%%%%%%%
\draw[thick,->] (-6.4,0.5)--(-6.4,3);\draw[thick,->] (-6.4,0.5)--(-3.8,0.5);
\node at (-8.4,2.7) {$z$};
%CAPTION
\node[text width=16cm, text justified] at (-2,-3){
\small {\hypertarget{Fig:7} \bf Fig.~7}: 
\sffamily{Weyl transforming the plane into a cylinder}}; 
\end{tikzpicture}
\end{center}
A Lorentzian theory on a cylinder would have an open coordinate $-\infty < x^0 <\infty$ and a compact one $x^1 \sim x^1 + 2\pi$. Therefore to go to Euclidean, we would have to compactify the time coordinate. We introduce the complex coordinate
\begin{align}
w = x^0 + ix^1~,\qquad w\sim w+2\pi i~. 
\end{align}
The cylinder can now be mapped to the complex plane (modulo the origin) by a conformal transformation. The origin corresponds to a point at infinity on the cylinder, by this integration on $z$ corresponds to contour integration around the origin . This also maps the generators \eqref{cngen} to space and time translations
\begin{align}\label{transla}
H = \ell_0 + \bar \ell_0~,\qquad P = i(\ell_0 - \bar \ell_0)~,
\end{align}
time evolution becomes radial rescaling and space translations become rotations. 
\\

Over the cylinder we can Fourier expand the fields using the periodicity of the $w-$coordinate 
\begin{align}\label{weylcylinder}
\phi(w,\bar w) = \sum_{n=-\infty}^{\infty}\sum_{m=-\infty}^{\infty}\phi_{n,m}\exp\{-(nw + m\bar w)\}~.
\end{align}
For primary fields \eqref{primtrans} the conformal transformation from the plane to the cylinder reads
\begin{align}
\phi(z,\bar z) = \left(\frac{dz}{dw}\right)^{-h}\left(\frac{d\bar z}{d\bar w}\right)^{-\bar h}\phi(w,\bar w)~,
\end{align}
and \eqref{weylcylinder} we get
\begin{align}\label{laurenprimary}
\phi(z,\bar z) = \sum_{n=-\infty}^{\infty}\sum_{m=-\infty}^{\infty} \phi_{n,m}z^{-n-h}\bar z^{-m-\bar h}~,
\end{align}
which corresponds to a Laurent series. Using Cauchy integration we can obtain the coefficients of the expansion as
\begin{align}\label{coeff}
\phi_{n,m} = \oint \frac{dz}{2\pi i}\oint \frac{d\bar z}{2\pi i}z^{n+h-1}\bar z^{m+\bar h h-1}\phi(z,\bar z)~.
\end{align}
To see the asymptotic states into the new coordinates we can use the fact that in CFT's the set of local operators and the space of states are isomorphic to each others. The states that are created in the infinite past corresponds to a single point $z =0$ on the complex plane. Therefore we can define the asymptotic states with the primary field as
\begin{align}
\ket{\phi} = \lim_{z,\bar z \rightarrow \infty} \phi(z,\bar z)\ket{0}~.
\end{align}
And therefore by notice \eqref{coeff}, we can see that in order to have a well define asymptotic states we must requiere that the vacuum
\begin{align}
\phi_{n,m}\ket{0} = 0~, \qquad n>-h~,\qquad m>\bar h~.
\end{align}
The hermitian conjugation in Lorentzian signature just affect the time like coordinate. Then we must impose that $z \rightarrow 1/z*$ in order to cancel the factor of $i$ that appears in the continuation. A primary field is conjugated as
\begin{align}
\phi(z,\bar z)^\dagger = \bar z^{-h}z^{-\bar h}\phi(1/\bar z, 1/z)~. 
\end{align}
Laurent expanding the right hand side we get
\begin{align}
\phi(z,\bar z)^\dagger =  \sum_{n=-\infty}^{\infty}\sum_{m=-\infty}^{\infty} \phi_{-n,-m}z^{-n-h}\bar z^{-m-\bar h}~,
\end{align}
and therefore using \eqref{laurenprimary} we get
\begin{align}
\phi_{n,m}^\dagger = \phi_{-n,-m}~.
\end{align}
In cylinder coordinates, a conserved current is expanded as 
\begin{align}
\partial_\mu j^\mu = \partial \bar j + \bar\partial j = 0~,
\end{align}
with corresponding N\"other charge
\begin{align}
Q_j = \frac{1}{2\pi i}\oint dz~j_z + \frac{1}{2\pi i}\oint d\bar z~j_{\bar z}~,
\end{align}
and we can notice that each term is conserved separately, due to each of the currents is pure holomorphic and antiholomorphic respectively. 
\\

As we have seen above, the conserved currents in CFT corresponds to components of the stress-tensor. This can be seen by considering an infinite set of conserved quantities by $\epsilon(z) j$ and $\bar \epsilon (\bar z) \bar{j}$ with $\epsilon$ and $\bar \epsilon$ both arbitrary functions. For the stress-tensor
\begin{align}
\partial T_{\bar z  z} + \bar\partial T_{zz} = 0~,\qquad \partial T_{\bar z \bar z} + \bar\partial T_{z \bar z} = 0~.
\end{align}
Using the traceless condition $T_{z\bar z} = 0$, then we define $T(z) = T_{zz}$ and $\bar T(\bar z) = T_{\bar z \bar z}$. Expanding the $\epsilon$ parameter in a general form
\begin{align}
\epsilon(z) = \sum_{n=-\infty}^\infty \epsilon_n z^{n+1}~.
\end{align}
Replacing this leads to the conserved charges 
\begin{align}\label{virgen}
L_n = \frac{1}{2\pi i}\oint dz~z^{n+1}T~, \qquad \bar L_n = \frac{1}{2\pi i}\oint d\bar z~,\bar z^{n+1}\bar T~.
\end{align}
On the literature this are called \emph{Virasoro generators}. The stress tensor in terms of this generators are
\begin{align}\label{Tmode}
T = \sum_{n=-\infty}^\infty L_n~z^{n+2}~,\qquad \bar T = \sum_{n=-\infty}^\infty \bar L_n~\bar z^{n+2}~.
\end{align}
We now define the \emph{radial ordering opertaror} $\mathfrak R$, which acts over two different local operators evaluated at different points as
\begin{align}
\mathfrak R (\phi_1(z)\phi_2(w)) = \left\{ \begin{matrix} \phi_1(z)\phi_2(w) & |z| > w \\ \phi_2(w)\phi_1(z) & |w| > |z|  \end{matrix}~, \right.
\end{align}
which is the analogue to time ordering on the cylinder. We can now see that the integral of two radial ordered local operators evaluated at the same point becomes 
\begin{align}
\oint_{z=w}dz~\mathfrak R (u(z)v(w)) = \oint_{C_1} dz~u(z)v(w) - \oint_{C_2}dz~v(w)u(z)~,
\end{align}
where the contours are $C_1 := |z| = |w| + \varepsilon$ and $C_2 := |z| = |w| - \varepsilon$. Defining $U = \oint_{z = 0} dz~u(z)$ and taking the limit $\varepsilon \rightarrow 0$ the commutator takes the value
\begin{align}
[U,V] = \oint_{w =0} dw \oint_{z=w} dz~\mathfrak R (u(z) v(w))~.
\end{align}
Therefore under conformal transformations the variation of a primary field reads
\begin{align}\label{varprim}
\delta\phi(w) = [\phi(w),Q_\epsilon] = -\frac{1}{2\pi i}\oint_{z=w} dz~\epsilon(z) T(z)\phi(w)~.
\end{align}
Recalling the Cauchy integration for an arbitrary function $f(z)$ that is analytic at $z = w$ 
\begin{align}
\oint_{z=w}\frac{dz}{2\pi i}f(z)(z-w)^{-n} = \frac{f^{(n-1)}(w)}{(n-1)!}~,
\end{align}
now imposing \eqref{deltaphi}  we see that the product between $T$ and $\phi$ must behave as
\begin{align}
T(z)\phi(w) = \frac{h}{(z-w)^2}\phi + \frac{1}{(z-w)}\partial\phi + \text{regular terms}~.
\end{align}
This are called \emph{operator product expansion} (OPE), which allow to analyze the divergent terms on the product between two fields evaluated at different points, which is expanded as a series of operators at a single point.  For the stress-tensor the OPE between it self reads
\begin{align}
T(z)T(w) \sim \frac{c}{2(z-w)^4} + \frac{2}{(z-w)^2}T(w) + \frac{1}{(z-w)}\partial T(w)~,
\end{align}
where the term $c$ in the lead divergent term its known as the \emph{central charge} of the theory, and is related with the degrees of freedom on the theory. Therefore the transformation of the stress-tensor
\begin{align}
\delta_\epsilon T(w) ={}& \frac{1}{2\pi i}\oint_{z=w} dz~ \epsilon(z)\mathfrak R (T(w)T(z)) \nonumber \\ ={}& \frac{c}{12}\partial^3 \epsilon(w) + 2\partial \epsilon(w) T(w) + \epsilon(w)\partial T(w)~.
\end{align}
We can now see the finite form of this variation by using exponentiation to get
\begin{align}
T(w) \rightarrow T(w) (\partial f)^2 T(g(w)) + \frac{c}{12}Sc(f,w)~,
\end{align}
where $c$ is a constant defined with the factor $1/12$ by convention, $f(z)$ corresponds to the infinitesimal transformation on the coordinates \eqref{inftran}, and $Sc[f,z]$ is the Schwarzian derivative defined as
\begin{align}\label{Scder}
Sc[f,z] = \frac{\partial^3 f}{\partial f} - \frac{3}{2}\left(\frac{\partial^2 f}{\partial f}\right)^2~.
\end{align}
On the cylinder coordinates $z$ we get
\begin{align}
T^{cylinder}(z) = (\partial_z \omega (z))^2 T(w(z)) + \frac{c}{12}Sc[w,z] = -\left(w^2 T(w) - \frac{c}{24}\right)~,
\end{align}
and the mode expansion \eqref{Tmode} renders
\begin{align}
T(w) = \sum_{n=-\infty}^\infty w^{-n-2}L_n~,
\end{align}
then comparing both last we obtain
\begin{align}
T^{cylinder}(z) = -\left(\sum_{n=-\infty}^\infty L_n \exp\{-inz\} - \frac{c}{24}\right)~,
\end{align}
and analogue for the antiholomorphic counterpart. Also the translation generators \eqref{transla}  adopts a shift by this definition 
\begin{align}\label{translationsLo}
H ={}&  \frac{1}{2\pi}\int dx^1 T_{00} = L_0 + \bar L_0 - \left(\frac{c}{24} + \frac{\bar c}{24}\right)~, \nonumber \\ \qquad P ={}& \frac{1}{2\pi}\int dx^1 T_{01} = L_0 - \bar L_0 - \left(\frac{c}{24} - \frac{\bar c}{24}\right)~,
\end{align}
By computing the commutator between two different Virasoro operators \eqref{virgen} we get
\begin{align}\label{Virasoro}
[L_n, L_m] ={}& \oint_{w=0}\frac{dw}{2\pi i}\oint_{z=w} \frac{dz}{2\pi i} w^{m+1}z^{n+1}\mathfrak R (T(z)T(w)) \nonumber \\ ={}& \oint_{w=0}\frac{dw}{2\pi i}w^{m+1}\left\{ \frac{c}{12}(n+1)n(n-1)w^{n-2} + 2(n+1)w^n T(w) + w^{n+1}\partial T(w) \right\} \nonumber \\ ={}& \frac{c}{12}n(n^2-1)\delta_{n+m,0} + 2(n+1)L_{n+m} - (m+n+2)\oint_{w=0}\frac{dw}{2\pi i}w^{m+n+1}T(w)~,
\end{align}
and finally 
\begin{align}\label{Virasoro}
\boxed{[L_n, L_m] = (n-m)L_{n+m} + \frac{c}{12}n(n^2 -1)\delta_{n+m,0}}~.
\end{align}
This its known as the \emph{Virasoro algebra} which is also satisfied by the antiholomorphic Virasoro generators $\bar L_n$ with central charge $\bar c$. This algebra corresponds to the centrally extended Witt algebra \eqref{Witt}. For $n=0$ the generators satisfy
\begin{align}
[L_0,L_m] = -mL_m~,
\end{align}
and for $m = -1,0,1$ this forms the algebra of $SL(2,\mathbb R)$ which differs by a sign with the $SU(2)$ algebra.
The algebra by $L_n$ and $\bar L_n$ with $n=-1,0,1$ form $SL(2,\mathbb R)\times SL(2,\mathbb R) \sim SO(2,2)$.
\subsubsection{On a Torus}
As we have obtained some properties of CFTs by using radial quantization on a cylinder, we now can go to the particular case of a torus, which corresponds to a cylinder that has been compactified on the non-compact direction $x^0$ therefore is the same analysis for periodic boundary conditions on this direction. 
\vspace{1cm}
\begin{center}
\begin{tikzpicture}
\draw[thick] (0,0) ellipse (3 and 1.2);
%Hole
\begin{scope}[scale=.8]
\path[rounded corners=24pt] (-1.1,0)--(0,.6)--(1.2,0) (-1.2,0)--(0,-.56)--(1.2,0);
%\draw[rounded corners=28pt] (-2.4,.1)--(0,-.6)--(2.4,.1);
%\draw[rounded corners=24pt] (-2.5,0)--(0,.6)--(2.5,0);
\draw[thick] (1.5,0.3) arc[start angle=0,end angle=-180, x radius=1.65, y radius=0.7];
\draw[thick] (1.3,0.05) arc[start angle=30,end angle=150, x radius=1.65, y radius=0.7];

\end{scope}
%Cut 1
\draw[densely dashed,thick] (0,-1.2) arc (270:90:.2 and .46);
\draw[thick] (0,-1.2) arc (-90:90:.2 and .46);
%Cut 2
\draw[thick] (0,1.2) arc (90:270:.2 and .43);
\draw[densely dashed,thick] (0,1.2) arc (90:-90:.2 and .43);
%Caption~~~~~~~~~~~~~~~~~~~~
\node[text width=12cm, text justified] at (0,-2){
\small {\hypertarget{Fig:8} \bf Fig.~8}: 
\sffamily{A torus with two cuts representing the compactification of the cylinder.
}};
\end{tikzpicture}
\end{center}
The torus can be understood in terms of points in the complex plane that differ by a linear combination of two identical vectors. Choosing $x^0,x^1 \in [0,L]$ the area of the torus is just $L$. The torus can be parametrized by a single complex number $\tau = \tau_1 + i\tau_2$ with $\tau_2 \geq 0~$ where in this parametrization the number $\tau$ is known as the complex torus modulus or structure. We can identify points $z$ on the complex plane simply by
\begin{align}
z \rightarrow z + L~,\qquad z \rightarrow z +L~.
\end{align}
which can be viewed as
\vspace{1cm}
\begin{center}
\begin{tikzpicture}
\draw[thick,->] (-5,-0.8)--(-5,3);%y-axis
\draw[thick,->] (-5.8,0)--(0,0);%x-axis
\node[scale=0.8] at (-5.7,2.8) {$\text{Im}~z$};
\node[scale=0.8] at (-0.5,-0.5) {$\text{Re}~z$};
\node[scale=0.9] at (-5.2,-0.3) {$0$};
\node[scale=0.9] at (-2.5,-0.3) {$R$};
%\draw[thick,-] (-5,0)--(0,0);%firstline
%\draw[thick] at (-5,0) plot(\x,{1.9*(\x)});
\draw[thick,-] (-5,0)--(-4,1.8);
\draw[thick,-] (-4,1.8)--(-1.5,1.8);
\draw[thick,-] (-1.5,1.8)--(-2.5,0);
\node[scale=0.9] at (-3.9,2.1) {$\tau$};
\node[scale=0.9] at (-1.4,2.1) {$\tau+R$};
%Caption~~~~~~~~~~~~~~~~~~~~
\node[text width=12cm, text justified] at (-3,-2){
\small {\hypertarget{Fig:9} \bf Fig.~9}: 
\sffamily{Symmetries of the torus see it at the complex plane.
}};
\end{tikzpicture}
\end{center}
It can been seen that $\tau$ is the same torus as $\tau + R$, therefore corresponds to a symmetry. Also an inversion of this form $\tau \rightarrow \frac{\tau}{\tau + R}$  corresponds to a symmetry, therefore using the translation by $R$ we get
\begin{align}
\tau \rightarrow -\frac{1}{\tau}~.
\end{align}
These transformation corresponds just to the modular group
\begin{align}
\tau' = \frac{a\tau + b}{c\tau + d}~, \qquad a,b,c,d \in SL(2,\mathbb R)~,
\end{align}
and as we have seen the inversion of sign does not affect the transformations, and therefore it corresponds to modular invariance $PSL(2,\mathbb R) = SL(2,\mathbb R)/\mathbb Z_2$~.
Now let us see the partition function on the torus, for this we should compute the path integral of the Euclidean action of configurations on the torus. 
\begin{align}
\int \mathcal D \phi \exp\{-S_E[\phi]\}~,
\end{align}
this can be expressed in terms of the Hamiltonian of the theory, which is related to the zero Virasoro modes, for periodic boundary conditions is simply
\begin{align}
\int_{bc} \mathcal D \phi \exp\{-S_E[\phi]\} = \text{Tr}\exp\{-\beta H\}~,
\end{align}
where $\beta$ is the period of the euclidean time. The case of the torus this formula is valid, due to the compactification of the cylinder. If we would have $\tau_1 = 0$ (pure imaginary torus modulus), then for a torus of radius $R=1$, the periodicity is just proportional to the imaginary part of the modulus
\begin{align}
\beta = 2\pi \text{Im}~\tau H~.
\end{align}
In the case that we have $\tau_1 \neq 0$ then we have to consider a shift on the $x^1$-coordinate of the cylinder before gluing both extremes. To do this we must use momentum operator $P$ that generates translations onto the spatial direction. This will be shifted by just a factor of $2\pi \text{Re}\tau$, then the exponential obtains an extra correction of the form
\begin{align}
\int_{bc} \mathcal D \phi \exp\{-S_E[\phi]\} = \text{Tr}\exp\{-2\pi \text{Im}\tau H \}\exp\{iP(2\pi\text{Re}\tau)\}~,
\end{align}
and using \eqref{translationsLo} we obtain 
\begin{align}
\int_{bc} \mathcal D \phi \exp\{-S_E[\phi]\} = \text{Tr}\exp\{2\pi i\tau\left(L_0 - \frac{c}{24}\right) \}\exp\{-2\pi i\bar\tau \left(\bar L_0 - \frac{c}{24}\right)\}~,
\end{align}
%%%%%
\section{Cardy entropy}\label{Sec:Cardy}
Cardy has shown \cite{Cardy:1986ie} by manipulation of the partition function of a two dimensional CFT and using the modular invariance, that the number of states of the theory can be related to the central charge of it, this has been used particularly in the context of $AdS_3 / CFT_2$ in order to count the number of degrees of freedom that ensembles the BTZ black hole \cite{Banados:1992aa, Banados:1993aa}, and for extremal black holes in higher dimensions in the context of string theory \cite{Strominger:1996sh, Strominger:1997eq}.
\\

We will made use of this formula in order to describe the microscopic origin of the degrees of freedom the are enhanced due to the presence of a static observer on our universe. We will derive this formula by following \cite{Carlip:1998aa}. The partition function on the torus defined before in the microcanonical ensemble as 
\begin{align}
\mathcal Z(\tau,\bar\tau) = \text{Tr}\exp\{2\pi i(\tau L_0 - \bar\tau \bar L_0)\} = \sum \rho(\Delta,\bar\Delta)\exp\{2\pi i(\Delta\tau - \bar\Delta\bar\tau)\}~,
\end{align}
where $\rho$ corresponds to the number of state which have eigenvalue $L_0 = \Delta$ and $\bar L_0 = \bar\Delta$. Now let us use contour integration and isolate the number of states. 
\begin{align}
\rho(\Delta,\bar\Delta) = \frac{1}{(2\pi i)^2}\int \frac{dq}{q^{\Delta+1}}\frac{d\bar q}{\bar q^{\bar\Delta +1}}\mathcal Z(q,\bar q)~.
\end{align}
Where $q = \exp\{2\pi i \}, \bar q = \exp\{2\pi i \bar \tau\}$. For simplicity we will suppress the $\bar\tau$-dependance for the computation. Notice that
\begin{align}
\mathcal Z(\tau) = \exp\left\{ \frac{2\pi i}{24}c \tau\right\}\mathcal Z_0 (\tau)~.
\end{align}
Using the modular invariance of the torus, then 
\begin{align}
\mathcal Z(\tau) = \exp\left\{ \frac{2\pi i}{24}c \right\}\mathcal Z_0 (-1/\tau) = exp\left\{ \frac{2\pi i}{24}c\tau \right\}exp\left\{ \frac{2\pi i}{24}c\frac{1}{\tau} \right\}\mathcal Z(-1/\tau)~,
\end{align}
therefore
\begin{align}
\rho(\Delta) = \int d\tau \exp\{-2\pi i \Delta\tau\}exp\left\{ \frac{2\pi i}{24}c\tau \right\}exp\left\{ \frac{2\pi i}{24}c\frac{1}{\tau} \right\}\mathcal Z(-1/\tau)~.
\end{align}
If $\mathcal Z(-1/\tau)$ varies slowly near the extremum $\tau \sim i\sqrt{c/24\Delta}$ for large $\Delta$, substituting this we obtain the Cardy formula 
\begin{align}
\rho(\Delta,\bar\Delta\} \sim \exp\left\{2\pi\sqrt{\frac{c\Delta}{6}} + 2\pi\sqrt{\frac{c\bar \Delta}{6}}\right\}\mathcal Z(i\infty)~.
\end{align} 
Where we have restored the $\bar\Delta$-dependance. This is known as the Cardy formula and the logarithm describes the asymptotic density of states for two Virasoro generators of a two dimensional CFT. 
We need to check the saddle point approximation (see \autoref{App:B}),
\begin{align}
\mathcal Z (i/\epsilon) = \sum\rho(\Delta)\exp\{-2\pi \Delta/\epsilon\}~,
\end{align}
If the lowest eigenvalue of $L_0$ is $\Delta=0$, then $\lim_{\epsilon\rightarrow 0} \mathcal Z (i/\epsilon) \rightarrow constant$. But if not, then the approximation is not valid. This can be corrected by defining 
\begin{align}
\widetilde Z (\tau) = \sum \rho(\Delta)\exp\{2\pi i(\Delta - \Delta_0)\tau = \exp\{-2\pi i \Delta_0 \tau\}\mathcal Z(\tau)~,
\end{align}
which goes to a constant if $\tau \rightarrow i\infty$. Then
\begin{align}
\rho(\Delta) = \int d\tau \exp\{-2\pi i(\Delta\tau - \Delta_0/\tau)\}\exp\left\{\frac{2\pi i }{24}\left( c\tau + c/\tau\right)\right\} \widetilde Z(-1/\tau)~,
\end{align}
which can be evaluated on the saddle point approximation for a large $\Delta$, which renders
\begin{align}
\rho(\Delta)\sim\exp\left\{2\pi\sqrt{\frac{(c-24\Delta_0)\Delta}{6}}\right\}\rho(\Delta_0) = \exp\left\{2\pi\sqrt{\frac{c_{eff}\Delta}{6}}\right\}\rho(\Delta_0)~,
\end{align}
which is the generalization for the case of $\Delta_0\neq0$, for example Liouville theory has a non-zero $\Delta_0$, and therefore the effective central charge must be used. For logarithmic corrections to this formula see \cite{Carlip:2000aa}.
\\

If the theory has a chiral- and non chiral-sector, recovering the $\bar\Delta$-dependence, the Cardy formula is just
\begin{align}\label{MicroCardy}
\mathcal{S} = 2\pi\sqrt{\frac{c}{6}\Delta} + 2\pi\sqrt{\frac{\bar c}{6}\bar\Delta}
\end{align}
whith the numbers $c$ and $\bar c$ as the holomorphic and anti-holomorphic central charges of the CFT respectively. Using the definition of the associated temperatures for each of the movers we can transform the system into the canonical ensemble
\begin{align}
\left(\frac{\partial \mathcal{S}}{\partial \Delta}\right)_{\bar\Delta} = \frac{1}{T_L}, \quad \left(\frac{\partial \mathcal{S}}{\partial \bar\Delta}\right)_{\Delta} = \frac{1}{T_R}~,
\end{align}
which implies
\begin{align}\label{ElEr}
\Delta = \frac{\pi^2}{6}c T_L^2, \qquad \bar\Delta = \frac{\pi^2}{6}\bar c T_R^2~,
\end{align}
and finally, the Cardy formula adopts the form
\begin{align}\label{CanoCardyMovers}
\boxed{\mathcal{S} = \frac{\pi^2}{3}\left(c T_L + \bar c T_R\right)}~.
\end{align}
In \autoref{Chp:HEE} we will describe different entropies on quantum field theories, that later will be used together with Cardy entropy in order to do a microscopic description of the cosmological horizon.

%% file: Chapters/Liouville.tex
% Chapter 1

\chapter{Liouville Theory} % Main chapter title

\label{Chp:Liouville} % For referencing the chapter elsewhere, use \ref{Chapter1} 

%----------------------------------------------------------------------------------------

%----------------------------------------------------------------------------------------

The Liouville-Polyakov theory consist in a two-dimensional CFT with a dynamical metric which posses a continuous spectrum and has been fully solved in the sense that the three point function has been found it analytically on the sphere \cite{Dorn:1992aa, A.B.Zamolodchikov:1996aa}. It corresponds to quantum gravity in two dimensions which consist in a toy model for four-dimensional gravity. The theory has a critical and non-critical sector. 
\\

In this chapter we will made use of the wonderful different reviews \cite{Ginsparg:ab, Ginsparg:aa, Nakayama:2004aa} and subtract what will be used latter in \autoref{Chp:Microscopic} in order to understand the cosmological horizon in terms of the Liouville conformal field theory. 

Discuss $c=1$ class not enought (Martinec), see non constant central charge (banados embbedings).

\section{Classical Liouville theory} 
It is possible to perform a local analizys of the Liouville theory by choosing some reference metric $\hat g$ on a surface $\Sigma$ such that we define the Liouville scalar field $\phi$ in terms of the \textit{fiducial} metric
\begin{align}\label{fiducial}
g = \exp\{\gamma \phi\}\hat g~.
\end{align}
By this change, the Liouville action on complex coordinates reads \cite{Polyakov:1981rd, Polyakov:1987zb, Distler:1988jt, Knizhnik:1988ak}
\begin{align}\label{Lact}
S_{Liouville} = \frac{1}{4\pi}\int d^2 z \sqrt{-\hat g}\left( \frac{1}{\gamma}\phi \hat R  + \frac{1}{2}(\hat\nabla^2 \phi)^2 + \frac{\mu}{2\gamma^2}\exp\{\gamma\phi\} \right)~.
\end{align}
Where $\gamma^2 = \hbar$ is the only relevant coupling constant of the theory. The $\mu$-term is referred as the \textit{cosmological constant} term makes the different between the Liouville theory and the background charge for the case of pure imaginary background charge. Off course boundary terms can be added on top of \eqref{Lact} but these does not affect the local analysis performed here.
If $\mu > 0$ then the action \eqref{Lact} is bounded from below and the equation of motion respect the Liouville field (in terms of the original metric) $R[g] = -\mu/2$ implies that the metric \eqref{fiducial} has a constant negative curvature. 
The action \eqref{Lact} posses a Weyl invariance and defines a classical conformal field theories for a particular value of $Q$, the Weyl transformation
\begin{align}\label{conftrn}
\hat g \rightarrow \exp\{2\beta\}\hat g~,\qquad \gamma\phi \rightarrow \gamma\phi - 2\beta~,
\end{align}
implies that the Ricci scalar transform as
\begin{align}
R[\exp\{2\beta\}\hat g] = \exp\{-2\beta\} \left( \hat R - 2\hat\nabla^2 \beta\right)~,
\end{align}
using these transformations the action remains invariant only if 
\begin{align}
Q = 2/\gamma~,
\end{align}
and therefore a classical conformal field theory is defined. The stress-tensor $T_{\mu\nu} = -2\pi \frac{\delta S}{\delta g^{\mu\nu}}$ takes the form
\begin{align}
T_{zz} ={}& -\frac{1}{2}\partial_\mu \phi \partial^\mu  + \frac{1}{2}Q\partial^2\phi~, \nonumber \\ T_{\bar z \bar z} ={}& -\frac{1}{2}\bar\partial_\mu \phi \bar\partial^\mu  + \frac{1}{2}Q\bar\partial^2\bar\phi~,
\end{align}
and is traceless as its expected $T_{z\bar z} = 0$. As its seen in \eqref{fiducial} the Liouville field is part of the metric and therefore under conformal transformations $z \rightarrow v = f(z)$ the field change as
\begin{align}\label{phiconf}
\gamma\phi \rightarrow \gamma\phi + \log \bigg\rvert \frac{d v}{d z}\bigg\rvert^2~.
\end{align}
These leaves the element $\exp\{\gamma\phi\}dzd\bar z$ invariant. Under this transformation law eqref{phiconf} the stress-tensor transform as a tensor up to a Schwarzian derivative \eqref{Scder} $Sc[v,\gamma] = \frac{v'''}{v'} - \frac{3}{2}\left(\frac{v''}{v'}\right)$,
\begin{align}
T_{zz} \rightarrow \left( \frac{dv}{dz} \right)^2 T_{v v} + \frac{1}{\gamma^2} Sc[v ; \gamma]~,
\end{align}
which, by followin \autoref{CFT}, a Virasoro algebra with central charge $c = 12/\gamma^2$. For this, consider the system living on a flat cylinder with $0 \leq \theta < 2\pi$ the parametrization of the space direction, and time time coordinate $t$ to be non-compact. 
\\

Defining the conjugated momenta to $\phi$ 
\begin{align}
\Pi = \frac{\delta S}{\delta \dot\phi} = \frac{1}{4\pi}\dot{\phi}~.
\end{align}
Whose commutation relations (under classical Poisson brackets) reads
\begin{align}
\{ \phi(\theta, t), \Pi(\theta' , t')\} = \delta(\theta' - \theta)~.
\end{align}
Going to light-cone coordinates $x^{\pm} = \theta \pm t$~, the stress tensor remains traceless and it has an extra term due to the change of coordinates to the cylinder $z = \exp\{t + i\theta\}$
\begin{align}\label{Tpp}
T_{\pm\pm} ={}& \frac{1}{2}(\partial_\pm \phi)^2 - \frac{1}{\gamma}\partial_\pm^2 \phi + \frac{1}{2\gamma^2} \nonumber \\ ={}& \frac{1}{4}(4\pi \Pi + \phi')^2 - \frac{1}{2\gamma}(4\pi\Pi + \phi')^2 + \frac{\mu}{8\gamma^2}\exp\{\gamma\phi\} + \frac{1}{2\gamma^2}~.
\end{align}
Expanding these into Fourier modes
\begin{align}
L_n^\pm = \int_0^{2\pi} \frac{d\theta}{2\pi}T_{\pm\pm}\exp\{in\theta\}~,
\end{align}
defines two copies of the Virasoro algebra
\begin{align}
i\{L^\pm_n (t), L^\pm_m\} = (n-m)L_{n+m}^\pm + \frac{c^\pm}{12}n(n^2 -1)\delta_{n+m,0}~,
\end{align}
with same central charge 
\begin{align}\label{Lc}
\boxed{c^\pm =12/\gamma^2}~.
\end{align}
Also the commutator of $L_n$ and $\exp\{\alpha\phi\}$ shows that this last is a primary field (defined in \autoref{CFT} as the field annihilated by all $L_n$ for $n>0$) with conformal weight $\Delta = \alpha/\gamma$~.\\

Solutions to the Liouville theory can be obtained by the method developed in \cite{Martinec:aa} usinf the Uniformization theorem \cite{Farkas_1980} which says that any Riemann surface $\Gamma$ is conformally equivalent to the Riemann sphere ($CP^1$), the Poincare upper half plane ($\mathfrak H$) or a quotient of $\mathfrak H$ by a discrete subgroup $\Sigma \subset SL(2,\mathbb R)$ (acting as M\"{o}bius transformation).
\\

A constant negative curvature solution to the field equations for \eqref{fiducial} can be supported by $\mathfrak H$
\begin{align}
ds^2 = \exp\{\gamma\phi\}dzd\bar z = \frac{4}{\mu}\frac{1}{(\text{Im}~z)^2} dz d\bar z~.
\end{align}
These element is invariant under M\"{o}bius transformations \eqref{mobius}, i.e., the group $PSL(2,\mathbb R) = SL(2,\mathbb R)/\mathbb Z_2$, and thus descends to a metric
\begin{align}
ds^2 = \exp\{\gamma \phi\} dz d\bar z = \frac{4}{\mu}\frac{\partial A\bar\partial B}{(A - B)^2}dz d\bar z~,
\end{align}
on the Riemann surface $\hat{\mathfrak H} = \mathfrak H/\Sigma$ for arbitrary local holomorphic functions $A$ and $B$. These solutions are off course in the Euclidean case.  
\\
It can be show for these solution that theory can be described by a single free field \cite{gervais1985, GERVAIS198259, DHoker:1982aa} by expanding in a finite sum of products of holomorphic and antiholomorphic functions and using the B\"{a}cklund transformation \cite{Liouville1853}
\begin{align}
\exp\{-j\gamma\phi\} = \left(\frac{16}{\mu}\right)^{-j}\sum^{j}_{m=-j} \psi_m^j (z)\psi^{jm}(\bar z)~,
\end{align}
which under \eqref{mobius} transformations the holomorphic field transform like spin $j$ representations of $SL(2,\mathbb R)$. For $j=1/2$ the fields satisfy 
\begin{align}
\left(\partial^2 + \frac{\gamma^2}{2}T(z)\right)\psi_{\pm 1/2}(z) = 0~.
\end{align}
We can use the \textit{bosonization} method by writing the fields as
\begin{align}\label{freeLf}
\psi_{\pm 1/2} := \exp\left\{ \frac{-\gamma}{2}\Phi_\pm\right\}~,
\end{align}
which produces the stress-tensor
\begin{align}
T = -\frac{1}{2}(\partial\Phi_\pm)^2 + \frac{1}{\gamma}\partial^2 \Phi_\pm~.
\end{align}
It has been showed in \cite{gervais1985} that the fields $\Phi_\pm$ and their respective momenta satisfy free field Poisson brackets, but not the bracket of two different $\Phi_+$ and $\Phi_-$, therefore are not independent implying that one can be solved in terms of another and the full theory on this background is fully described by a singled field. 
\\

The Minkowskian case is find it by  setting for simplicity the Liouville field to constant $\phi = \phi_0(t)$ that is independent of the spatial coordinate $\theta$. This method is called \textit{mini-superspace approximation}, that defines the equation of a particle moving on a potential $V = \frac{\mu}{\gamma^2}\exp\{\gamma \phi_0\} + (2\gamma^2)^{-1}$ which is a particle with momentum $p>0$, with general solution
\begin{align}
\exp\{\gamma\phi_0(t)\} = \frac{16\gamma p^2}{\mu}\frac{\exp\{2p\gamma t\}}{(1-\exp\{2p\gamma t\})^2}~,
\end{align}
with no solution for $p = 0$ and invariant under $p = -p$. More general solutions with $\theta-$dependence are given by 
\begin{align}
\exp\{\gamma\phi\} = \frac{16}{\mu}\frac{A'(x^+)B'(x^-)}{(1-AB)^2}~,
\end{align}

\section{Quantum Liouville theory}\label{Sec:QMLiouv}
In this section we review the quantum structure of the Liouville theory. In this regime we are interested in compute the correlation functions
\begin{align}\label{correlation}
\expval{\Pi_i \exp\{\alpha_i \phi(\zeta_i)\}} = \int \mathcal D \phi \exp\{-S[\phi]\}\Pi_i\exp\{\alpha_i(\zeta_i)\}~.
\end{align}
The study of this correlation functions depends heavily on the expansion of $\gamma$, because expanding on the cosmological constant $\mu$ does not make sense, this because its value can be changed by shifting the Liouville field. Therefore the different sectors of the theory depends on the value of the coupling constant $\gamma$.\\

 Using semi-classical limit approximation $\gamma << 1$ has been found that the correlation functions can be obtained on this regime for fixed areas method \cite{Seiberg:1990eb} and show differences with the free bosonic conformal field theory and that \eqref{correlation} does not vanish for generic $\alpha_i$'s.
\\

For the quantum regime we must consider high-order corrections of $\gamma$ in the $1/\gamma$ expantion. We follow to made use of \textit{canonical quantization} following \cite{Braaten:1983np, Braaten:1982yn, Braaten:1982fr, Braaten:1983pz, Curtright:1982aa}. Constructing free field using normal ordering and checking commutation relation in order to achieve the quantum oscillator algebra, the expanded field reads
\begin{align}\label{qmop}
\phi(\beta, \tau) ={}& \phi_o(\tau) - \frac{1}{4\pi}\sum_{n\neq0}\frac{1}{in}\left(a_n (\tau)\exp\{-in\beta\} + b_n(\tau)\exp\{in\beta\}\right)~,\nonumber\\\ \Pi(\beta, \tau) ={}& p_o(\tau) + \frac{1}{4\pi}\sum_{n\neq0}\left(a_n (\tau)\exp\{-in\beta\} + b_n(\tau)\exp\{in\beta\}\right)~,
\end{align} 
where $a_n^\dagger = a_{-n},~ b_n^\dagger = b_{-n}$. Now it follows to impose the standard relation on the equal time commutator
\begin{align}
[\phi(\beta', \tau), \Pi(\beta, \tau)] = i\delta(\beta' - \beta)~. 
\end{align}
Which implies that the $a_n$ and $b_n$ are creation operator if $n<0$ and annihilation operators if $n>0$.
\begin{align}
[b_n , b_m] = [a_n , a_m] = n\delta_{n,-m}~,\qquad [a_0, a^\dagger_0] = 1~, \qquad [a_n,b_m] = 0~.
\end{align}
By introducing a new parameter $Q = 2/\gamma + \mathcal O(1)$ the stress-tensor \eqref{Tpp} its modified as 
\begin{align}
T_{+-} ={}& 0 \nonumber \\ T_{++} ={}& \frac{1}{8}(4\pi\Pi + \phi')^2 - \frac{Q}{4}(4\pi\Pi + \phi')' + \frac{\mu}{8\gamma^2}\exp\{\gamma \phi\} + \frac{Q^2}{8}~.
\end{align}
And also modifying the action \eqref{Lact} as
\begin{align}\label{QMLact}
S_{Liouville} = \frac{1}{4\pi}\int d^2 z \sqrt{-\hat g}\left( \frac{Q}{2}\phi \hat R  + \frac{1}{2}(\hat\nabla^2 \phi)^2 + \frac{\mu}{2\gamma^2}\exp\{\gamma\phi\} \right)~.
\end{align}
In order to the quantum action to be conformal invariant, we can check the that the Virasoro algebra is satisfied in the equal time commutator relation of $T_{++}$ (using the operators \eqref{qmop}) and that $[T_{++},T_{--}] = 0$, these are both satisfied only if 
\begin{align}\label{conformalQ}
\boxed{Q = \frac{2}{\gamma} + \gamma}~.
\end{align}
The value of the central charge in terms of $Q$ is 
\begin{align}
c = 1 + 3Q^2~.
\end{align}
The quantization also can be carried out by using the free fields \eqref{freeLf} and is more straightforward \cite{Braaten:1983np, Braaten:1982yn, Braaten:1982fr, Braaten:1983pz, Curtright:1982aa}.
\\

The study of the spectrum of the quantum theory can be carried out by analyzing the $\phi_0$ mode \cite{DHoker:1982wmk}. The Schr\"{o}dinger equation for the zero mode is simply
\begin{align}
H\psi = \left(\frac{1}{2}p_0^2 + \frac{\mu}{8\gamma^2}\exp\{\gamma\phi_0\} + \frac{Q^2}{8}\right)\psi = \Delta \psi~,
\end{align}
where $p = -i\frac{\partial}{\partial \phi_0}$~, is the hermitian conjugated momenta to the zeroth mode, which defines normalizable wave function (of the delta form) if this is real. In the limit of $\phi_0 \rightarrow -\infty$, the potential term $V = \frac{\mu}{8\gamma^2}\exp\{\gamma\phi_0\} + \frac{Q^2}{8}$ is very small and the wave function can be solved and corresponds to a linear combination plane wave for the momenta and the zeroth mode field. In the mini-super space approximation, the wave function corresponding to $\mathcal{O} = \exp\{\alpha\phi\}$ in the limit of $\phi_0 \rightarrow -\infty$ like
\begin{align}\label{Lstates}
\psi_{\mathcal O} = \exp\left\{ -\phi_0\left(\frac{Q}{2} - \alpha\right)\right\}~,
\end{align}
which diverges in the analyzed limit and its not normalizable. This states can be regularized by putting a regulator that cuts at some $\phi_0$, such that when it is removed we can keep the finite norm and have $\psi(\phi_0) \rightarrow 0$ at any finite $\phi_0$. 
\\

The limit $\phi_0 \rightarrow -\infty$ will be mapped to the tensionless limit \eqref{tensionless} on the de Sitter case.

%% file: Chapters/HEE.tex
% Chapter 1

\chapter{Entanglement Entropy}
\label{Chp:HEE} % For referencing the chapter elsewhere, use \ref{Chapter1} 

%----------------------------------------------------------------------------------------

%----------------------------------------------------------------------------------------
As it is previously said it, we will made use of some quantum information tools in order to describe the underlying degrees of freedom on the cosmological horizon $\mathcal H$. In this chapter we will made some reviews mainly on this quantities. We first start with the classic and quantum definition of entropy and there defining Relative and Entanglement Entropy. Finally we give some short inside on how to obtain the Entanglement Entropy for CFT's by using the AdS/CFT correspondence.
\section{Entropy on Quantum and Classical Field Theories}
Entropy is a fundamental concept of thermodynamics and statistical mechanics, which can be understood as the average thermodynamic macroscopic properties of the constituents of a given system, and Ludwig Boltzmann has defined in terms of the number of microstates $\Omega$ of a system $S = k_b \log \Omega$. 
 From the quantum information theory point of view, Entropy is a measure of how much uncertainty exist on a state giving a physical system. In the classical regime, Claude Shannon has define the Entropy in terms of probabilities. If we lear the value of a random variable $\alpha$, the \emph{Shannon Entropy} $H$ of $\alpha$ measures how much information we obtain on average when we learn that particular value, said it in the other way; this quantifies how much \emph{uncertainty} about $\alpha$ we have before learn the value of it. This is defined in terms of the probability distribution $p_i$ as
 \begin{align}\label{Shannoncl}
\boxed{ H(\alpha) \equiv -\sum_{i}p_i \log p_i}~.
 \end{align}
It can be notice that for $p_i = 0$ there is an ambiguity, but if some event has $p_i = 0$, therefore this does not contribute to the entropy, then we can use $\lim_{p \to 0} p\log p = 0$. If we have two random variable, this is some times referred as \emph{binary entropy}, which is defined as
\begin{align}
\boxed{H_{Bin}(p) \equiv -p\log p - (1-p)\log(1-p)}~,
\end{align}
where $p$ is the probability of the variables. Notice that $H_{Bin}(p) = H_{Bin}(1-p)$ and the maximum value is $H_{Bin}(p = 1/2) = 1$~. This is really useful to understand the behavior of Entropy when we have  mix more probabilities distribution. In the quantum description of entropy this is way harder to understood. There is a way to rewrite the uncertainty principle in terms of enttropy as
\begin{align}
H(\mathcal O_a) + H(\mathcal O_b) \geq -2\log f(\mathcal O_a, \mathcal O_b)~,
\end{align}
with $\mathcal O_a, \mathcal O_b)$ some different observable operators and $f(\mathcal O_a, \mathcal O_b) \equiv \max_{a,b} |\braket{a b}|$ by using $\mathcal O_a = \sum_a a\ket{a}\bra{a}$, and completely analogue for $\mathcal O_b$. Proofs and more deeper understanding in the context of quantum information theory can be found i.e. \cite{Nielsen_2009}.
%%%%%%%%%%%%%%%%%%%%%%%
\subsection{Classical Relative Entropy}\label{Sec:RelE}
We will now define \emph{Relative Entropy} which is a measure of how far (or close) are two probability distribution over the same index set. The quantum version of this measures how different are two quantum states, and this will be used in \autoref{Chp:Microscopic}. The classical version is defined as
\begin{align}
\boxed{H(p(x)||q(x)) = \sum_x p(x)\log \frac{p(x)}{q(x)} \equiv -H(X) - \sum_x p(x)\log q(x)}~.
\end{align}
This entropy is always positive define, $H(p(x)||q(x)) \geq 0$ where the equality is only if $p(x) = q(x)$, for all $x$. To prove this we use of the following inequality $-\log x \geq 1-x$~, therefore
\begin{align}
H(p(x)||q(x)) ={}& - \sum_x p(x) \log\frac{q(x)}{p(x)} \nonumber \\ \geq{}& \sum_x p(x)\left( 1 - \frac{q(x)}{p(x)}\right) \nonumber \\ ={}& \sum_x (p(x) - q(x)) \nonumber \\ ={}& (1-1) = 0~.
\end{align}
Which is the minimal value that occurs at $q(x) = p(x)$. This also follows the subadditivity condition
\begin{align}
H(p(x,y)||p(x)p(y)) = H(p(x)) + H(p(y)) - H(p(x,y))~,
\end{align}
which implies $H(X,Y) \leq H(X) + H(Y)$~. 
\\

Now we will define the \emph{joint entropy}. Giving a pair of random variables $X$ and $Y$, we can obtain the information content of $X$ related to the information content of $Y$ by the following definition
\begin{align}
\boxed{H(X,Y) \equiv -\sum_{x,y}p(x,y)\log p(x,y)}~,
\end{align}
adn with this define the condition entropy, whose measure of how uncertain we are, on average, about the value of $X$, given the value for $Y$
\begin{align}
\boxed{H(X|Y) \equiv H(X,Y) - H(Y)}~.
\end{align}
And finally we can also define the \emph{mutual information} which corresponds to subtract the joint information of the pair with the sum of both, \emph{viz}
\begin{align}
H(X:Y) \equiv H(X) + H(Y) - H(X,Y)~.
\end{align}
All this can be summarized with the "entropy Venn diagram"
\vspace{1cm}
\begin{center}
\begin{tikzpicture}
\draw[red,fill=red!80,opacity=0.4] (-2,0) circle (2.1cm);
\draw[red,fill=blue!80,opacity=0.4] (0,0) circle (2.1cm);
%\draw[thick,->] (-5,-0.8)--(-5,3);%y-axis
%\draw[thick,->] (-5.8,0)--(0,0);%x-axis
\node[scale=1] at (-2.2,2.5) {$H(X)$};
\node[scale=1] at (0.2,2.5) {$H(Y)$};
\node[scale=1] at (-3,0) {$H(X|Y)$};
\node[scale=1] at (1,0) {$H(Y|X)$};
\node[scale=1] at (-1,0) {$H(X:Y)$};

\draw[thick] (-2,0) circle (2.1cm);
\draw[thick] (0,0) circle (2.1cm);

%Caption~~~~~~~~~~~~~~~~~~~~
\node[text width=12cm, text justified] at (0,-4){
\small {\hypertarget{Fig:10}{\bf Fig.~10}}: 
\sffamily{Entropy Venn diagram which allows that allow us to obtain the relations between different entropies}};
\end{tikzpicture}
\end{center}
This provides a schematic guide to the properties of entropy.

\subsection{Quantum Relative entropy}\label{SubSec:QRE}
We can obtain the analogue of the classical Shannon entropy \eqref{Shannoncl}  in the quantum regime, in order to measure the uncertainty associated to quantum states. This based on the fact that states are defined in terms of density operators. The quantum analogue is known as \emph{Von Neumann entropy}, defined in terms of a quantum state $\rho$ as
\begin{align}\label{VonNeum}
\boxed{\mathcal S(\rho) \equiv -\Tr(\rho \log \rho)}~.
\end{align}
If the eigenvalues of $\rho$ are $\lambda_i$, then
\begin{align}
\mathcal S(\rho) = -\sum_i \lambda_i \log \lambda_i ~,
\end{align}
which is useful to calculate this entropy. 
\\

With this we can define the \emph{quantum relative entropy}, which defines the difference between two states 
(formally thought of as density matrices) $\rho$ and $\sigma$.  as
\begin{align}
\boxed{\mathcal S(\rho||\sigma) \equiv \Tr\left[\rho(\log\rho - \log\sigma)\right]}~.  
\end{align}
This entropy satisfy the Klein's inequality, which defines the positivity of this amount
\begin{align}
\mathcal S(\rho||\sigma) \geq 0~,
\end{align}
which can be proved by expanding the states in terms of density operators and the equality is obtained at $\rho = \sigma$.
Essentially, the relative entropy is a measure of how difficult is
to \emph{distinguish} the state $\rho$ from the state $\sigma$.
Statistically, in the particular case where $\rho$ and 
$\sigma$ are both thermal density matrices, 
with $\sigma$ the equilibrium state. This can be rewritten in terms of the free energy \cite{Donald_1987, Wilming:2017aa} as 
$$
\boxed{\mathcal S(\rho| |\sigma)= \Delta \mathcal F}~,
$$
where $\Delta \mathcal F$ measures how far $\rho$ 
is from the equilibrium state $\sigma$~. Relative entropy has the following properties \cite{ohya2004quantum, Vedral:aa}~ 
\begin{align}
\mathcal S(\rho| |\rho) ={}& 0~, \\ \mathcal S(\rho_1 \otimes \rho_2 | |\sigma_1 \otimes \sigma_2) =& \mathcal S(\rho_1 | |\sigma_1) + \mathcal S(\rho _2| |\sigma_2) \\ \mathcal S(\rho| |\sigma) \geq{}& \frac{1}{2}|| \rho - \sigma||^2~, \\ \mathcal S(\rho| |\sigma) \geq{}& \mathcal S(\Tr\rho| |\Tr \sigma)~,
\end{align}
where the last trace corresponds to tracing out respect some arbitrary subsystem. Also norm is defined as
\begin{align}
||\rho|| = \Tr \sqrt{\rho^\dagger \rho}~.
\end{align}
In terms of an operator $\mathcal O$ expectation value respect some density matrix $\expval{O}_\rho$, and using Schwarz inequality, it can be rewritten as 
\begin{align}
\mathcal S(\sigma|| \rho) \geq \frac{1}{2}\frac{(\expval{\mathcal O}_\rho - \expval{\mathcal O}_\sigma)^2}{||\mathcal O||^2}~.
\end{align}
It will be made used of relative entropy in \autoref{Chp:Microscopic} to conformally describe the de Sitter horizon.
\section{Entanglement Entropy}\label{Sec:EE}
Another important concept in quantum field theory is the one of \emph{Entanglement}, this does not exist in classical field theory. Therefore allow us to see how quantum a system is. Giving a Hilbert space that can be decompose as a tensor product 
\begin{align}\label{Hilbert}
\mathcal H_{tot} =  \mathcal H_a \otimes \mathcal H_b~,
\end{align}
a quantum system described by a wave function $\ket{\Psi}$ has a total density matrix of the form
\begin{align}
\rho_{tot} = \ket{\Psi}\bra{\Psi}~.
\end{align}
By this, the Von Neumann entropy \eqref{VonNeum} is zero. But we can use the decompose Hilbert space \eqref{Hilbert} and we can obtain the entropy of only one of the system, let us say system $a$, by tracing out the degrees of freedom of the region $b$ by defining a reduced density matrix as 
\begin{align}
\rho_{a} = \Tr_{b}\rho_{tot}~.
\end{align}
And the Von Neumann entropy \eqref{VonNeum} renders
\begin{align}\label{SEEq}
\boxed{\mathcal S_E = -\Tr_{b}(\rho_a \log \rho_a)}~.
\end{align}
This is known as \emph{entanglement entropy}, which provides how to measure how closely entangled a given function $\ket{\Psi}$ is (see \autoref{App:C} for an example). It can be seen as the entropy due to lack of information of an observer in $b$ who can not access degrees of freedom in $a$
\vspace{1cm}
\begin{center}
\begin{tikzpicture}
\draw[smooth cycle, tension=0.5, fill=white, pattern color=red, pattern=north west lines, opacity=0.7] plot coordinates{(0,2) (-2.5,0) (1,-2) (3,1)};
\draw[smooth cycle, tension=0.8, fill=white, pattern color=blue, pattern=north west lines, opacity=0.7] plot coordinates{(1,1.3) (-1.5,0.5) (1.3,-1) (2.2,0.5)};
\node[scale=1] at (0,1.6) {$\textbf{a}$};
\node[scale=1] at (1,0) {$\textbf{b}$};
%Caption~~~~~~~~~~~~~~~~~~~~
\node[text width=12cm, text justified] at (0,-4){
\small {\hypertarget{Fig:11}{\bf Fig.~11}}: 
\sffamily{Schematic representation of two entangled systems with one the complement of the other $b = a^c$.}};
\end{tikzpicture}
\end{center}
This can be seen as an analogue to black hole horizons, where some observer sitting outside the horizon $a$ is disconnected with the subsystem at $b=a^c$. 
\\

Another important quantity to define the Entanglement Entropy is the \emph{R\'enyi entropy} defined \cite{renyi1961} as
\begin{align}\label{renyi}
\mathcal S_a^q = \frac{1}{1-q}\log \Tr_{a}\rho^q_a~,
\end{align}
where $q\in \mathbb Z_+$, and \eqref{renyi} recovers the Von Neumman entropy \eqref{SEEq} in the limit $q\rightarrow 1$.
\begin{align}\label{SEren}
\boxed{\mathcal S_E = \lim_{q\to1}\mathcal S^q_a}~.
\end{align}
All this can be define in terms of thermal propeties of the density matrix, which in terms of $\beta$ as the inverse of the temperature is defined as
\begin{align}
\rho_a = \exp\{-\beta \mathcal H_a\}~.
\end{align}
Sometimes $\mathcal H_a$ is called \emph{modular Hamiltonian}. This definition satisfy that if the total density matrix is pure, ($T=0$ system), then we have
\begin{align}
\mathcal S_E(a) = \mathcal S_E(a^c)~,
\end{align}
which is violated at finite temperature. Also it satisfy the strong subadditivity relation
\begin{align}
\mathcal S_{a \cup b \cup c} + \mathcal S_{a} \leq{}& \mathcal S_{a \cup b} + \mathcal S_{b \cup c}~, \\  \mathcal S_a +  \mathcal S_c \leq{}& \mathcal S_{a\cup b} + \mathcal S_{b\cup c}~.
\end{align}
This conditions allow to derive all the other conditions for entanglement entropy. 
The thermal partition function 
\begin{align}
\boxed{\mathcal Z(\beta) \equiv \Tr_a \rho_a^q = \Tr_a \exp\{-\beta\mathcal H_a\}}~,
\end{align}
allow us to obtain the analogue of thermodynamical quantities in terms of the modular hamiltonian
\begin{align}
E(\beta) \equiv{}& -\partial_q\log \mathcal Z(\beta)~, &&& \text{\emph{modular energy}}~, \\ \widetilde{\mathcal S} (\beta)\equiv{}& (1-\beta\partial_\beta)\log\mathcal Z(\beta)~, &&&\text{\emph{modular entropy}}~, \label{ModS}\\ C(\beta) \equiv{}& \beta^2\partial^2_\beta \log \mathcal Z(\beta)~, &&& \text{\emph{modular capacity}}~.
\end{align}
And they are all positive defined.  
\\

Actually R\'enyi entropy and modular entropy are related by the equation
\begin{align}
\widetilde{\mathcal{S}} = q^2\partial_q \left(\frac{q-1}{q} S_q\right)~,
\end{align}
which can be inverted by
\begin{align}
\boxed{\mathcal S_q = \frac{q}{q-1}\int_1^q d\widehat{q}~ \frac{\widetilde{\mathcal S}_{\hat{q}}}{\hat{q}^2}}~,
\end{align}
and obtain the R\'enyi entropy in terms of the modular entropy. Also it is possible to define the \emph{relative R\'enyi entropy} as \cite{Muller-Lennert:2013aa}
\begin{align}
\mathcal S_q (\rho||\sigma) = \frac{1}{q-1}\log\left[\Tr\left(\sigma^{\frac{1-q}{2q}}\rho \sigma^{\frac{1-q}{2q}}\right)^q\right]~,
\end{align}
which is a monotonic function with respect to the parameter $q$.
\subsubsection{Replica Trick}
One of the most powerful tools to calculate the entanglement entropy is the so-called \emph{replica trick}, developed in the spin-glasses physics \cite{0305-4608-5-5-017}, which consist in compute the R\'enyi entropy by considering a functional integral in some generated branched cover geometry and analytically continuing the $q$-parameter we obtain the Entanglement entropy by using \eqref{SEren}. 
\\

The definition of entanglement entropy \eqref{VonNeum} together with \eqref{renyi} and \eqref{SEren} can be rewritten as
\begin{align}\label{SEErep}
\boxed{\mathcal S_E = -\lim_{q\to 1}\frac{\log \Tr_a \rho_a^q}{q-1} = -\lim_{q\to 1} \partial_q \log \Tr_a \rho_a^q~.}
\end{align}
where in the second equality an analytic continuation of the $q-$parameter to the real numbers. This is called the replica trick, which allow us to perform calculations of entanglement entropy in quantum field theories. 
\\

Cardy and Calabrese \cite{Calabrese:2004eu} have shown that the representation of the $q^{th}$ power of the reduced density matrix can be obtained by the partition function $\mathcal Z_q$ on the $q-$fold cover $\mathcal M_q$ manifold of the original spacetime, which is constructed by gluing $q$-copies of the sheet with a cut along $a$ (for some nice reviews and explanation of this method can be found for example in \cite{Nishioka:2009un, Ryu:2006ef, Rangamani:2016dms, 500000974160}).  By substituting this into \eqref{SEErep} we obtain 
\begin{align}
\boxed{S_E = -\lim_{q\to 1}\partial_q \left[ \log \mathcal Z_q - q\log \mathcal Z\right]~,}
\end{align}
where the $q-$folded cover manifold $\mathcal M_q$ has a conical singularity with deficit angle $\Delta\phi = 2\pi(1-q)$ which cast the inverse of the temperature as $\beta = 2\pi q$, which defines the partition function on this manifold as
\begin{align}
\mathcal Z(\beta) = \frac{\mathcal Z_q}{\mathcal Z^q}~.
\end{align}
In two-dimensional CFT it has been obtained the relation between entanglement entropy and the central charge of the theory with the universal divergent part of the entanglement entropy \cite{Calabrese:2004eu, Holzhey:1994aa, B.-Q.Jin:aa}
\begin{align}
\mathcal S_E = \frac{c}{3}\log \frac{L}{\epsilon} + \dots~,
\end{align}
where $\epsilon$ is some UV cutoff, $L$ is the size of the entanglement surface and the dots means finite parts. This also has been extended to four-dimensional CFT \cite{Fursaev:2013fta, Solodukhin:2008dh}. 
\\

The entanglement is an important property of quantum systems that has give deep inside in the understanding of quantum field theories from different point of views, but it sometimes hard to calculate. 
\subsubsection{Holographic Entanglement Entropy}\label{SubSub:HEE}
In 2006 Ryu and Takayanagi \cite{Ryu:2006bv} have conjecture how to calculate this quantity in conformal field theories by using the holographic gravity dual by the AdS/CFT correspondance, which by using the GKPW relation \cite{Gubser:1998aa, Witten:1998aa}, (which relates the partition function from a bulk gravity theory on an asymptotically AdS background, and the partition function of a conformal field theory living at the boundary of it) produces the entanglement entropy as 
\begin{align}\label{SEwit}
\mathcal S_E  = \lim_{q\to 1}\partial_q\left( I^E[\mathcal B_q] - qI^E[\mathcal B]\right)~,
\end{align} 
where $I^E[\mathcal B_q]$ corresponds to the Euclidean on-shell action of some solution $\mathcal B_q$. Ryu and Talayanagi have proposed that the holographic entanglement entropy is 
\begin{align} 
\boxed{\mathcal S_E = \frac{\mathcal A_\gamma}{4G_{d}}~,}
\end{align}
where $\mathcal A_\gamma$ is the area of some minimal surface $\gamma$ in $\mathcal B$ which is attached on the entangled surface. It is provided that $\mathcal M_q$ is invariant under $\mathbb Z_q$ symmetry that shifts the modular time by a factor of $2\pi$ for integer $q$. For this proposal it must be proposed that the bulk geometry $\mathcal B_q$,  must be a solution of Einstein equations that is invariant under the replica $\mathbb Z_q$ symmetry which allows to define the orbifold
\begin{align}
\widehat{\mathcal B}_q \equiv \mathcal B_q/\mathbb Z_q~,
\end{align}
which by construction have a conical singularity with deficit angle $\Delta\phi = 2\pi\left(1-\frac{1}{q}\right)$. The original bulk solution $\mathcal B_q$ must be attached to the singular boundary $\mathcal M_q$, but the boundary of the orbifold is regular and corresponds to the original CFT manifold
\begin{align}
\partial \widehat{\mathcal{B}}_q = \partial \mathcal B = \mathcal M~.
\end{align} 

%%%%%%%%
\begin{center}
\begin{tikzpicture}[scale=0.6]
\clip(-7,-5.75) rectangle (28,8.79);
%bottom ads
\draw[thick,fill=wwccqq,opacity=0.3,line width=2pt] (3.57,6.2) arc[start angle=90,end angle=-90, x radius=4, y radius=2.2];
\draw[thick,line width=1pt, dotted] (1,-0.3) arc[start angle=140,end angle=50, x radius=5.5, y radius=5];
\draw[thick,line width=1pt, dotted] (6,2) arc[start angle=140,end angle=50, x radius=5.5, y radius=5];
\draw [line width=1pt] (1,6)-- (1,-0.3);
\draw [line width=1pt] (1,-0.3)-- (6,2);
\draw [line width=1pt] (1,6)-- (6,8);
\draw [line width=1pt] (6,8)-- (6,2);
\draw [rotate around={-89.7303747925932:(3.57,4.005)},line width=2pt,dash pattern=on 1pt off 1pt,color=wvvxds!70,fill=wvvxds!70,fill opacity=0.8] (3.57,4.005) ellipse (2.2257624666640936cm and 0.662037429463502cm);
%top ads
\draw[thick,line width=1pt,dotted] (1,6) arc[start angle=-140,end angle=-50, x radius=6, y radius=3];
\draw[thick,line width=1pt, dotted] (6,8) arc[start angle=-140,end angle=-50, x radius=5, y radius=3];
%nodes
\node[scale=1.4,text=wwccqq,thick] at (7,5.7) {$\gamma$};
\node[scale=1.4,text=black,thick] at (3.5,4) {$a$};
\node[scale=1.2,text=black] at (8,-1.3) {$r$};
\draw[thick,->] (1,-0.8)--(8,-0.8);
%Caption~~~~~~~~~~~~~~~~~~~~
\node[text width=16cm] at (8,-4) {\small {\hypertarget{Fig:11}{\bf Fig.~11}}: 
\sffamily{Holographic entanglement entropy is computed as the area of  the minimal \\ surface $\gamma$ (in green) which is attached at the boundary of  the region $a$ of certain CFT$_d$. \\ This surface lives at some asymptotically AdS$_{d+1}$ spacetime (represented with the \\ dotted lines), whose holographic emergent coordinate is labeled by $r$.}};
\end{tikzpicture}
\end{center}
%%%%%%%%
Also can be defined the \emph{bulk-per-replica action} for the orbifold as
\begin{align}
\widehat{I}[\widehat{\mathcal{B}}_q] = I[\mathcal B]/q~,
\end{align}
and \eqref{SEwit} becomes
\begin{align}
\boxed{\mathcal S_E = \partial_q \widehat{I}[\widehat{\mathcal{B}}_q] \rvert_{q = 1}~.}
\end{align}
Due to the conical singularity at the orbifold $\widehat{\mathcal B}_q$ the Einstein action evaluated at this obtains an extra co-dimension $2$ contribution in order to solve the field equations \cite{Fursaev:1995ef}. To see this notice that the Ricci tensor in the presence of this deficits becomes
\begin{align}
R\rvert_{\widehat{\mathcal B}_q} = R\rvert_{\mathcal B} + 4\pi\left(1-\frac{1}{q}\right)\delta_{\gamma}~, 
\end{align}
with $\delta_\gamma$ projector onto the $\gamma-$surface, \emph{viz.} 
$\int_{\widehat{\mathcal B}_q} f \delta_{\gamma}=\int_{\gamma_q} f|_{\gamma_q}$. This leads to a guaranteed minimal $\gamma_1$-surface which corresponds to a cosmic brane with tension $\frac{\Delta\phi}{8\pi G}$ \cite{Dong:2016fnf}. 
\\

The Ryu-Takayanagi formula has been proved to satisfy the strong-subadditivity inequality \cite{Headrick:2007km} and the monogamy of mutual information \cite{Hayden:2011ag}. Also a derivation of holographic entanglement entropy for spherical entangling surfaces has been give in \cite{Casini:2011aa}, and generalized in \cite{Lewkowycz:2013aa}.
\\
To summarize, in the holographic prescription the replica trick consist on replicates $q-$times the original boundary $\mathcal B$ along the entangling region $\mathcal B' \subseteq \mathcal B$, which construct a q-fold branched cover $\mathcal B_q$. This last must admit a $\mathbb Z_q$ action whose fixed points are the entangling surface $\partial \mathcal B'$. 

%% file: Chapters/KerrCFT.tex
% Chapter 1

\chapter{The Kerr/Conformal Field Theory Correspondance} % Main chapter title

\label{KerrCFT} % For referencing the chapter elsewhere, use \ref{Chapter1} 

%----------------------------------------------------------------------------------------

%----------------------------------------------------------------------------------------

Since the black hole thermodynamics defined mainly by Bardeen, Bekenstein, Carter and Hawking \cite{Bekenstein:1974ax, Bekenstein:1973ur, Bekenstein:1972tm, Bardeen:1973gs, Hawking:1974sw, Hawking:1976de}, which corresponds to a macroscopic description, the entropy area law
\begin{align}
S = \frac{A_{\mathcal H}}{4\hbar G}~,
\end{align}
has been relevant to understand the microscopic underlying degrees of freedom, which seems imperative to be defined in this form in order to have a well defined thermodynamics and understand what really this degrees of freedom actually are. For a particular black hole geometry, this problem has been solved using string theory \cite{Strominger:1996sh} and showed that an unitary quantum gravity theory that contain this black hole solutions must resemble in the same way \cite{Strominger:1997eq}~.

In \cite{Guica:2008mu} has been proposed that quantum gravity near the extreme Kerr horizon is dual to a two-dimensional CFT, which corresponds to a proposal to realize the holographic principle for a realistic gravitational setting. These allow to count the microscopic degrees of freedom of an extremal rotating black holes in four dimensions by analyzing the asymptotic symmetries of the background. Extensions of this correspondence has been obtained during the years. 

In this chapter we will give a short review of the original paper of the Kerr/CFT correspondence that will be used to describe the cosmological horizon microscopically 

\section{Kerr black hole and the extremal limit}\label{Sec:Kerr}
The assymptoticall flat rotating solution to General Relativity were found by Roy Kerr in \cite{Kerr:1963ud}, which in Bonyer-Liquidist form reads
\begin{align}
ds^2 = -\frac{\Delta}{\rho^2}(dt - a\sin^2\theta d\phi)^2 + \frac{\rho^2}{\Delta}dr^2 + \frac{\sin^2\theta}{\rho^2}\left( (r^2 + a^2)d\phi - adt\right)^2 + \rho^2d\theta^2~,
\end{align}
\begin{align}\label{metricfunc}
\Delta = r^2 - 2Mr + a^2~, \qquad \rho^2 = r^2 + a^2 \cos^2\theta~,
\end{align}
with $a = \mathcal J /M$ where $M$ corresponding to the mass and $\mathcal J$ the angular momentum. The black hole horizon is defined where the function $\Delta(r)$ becomes equals to zero, there are two solutions for this, the outer horizon denoted by $r_+$ and the inner horizon by $r_-$:
\begin{align}
r_\pm = M \pm \sqrt{M^2 - a^2}~.
\end{align}
We can immediately notice that if $a^2 > M^2$, then the root becomes imaginary, and therefore are no horizons but a naked singularity, which violates the cosmic censorship. This implies that the angular momenta is bounded for below, i.e. $\mathcal J \leq M^2$, this also because the angular velocity of the horizon is equal to the speed of light at the extremal limit $\mathcal J = M^2$ or $a = M$.  The Hawking Temperature reads 
\begin{align}
T_H = \frac{r_+ - M}{4\pi M r_+} = 0~,
\end{align}
where the second equality corresponds to the extremal case. So for maximum rotation there is no Hawking radiation, and the extremal black hole can be seen as ground states. And the Bekenstein Entropy becomes
\begin{align}
\mathcal S = 2\pi Mr_+ \implies \mathcal S_{Ext} =  2\pi \mathcal J~,
\end{align}
where we have use that in the extremal case $r_+ = M = \sqrt{\mathcal J}$.

Also the metric functions \eqref{metricfunc} are reduced to 
\begin{align}
\Delta_{Ext} = (r-a)^2~,\qquad \rho^2_{Ext} = r^2 + a^2 \cos^2\theta~.
\end{align}
\subsection{Near-Horizon Limit of Extreme Kerr}\label{SubSec:ExtremeKerr}
In the near-horizon region for the extremal Kerr solution matter must rcorotate with the black hole does at the speed of light. Hence, only chiral modes appears. In this sense, Quantum Gravity near this region is expected to simplify drastically. 

The Near-Horizon limit of extreme Kerr is obtained by introducing the scaling parameter $\lambda$ in the extremal case $a = M$. Changing coordinates as in \cite{Bardeen:1999px}:
\begin{align}\label{extcoord}
\hat r = \frac{r - M}{\lambda M}~,\qquad \hat t = \frac{\lambda t}{2M}~, \qquad \hat\phi = \phi - \frac{t}{2M}~.
\end{align}
Such that any value of $\hat r$ is forced to be near to the horizon $r_+ = M$ when $\lambda \rightarrow 0$. The resulting geometry adopts the angular momentum as an overall factor, and the metric reads
\begin{align}\label{NHEKerr}
ds^2 = 2\Omega^2 J\left[  - (1+\hat r^2) d\hat t^2  + \frac{d\hat r^2}{1+\hat r^2} + d\theta^2 + \Lambda^2 (d\hat\phi + \hat r d\hat t)^2\right]~.
\end{align}
With
\begin{align}
\Lambda = \frac{2\sin\theta}{1 + \cos^2\theta}~,\qquad \Omega^2 = \frac{1+ \cos^2\theta}{2}~.
\end{align}
The metric \eqref{NHEKerr} is referred to Near Horizon Extremall Kerr geometry (NHEK), which remains as a solution of Einstein equations, because is a limit of a coordinate transformation of Kerr solution. The asymptotics $\hat r\rightarrow \infty$ is very peculiar, actually \eqref{NHEKerr} is not asymptotically flat nor asymptotically AdS.

For fixed values of $\theta$ we obtain a warped version of $AdS_3$ geometry. In a particular value $\theta_0$ the function $\Lambda(\theta_0) = 1$ and the line element is exactly that of $AdS_3$
\begin{align}
ds^2 = 2\Omega^2 J\left[  - (1+\hat r^2) d\hat t^2  + \frac{d\hat r^2}{1+\hat r^2} + (d\hat\phi + \hat r d\hat t)^2\right]~.
\end{align}
It was shown by Brown and Henneaux \cite{Brown:1986nw}, that gravity on $AdS_3$ background always has a conformal symmetry. The warped version of $AdS_3$ has been study mainly in the Topological Massive Gravity and String theory context \cite{DHoker:2010aa, Compere:2009ab, Compere:2009aa, Anninos:2009ac, Anninos:2009ab, Anninos:2009aa, Herzog:2008aa, Bouchareb:2007aa, Rooman:1998aa, Maldacena:1998aa}. 
This warped $AdS_3$ is the Lorentzian analogue of the squashed $S^3$. Due to the warp factor, the $S^3$ isometry group is break from $SU(2)\times SU(2)$ (this can be seen easily by the fact that the $S^3$ can be seen as a Hopf fibatrion of $S^1$ over $S^2$) down to $SU(2)\times U(1)$. In the $AdS_3$ case, the warp factor breaks the isometry group $SL(2,\mathbb R)_R \times SL(2,\mathbb R)_L$  down to $SL(2,\mathbb R) \times U(1)$, which will corresponds to the isometry group at each fix $\theta$. 
\\

\subsection{Temperature and Chemical Potentials}\label{SubSec:Temperature}
We have seen that at extremality the Hawkin temperature is zero, but that does not mean that the quantum states outside the event horizon are in pure states.
Defining the chemical potential
\begin{align}
\left(\frac{\partial \mathcal S_{Ext}}{\partial \mathcal J} \right) = \frac{1}{T_\phi}~,
\end{align}
And the potentials must satisfy the equation
\begin{align}
\delta \mathcal S_{Ext} = \frac{1}{T_\phi}\delta\mathcal J~.
\end{align}
We can now obtain the potentials by using the first law of Black Hole Thermodynamics 
\begin{align}\label{DeltaSext}
\delta\mathcal S_{Ext} = \frac{1}{T_H}\left(\delta M - \Omega_{J}\delta\mathcal J \right)~,
\end{align}
Where $\Omega_{Ext}$ is the angular velocity. This at extramility becomes
\begin{align}
T_H\delta\mathcal S_{Ext} = \delta M - \Omega_{Ext}\delta\mathcal J = 0~.
\end{align}
We can take the extremal limit by vanishing the Hawking Temperature, and comparing with \eqref{DeltaSext}, then
\begin{align}\label{FrolovTemp}
T_\phi = \lim_{T_H \rightarrow 0}\,\frac{T_H}{\Omega_{Ext} - \Omega_{J}} = -\frac{\partial T_H/\partial r_+}{\partial \Omega_J /\partial r_+}\Big\rvert_{r_+=r_{Ext}} = \frac{1}{2\pi}~.
\end{align}
Which corresponds to the quantum states near the horizon. To compute explicitly this value we must interpret this chemical potential in the context of quantum fields living on curved spacetimes. A first example for this, is the Hartle-Hawking vacuum for a Schwarzschild Black Hole, which is restricted to a region outside the horizon is defined as the diagonal density matrix
\begin{align}
\rho = \exp\left\{-\hbar\frac{\omega}{T_H}\right\}~.
\end{align}
In the NHEK case, a global timelike Killing vector is not defined, then there is no quantum states with all the desire properties of a vacuum, and the Hartle-Hawking vacuum do not exist.
Anyhow, Frolov and Thorne define the vacuum for extreme Kerr geometry by using the generator of the horizon which is timelike from the horizon out to the surface where an observer must move at the speed of light to corotate with the extremal black hole \cite{Frolov:1989jh}. 
The Frolov-Thorne vacuum is the Kerr analogue of the Schwarzschild Hartle-Hawking vacuum. 

By expanding the quantum fields in eigenmodes of the angular momentum $m$ and the asymptotic energy $\omega$, for example for the scalar field, we have
\begin{align}
\Phi = \sum\limits_{\omega, m, \ell} \phi_{\omega m \ell}\exp\{-I(\omega  t - m\phi)\}f_\ell (r,\theta)~.
\end{align}
This defines the Frolov-Thorne vacuum as the diagonal density matrix
\begin{align}
\rho = \exp\left\{-\hbar\frac{\omega - \Omega_J}{T_H}\right\}~,
\end{align}
which reduces to the Hartle-Hawking vacuum in the non-rotating case. Now, by using the Near-Horizon coordinates \eqref{extcoord} and following Frolov-Thorne we get 
\begin{align}
\exp\{-i(\omega t - m\phi)\} = \left\{\frac{-i}{\lambda}(2M\omega - m)\hat t  + im\hat\phi \right\} = \exp\{-I n_R \hat t + i n_L \hat\phi\}~,
\end{align}
where $n_L \equiv m$ and $n_R \equiv (2M\omega -m)/\lambda$ are the left and right Frolov-Thorne charges associated to $\partial_{\hat\phi}$ and $\partial_{\hat t}$ respectively. This defines the Frolov-Thorne temperatures by rewriting the Boltzman factors in terms of this new variables
\begin{align}
\exp\left\{ -\hbar\frac{\omega - \Omega_J m}{T_H} \right\} = \exp\left\{-\frac{n_L}{T_L} - \frac{n_R}{T_R} \right\}~,
\end{align}
where the left and right temperatures are
\begin{align}
T_L = \frac{\hat r_+ - M}{2\pi(\hat r_+ - a)}~, \qquad T_R = \frac{\hat r_+ - M}{2\pi\lambda r_+}~.
\end{align}
Which in the extremal limit reduces to 
\begin{align}\label{temper}
T_L = \frac{1}{2\pi}~,\qquad T_R = 0~.
\end{align}
And we recognize $T_{\hat\phi} = T_L = (2\pi)^{-1}$ and the quantum fields near the horizon are not in a pure state, even if $T_H = 0$.

\section{Asymptotic Symmetry Group}
It has been pointed out that the symmetries of the classical theory are enough to determine the states of a quantum gravity theory \cite{Carlip:1999cy}. Therefore, we must specify the boundary conditions if we want make sense of quantum gravity on the NHEK, and how has been pointed out in \autoref{SubSec:ExtremeKerr} the asymptotic of \eqref{NHEKerr} at $r\rightarrow\infty$ is not flat nor $AdS$, therefore the set of boundary conditions is not trivial to achieve. For every set of boundary conditions there is an associated Asymptotic Symmetry Group (ASG) which is defined as the set of Allowed symmetry transformations modulo the set of trivial symmetries \cite{Guica:2008mu}
\begin{align}\label{ASG}
\text{ASG} = \frac{\text{Allowed diffeomorphism}}{\text{Trivial diffeomorphism}}~.
\end{align}
The allowed diffeomorphisms are defined as the transformations that are consistent with the specified set of boundary conditions. This means that given boundary conditions, the allowed diffeomorphism $\zeta$ produces a variation of the metric $\delta_\zeta g$ that is a allowed metric, such that the metric variations must asymptotically  vanish in order to not violate boundary conditions. 
\\

For the trivial diffeomorphisms means that the generators of transformations produces only vanishing conserved charges. Therefore, if we impose boundary conditions that are too strong, this would render the theory trivial, i.e. only the vacuum is an accepted state. 
And also if we try to demand boundary conditions that are too weak the symmetry generators would be all divergent. Then, the boundary conditions must be chosen such that the theory is not all divergent nor trivial. 
\\

The trivial diffeomorphism must transform in the ASG representation and should annihilate the states in the quantum theory. In the Kerr/CFT correspondence has been obtained a set of boundary conditions which gives the Virasoro algebra as the generator of the  ASG and the resulting theory has states transforming in the two-dimensional conformal group representations. 

\subsection{Boundary Conditions}
The set of boundary conditions that defines the ASG for the NHEK were obtained in \cite{Guica:2008mu}. A deviation $h$ of the full metric \eqref{NHEKerr} $\delta g = h$ require the set of boundary conditions 
\begin{alignat}{4}
\label{BC}
h_{\hat t \hat t} &\sim \mathcal O (\hat r^{2})& ~,~~
h_{\hat t \hat{r}} &\sim \mathcal O (\hat r^{-2})& ~,~~
h_{\hat t \theta} &\sim \mathcal O (\hat r^{-1} )& ~,~~ 
h_{\hat t \hat\phi} &\sim \mathcal O (\hat r) ~,~~ \\ \nonumber
&&
h_{\hat{r} \hat{r}} &\sim \mathcal O (\hat r^{-3})& ~,~~
h_{\hat{r} \theta} &\sim \mathcal O (\hat r^{-2})& ~,~~
h_{\hat{r} \hat\phi} &\sim \mathcal O (\hat r^{-1}) ~,~~ \\ \nonumber
&&&&
h_{\theta \theta} &\sim \mathcal O (\hat r^{-1} )& ~,~~
h_{\theta \hat\phi} &\sim \mathcal O (\hat r^{-1}) ~,~~  \\ \nonumber
&&&&&&
h_{\hat\phi \hat\phi} &\sim \mathcal O (1) ~.
\end{alignat} 
This set of boundary conditions implies that the energy $\mathcal E = Q[\partial_{\hat t}]$ is equal to zero at extremality, i.e. $\mathcal E = M^2 - \mathcal J = 0$. 

The most general diffeomorphisms which preserve \eqref{BC} have the form
\begin{align}
\xi = \left( C + \mathcal O(\hat r^{-3})\right)\partial_{\hat t} + \left(-\hat r\varepsilon'(\hat\phi) + \mathcal O(1)\right)\partial_{\hat r} + \left(\varepsilon(\hat\phi) + \mathcal O(\hat r^{-2})\right)\partial_{\hat \phi} + \mathcal O(\hat r^{-1})\partial_{\hat\theta}~,
\end{align}
with $\varepsilon(\hat\phi)$ is an arbitrary smooth function and $C$ is some constant. Computing the generators, the subleading terms corresponds only to trivial diffeomorphisms and do not contribute. Then the ASG generators its 
\begin{align}\label{zetas}
\boxed{\zeta_\varepsilon = \varepsilon(\hat\phi)\partial_{\hat\phi} - \hat r\varepsilon'(\hat\phi)\partial_{\hat r}}~.
\end{align}
Since $\hat\phi \sim \hat\phi + 2\pi$, we can expand the function $\varepsilon(\hat\phi)$ in Fourier modes $\varepsilon_n(\hat\phi) = -\exp\{-in \hat\phi \}$. In this basis the ASG generator reads
\begin{align}\label{differs}
\zeta_n \equiv \zeta_{\varepsilon_n} = -\exp\{-in \hat\phi \} \left(\partial_{\hat\phi} + in\hat r\partial_{\hat r}\right)~.
\end{align}
We can recognize the $U(1)$ generator  as the zero mode $\zeta_0 = -\partial_{\hat\phi}$.

The (classical) Lie brackets of the Fourier-expanded generators form the so-called Witt algebra defined in \autoref{SubSec:Generators2D}
\begin{align}
i[\zeta_n , \zeta_m] = (n-m)\zeta_{n+m}~.
\end{align}
This corresponds to only one copy of the conformal group in two dimensions on the circle, but \eqref{zetas} does not contain the $SL(2,\mathbb R)$ generator, only the $U(1)$.

\subsection{Asymptotic Charges}
The generators of the diffeomorphisms \eqref{differs} associated with asymptotic symmetries obey, under Dirac brackets, the same algebra as the symmetries themselves, up to a possible non-zero central term given by the Jacobi identity. 
For pure Einstein gravity with cosmological constant, the associated $d-2$ form has been obtained in \cite{Barnich:2001jy}
\begin{align}\label{surfchrg}
k_\xi [h,g] = -\delta k_\zeta^K[g] - \zeta \cdot \Theta [h,g] - k_{\mathcal{L}_{\zeta}g}^S[h,g]~,
\end{align}
where the first two terms have been obtained in \cite{Wald:1999wa, Iyer:1994ys} using covariant phase space methods
\begin{align}
k_\xi^K [g] ={}& \frac{\sqrt{-g}}{16\pi G}\left( \nabla^\mu \zeta^\nu - \nabla^\nu\zeta^\mu \right) d^{d-2} x_{\mu\nu} \\ \Theta[h,g] ={}& \frac{\sqrt{-g}}{16\pi G}\left(\frac{1}{2}g^{\mu\alpha}\nabla^\beta h_{\alpha\beta} - g^{\alpha\beta}\nabla^\mu h_{\alpha\beta}\right) d^{d-2}x_\mu~.
\end{align}
And the supplementary term vanishes for an exact Killing vector of the background metric $g$, but not necessarily for the asymptotic Killing vectors
\begin{align}
k_{\mathcal L_\zeta g}^S [h,g] = \frac{\sqrt{-g}}{16\pi G}\left(\frac{1}{2}g^{\mu\alpha}h_{\alpha\beta}(\nabla^\alpha\zeta^\beta + \nabla^\beta \zeta^\alpha) - (\mu \leftrightarrow \nu)\right) d^{d-2}x_{\mu\nu}~,
\end{align}
for asymptotically AdS spacetimes the expression \eqref{surfchrg} coincides with the one obtained firstly by \cite{Abbott:1981ff}.  By construction \eqref{surfchrg} is a closed form (i.e., $dk_\xi = 0$), and its integral defines the surface charges 
\begin{align}
\delta Q^{Eins}_\zeta [g] = \frac{1}{8\pi G} \int_{\partial\Sigma} k_\zeta [h, g]~,
\end{align}
Where the integral is over the boundary of a spatial slice and 
\begin{align}\label{twoform}
k_\zeta [h, g] =  -\frac{1}{4} \epsilon_{\alpha\beta\mu\nu} \Big[ 
\zeta^\nu \nabla^\mu h - \zeta^\nu \nabla_\sigma h^{\mu\sigma} 
&+ \zeta_\sigma\nabla^\nu h^{\mu\sigma} +\frac{1}{2} h \nabla^\nu \zeta^\mu
-h^{\nu\sigma}\nabla_\sigma \zeta^\mu \nonumber \\ 
&+ \frac{1}{2} h^{\sigma\nu} ( \nabla^\mu\zeta_\sigma+ \nabla_\sigma\zeta^\mu)
\Big] dx^\alpha \wedge dx^\beta ~.
\end{align}
Following the method of \cite{Compere:2012jk, Barnich:2001jy, Barnich:2006av, Compere:2009zj, Barnich:2003xg, Compere:2007az} we compute the central charge of the dual CFT by varying the charges to obtain the Dirac bracket
\begin{align}\label{qmalgebra}
\left\{Q_{\zeta_{m}},Q_{\zeta_{n}}\right\} = -i(m-n)Q_{\zeta_{m+n}} + \delta Q^{Eins}_{\zeta_m}[\mathcal L_{\zeta_n}g, g]~.
\end{align}
Here, $\mathcal{L}_{\zeta_{m}}g$ is the Lie derivative of the metric along $\zeta_{m}$ and $k_{\zeta_{m}}$ is the two form defined in Eq. \eqref{twoform}.
In the case of the NHEK geometry the Lie derivatives reads
\begin{align}
\left(\mathcal L_{\zeta_n}g\right)_{\hat t \hat t} ={}& 4G\mathcal J \Omega^2 (1-\Lambda^2)\hat r^2in\exp\{-in\hat\phi\}~, \nonumber \\
\left(\mathcal L_{\zeta_n}g\right)_{\hat \phi \hat \phi} ={}& 4G\mathcal J \Omega^2 \Lambda^2 in\exp\{-in\hat\phi\}~, \nonumber \\
\left(\mathcal L_{\zeta_n}g\right)_{\hat r \hat r} ={}& -4G\mathcal J\Omega^2(1+\hat r^2)^{-2}in\exp\{-in\hat\phi\}~, \nonumber \\
\left(\mathcal L_{\zeta_n}g\right)_{\hat r \hat\phi } ={}& -2G\mathcal J\Omega^2(1+\hat r^2)^{-1}\hat r n^2\exp\{-in\hat\phi\}~.
\end{align}
Which turns 
\begin{align}
\frac{1}{8\pi G}\int_{\partial\Sigma} k_{\zeta_m}[\mathcal L_{\zeta_n} g, g] = -i\mathcal J (m^3 + 2m)\delta_{m+n,0}~.
\end{align}
In the quantum version, the extension from the Dirac brackets to commutators ($\{ \, , \, \}\rightarrow \frac{1}{i\hbar}[\, , \,]$) we can define the quantum counterpart of the symmetry generators, with a shift on the zero mode 
\begin{align}
\hbar L_m \equiv Q_{\zeta_m} + \frac{3}{2}\mathcal J \delta_{m,0}~.
\end{align}
In general we can always shift the zero mode of the Virasoro generators and therefore the linear term on the central extension of the Witt algebra does not play a role. 
Hence the central charge is given by the $m^3$ factor of the Virasoro algebra, which for Einstein gravity can be calculated as
\begin{align}
c = \frac{12i}{\hbar} \lim_{\hat r \rightarrow \infty} Q^{Eins}_{\zeta_m}[\mathcal L_{\zeta_{-m}}g, g]\Big\rvert_{m^3}~.
\end{align}
This defines the asymptotic algebra \eqref{qmalgebra} into the Virasoro algebra
\begin{align}
[L_m , L_n] = (m-n)L_{m+n} + \frac{c}{12}(m^3 - m)\delta_{m+n,0}~,
\end{align}
This corresponds to one copy of the Virasoro algebra which defines a chiral two dimensional CFT near the horizon of an extremal Kerr Black Hole with central charge
\begin{align}\label{charge}
\boxed{c = \frac{12\mathcal J}{\hbar}}~.
\end{align}
Then the central charge depends only in the extremal angular momenta divided by $\hbar$, this is a huge number that the Event Horizon Telescope expect to reproduce \cite{Doeleman:2017nxk}, in {\cite{Gates:2018hub, Gralla:2017ufe} its proposed by the study of symmetries of the NHEK to predict the  polarized near-horizon emissions to be seen at the Event Horizon Telescope.

\section{Microscopic Entropy of Extremal Kerr Black Hole}
As we show in \autoref{SubSec:Temperature} the quantum theory in the Frolov-Thorne vacuum has only left-moving temperature and by the Virasoro algebra we identify the CFT as chiral. 
We now can make use of the Card formula derived in \autoref{Sec:Cardy}, which relates the entropy for a unitary CFT on the canonical ensemble with its central charge and temperature
\begin{align}
\mathcal S = \frac{\pi^2}{3}(c T_L + \bar c T_R)~.
\end{align}
In the NHEK case the CFT is only chiral with temperature \eqref{temper} and central charge \eqref{charge}, which combines to reproduce
\begin{align}
\boxed{\mathcal S_{Ext} = \frac{2\pi}{\hbar}\mathcal J = \frac{\mathcal A_{\mathcal H}}{4\hbar G}}~.
\end{align}
Which match exactly with the Bekenstein-Hawking entropy of extreme Kerr Black Hole. 
\\

Then quantum gravity in the region near the horizon of an extreme Kerr black hole is considered by the study of diffeomorphisms, which has give one copy of the Virasoro algebra which allows us to identify the quantum states near the horizon with those of a chiral two-dimensional CFT, which by assuming unitarity, the tools of 2d CFT allows to compute the microscopic degrees of freedom via the Cardy's formula \cite{Cardy:1986ie, Bloete:1986qm} which reproduces exactly the macroscopic Bekenstein-Hawking entropy. 
Therefore was conjectured in the original Kerr/CFT paper \cite{Guica:2008mu} that the near-horizon region of an extreme Kerr black holes are holographically dual to a chiral two-dimensional CFT with central charge \eqref{charge}.
\\

This mechanism has been extended to extremal non-rotating black holes or extremal charged-rotating black holes even in higher dimensions. \cite{Compere:2012jk, Lu:2008jk, Astorino:2015naa}. Also for Kerr-AdS black hole in higher dimensions \cite{Lu:2008jk}, but for asymptotically de Sitter spacetimes this has not been achieved. An attempt to reproduce the results of the Kerr/CFT correspondance for asymptotically $dS$ spacetime was given in  \cite{Anninos:2009yc}, in where the Kerr-$dS$ black hole is extremal in the sense that the cosmological horizon coincides with the horizon of the black hole. 
\\

Latter on we will made use of this correspondence and made use of the mechanism by mimicking the metric structure of \eqref{NHEK} on the $dS_4^q$. It is shown that using this mechanics, we can describe the branes by a CFT and associated the resulting Gibbons-Hawkin entropy with the \emph{relative entropy}.

%% file: Chapters/Entanglement.tex
% Chapter 5

\chapter{De Sitter Entropy from Entanglement} % Main chapter title
\label{Chp:dSEntanglement} % For referencing the chapter elsewhere, use \ref{Chapter1} 
We will now made use of the extended static patch \eqref{metric} which can be re written as a fibration of $dS_2$ and a two-sphere $S^2\times_w dS^2$, where the warp factor $w$ is given in terms of the polar coordinate of the sphere and $dS_2$ is the radially extended standard de Sitter in two dimensional space. As it is mentioned in \autoref{dS}, this extended patch covers both Rindler wedges that describes the word-line of two causally disconnected antipodal observers. \\

Another important fact is that in this patch the horizon $\mathcal H$ is described as the $S^2$ factor. It can be constantly deformed by identifications of the azimuthal coordinate, which generates a conical deficit $\Delta\phi = 2\pi(1-1/q)~$ which is defined through the orbifold $S^2/\mathbb Z_q$ whose fixed points corresponds to two antipodal two-dimensional defects with local $dS_2$ geometry that can be described with a fixed Nambu-Goto metric coupled through tension $\mathcal T_q \sim (1-1/q)$ for each defect.  
\\

On this chapter we will made use of this extended patch in order to perform explicit calculations of entanglement entropy, between two disconnected observers that live at the past- and future-infinity, and also between two antipodal bulk observers. For this we will assume the existence of an holographic in which quantum gravity on de Sitter background is dual to two copies of a certain conformal field theory living one on each of the past or future infinity boundaries of de Sitter. This bulk observers is then described in terms of a thermofield double states (\autoref{App:C}) in the tensor product of the field theory Hilbert spaces. Therefore the resultant density matrix is thermal. 
\\

For the case of entanglement between disconnected regions we do not need an holographic duality but just the assumption that exist a quantum gravity action whose on-shell semiclassical limit is described by the euclidean Einstein gravity $\mathcal Z_{QG} \approx \exp\{-I_E\}$~.\\

We finally obtain a formula to describe the entanglement entropy which follows the same form as the Bekenstein-Hawking Entropy but corresponds to the area of the set of fixed points $\mathcal F$ of the bulk action
\begin{align}
\boxed{S_E = \frac{\text{Area}(\mathcal F)}{4G_4}~.}
\end{align}
\section{Entaglement from two disconnected boundaries}
In 2001, Strominger has postulate it that exist a duality between certain conformal field theories and quantum gravity on a de sitter background \cite{Strominger:2001pn}. This describes the correspondence of quantum gravity on $dS_d$ to be dual to a conformal field theory on a lower dimensional sphere $S^{d-1}$. This is supported by explicit calculation of the correlation functions of a massive scalar field.  This conformal field theory lives at the future $\mathcal I^+$ or past $\mathcal I^-$ infinity, which corresponds to two disconnected conformal boundaries, therefore this CFTs can be Euclidean and possibly non-unitary. Anyhow this correspondence has been criticized for the lack of a supergravity description on a de Sitter background  \cite{Obied:2018aa, Kachru:2003aw}, and using a timelike T-duality generates a supergravity theory which has ghost \cite{Hull:1998vg}. And it related AdS/CFT formulations relies heavily on a string theory background. Also some criticism in terms of the two point function for thermal states can be found in \cite{Dyson:2002aa} which has been argued that can be solved in three dimensional case in \cite{Klemm:2002aa}, where they map the $dS_3$ entropy to the Liouville momentum and does not acquire the thermal properties of the bulk theory, also in three dimensions has been proposed that the dual $dS$ entropy corresponds to the mutual entropy (see \autoref{Chp:HEE}) of information theory \cite{Kabat:2002aa}.

Also for this correspondence has  been found a consistent higher spin realization \cite{Anninos:2011aa} between Vasiliev theory on a de Sitter background with a fermionic $Sp(N)$ model, and another proposal on this correspondence in four dimensions is the mapping between the monotonic decrease of the Hubble parameter and an RG flow between two conformal fixed points of a three-dimensional Euclidean field theory. 
\\

In the Bousso approach \cite{Bousso:1999cb, Bousso:1999xy, Bousso:2001mw, Bousso:2002ju} to holography, this conformal boundaries as seen as \emph{holographic screens} which are two special hypersurfaces embedded on the boundary of a given spacetime which stores all the bulk information of a manifold with a certain boundary. For $dS_d$ any null geodesic that begins on a certain point defined at past infinity, will end up on a different point on the future infinity. Therefore it can be seen that an inertial observer on $dS_d$ would be characterized simply by this pair of points, and that half of the bulk region can be holographically projected along light rays onto the past infinity and likewise the other half into the future infinity. In the Maldacena's reconstruction to dS/CFT  \cite{Maldacena:2003aa} the partition function of the future infinity conformal field theory can be described in terms of the late-time Hartle-Hawking wave equation of the universe \cite{Hartle:1983aa} leading to a negative central charge for the three-dimensional CFT dual, in agreement with the higher spin realization of the correspondence pointed out before.
\\

This two realize the existence of duality between $dS$ space and two copies of a certain conformal field theory (one for each boundary), such that the Hilbert space corresponding to the bulk Rindler wedge $\mathfrak R_S$ (defined in \autoref{Sub:maxext}) is equivalent to the tensor product of the past and future CFTs Hilbert spaces
\begin{align}
\mathcal H_{\mathfrak R_S} = \mathcal H_{\mathcal I^+} \otimes \mathcal H_{\mathcal I^-}~,
\end{align} 
and therefore the partition function of a certain quantum gravity theory on $dS$ background can be written as 
\begin{align}\label{dsZZ}
\boxed{\mathcal Z_{QG}[\mathfrak R_S] = \mathcal Z_{CFT}[\mathcal I^+]\times \mathcal Z_{CFT}[\mathcal I^-]~.}
\end{align}
Now we will assume the existence of an infrared limit in which the quantum gravity action can be described in terms of Einstein gravity, and the partition function admits a saddle point approximation such that
\begin{align}
\mathcal Z_{QG}[\mathfrak R_S] \approx \exp\left\{-I_E[\mathfrak R_S]\right\}~,
\end{align}
therefore \eqref{dsZZ} renders
\begin{align}
\mathcal Z_{CFT}[\mathcal I^+] \approx \mathcal Z_{CFT}[\mathcal I^-] = \approx \exp\left\{-\frac{1}{2}I_E[\mathfrak R_S]\right\}~.
\end{align}
\\

Now, another important assumption that we will made is that the density matrix for an observer is thermal 
\begin{align}
\rho \equiv \ket{\mathcal O_S}\bra{\mathcal O_S} \sim \exp\{-\beta \mathcal H\}~,
\end{align}
where the observer state $\ket{\mathcal O_S}$ can be holographically described by the thermofield double (\autoref{App:C})
\begin{align}
\boxed{\ket{\mathcal O_S} \sim \sum_n \exp\left\{-\frac{1}{2} \beta E_n\right\}\ket{n}_{\mathcal I^-}\otimes \ket{n}_{\mathcal I^+}~,}
\end{align}
with boundary hamiltonian $\mathcal H \ket{n}_{\mathcal I^\pm} = E_n \ket{n}_{\mathcal I^{\pm}}$~.
This is motivated by the porposal give in \cite{Jatkar:2018aa, Fernandes:2019ige, Narayan:2019pjl, Narayan:2017xca}, in which the $dS$ spacetime is dual to a dual to an entangled thermofield-double type state of generic ghost-spin. This is based on the extremal surfaces existing between the past and future boundary on the $dS_4$ spacetime, but also is based on the reformulation of the $dS$/CFT reformulation proposed by Maldacena in \cite{Maldacena:2003aa} in which the partition function of a certain CFT living at future infinite $\mathcal I^+$ is related to the late-time Hartle-Hawking \cite{Hartle:1983aa} wave function of the universe $\Psi_{dS}$  \footnote{We thank to K. Narayan for pointed this out and some discussion on the dS/CFT dictionary}.
\begin{align}
\mathcal Z_{CFT} = \Psi_{dS}~.
\end{align}

The previous ingredients provides a holographic framework from which the area formula for the entropy can be derived from the entanglement between both boundaries. The on-shell Euclidean gravity action for the southern Rindler wedge has the value
\begin{align}\label{Ract}
I_E[\mathfrak R_S] ={}& -\frac{1}{16\pi G_4}\int_{\widehat{dS}_4/\Sigma_S}d^4 x \sqrt{g} \left( R - 6\ell^{-2}\right) + \mathcal T_q \int_{\Sigma_S}d^2 y \sqrt{h} \nonumber \\ \approx {}& -\frac{3}{4q G_4}\int^{\pi}_{\pi/2} \sin\theta\cos^2\theta~d\theta + \mathcal T_q A_{\Sigma_S} \nonumber \\ \approx{}& \left(1-\frac{2}{q}\right)\frac{\pi\ell^2}{G_4}~.
\end{align}
Which in the tensionless limit recovers $I_E \approx -\mathcal S_{dS}$. The limits of the integration corresponds to the ones where the southern Rindler wedge is defined, therefore we must exclude the non-regular part that consist on the southern defect. With this we can perform the holographic computation of entanglement entropy defined in \autoref{Chp:HEE} between two conformal field theories at $\mathcal I^\pm$ as two entangled complementary subsystems 
\begin{align}
\mathcal S_E = - \Tr[\widehat\rho_{\mathcal I^+}\log \widehat\rho_{\mathcal I^-}]~,
\end{align}
where we use the normalized density matrix $\Tr\widehat \rho = 1$. Then by using \eqref{SEErep}
\begin{align}\label{SEds}
\mathcal S_E = -\partial_q \log \Tr \widehat\rho_{\mathcal I^+}^q \rvert_{q=1}~.
\end{align} 
The holographic prescription summarized on \autoref{SubSub:HEE} can be extended to $dS$ boundaries by first using the fact that null geodesics induce the antipodal map \cite{Strominger:2001pn, Witten:aa} of the form
\begin{align}
\pi:S^3_- \rightarrow S_+^3~,
\end{align}
that send every point on the past sphere to an antipodal point of the future sphere. On the embedding coordinates, the antipodal map corresponds to the decomposition of $d+1$ reflections, one on each coordinates 
\begin{align}
(X^0,X^1,\dots,X^{d+1}) \stackrel{\pi}\longmapsto (-X^0, -X^1,..., -X^d)~.
\end{align}
The boundary CFT manifold can be defined by
\begin{align}\label{Ban}
\mathcal B \equiv (\mathcal I^+ \cup \mathcal I^-)/\pi~ \cong S^3~.
\end{align}
This boundary metric admits an obvious $\mathbb Z_q$ action that can be naturally treated as boundary replica symmetry. This simply given by azimuthal identification as it is described in \autoref{Sec:conical}, this implies that the set of fixed points have an $S^1$ topology. The boundary 3-sphere metric
\begin{align}
ds^2_{\mathcal B} = d\Theta^2 + \sin^2\Theta\left(d\chi^2 + \sin^2\chi \Phi^2\right)~,
\end{align}
under the discrete $\mathbb Z_q$ azimuthal identification $\Phi \sim \Phi + \frac{2\pi}{q}$ has invariant points that form a ring defined by $\chi = 0,\pi$. Also we can notice that the boundary manifold \eqref{Ban} can be assembled as
\begin{align}
\mathcal B = (\mathcal I^+/\pi)\cup (\mathcal I^-/\pi)~,
\end{align} 
and the antipodal symmetry $\pi$ acts separately on each boundary, which give rise two different hemispheres for the azimuthal angle $\Phi$. They both can be glued together along the $S^1$-fixed points, and the resulting $S^3$ has a natural replica symmetry $\mathbb Z_q\subseteq U(1)$ and by an stereographic projection of the boundary manifold, one hemisphere becomes the entangled region with boundary of set points of the replica symmetry respects to the other. 
%%%%%%%%%%%%%%%%%%
\vspace{1cm}
\begin{center}
\begin{tikzpicture}
%I+I-
\draw[thick] (-6,1)arc[start angle=90, end angle=270, x radius=1, y radius=1];
\draw[thick, darkred] (-6,1) 
arc[start angle=90, end angle=-90, x radius=0.2, y radius=1];
\draw[thick, darkred] (-6,1) 
arc[start angle=90,end angle=270, x radius=0.2, y radius=1];
%%%
\draw[thick] (-5,1)arc[start angle=90, end angle=-90, x radius=1, y radius=1];
\draw[thick, darkred] (-5,1) 
arc[start angle=90, end angle=-90, x radius=0.2, y radius=1];
\draw[thick, darkred] (-5,1) 
arc[start angle=90,end angle=270, x radius=0.2, y radius=1];
\node at (-7,1.1) {$\mathcal I^-/\pi$};
\node at (-3.9,1.1) {$\mathcal I^+/\pi$};
\node at (-2.5,0.3) {\footnotesize Gluing};
\draw [thick, ->] (-3,0)--(-2,0);
%mathcalB
\draw [thick] (0,0) circle (1cm);
\draw[thick, darkred, dashed] (0,1) 
arc[start angle=90, end angle=-90, 
x radius=0.2, y radius=1];
\draw[thick, darkred] (0,1) 
arc[start angle=90,end angle=270, 
x radius=0.2, y radius=1];
\node at (0,-1.5) {$\mathcal B= (\mathcal I^- \cup\mathcal I^+)/\pi$};
\node at (0,1.3) {\footnotesize\color{darkred}Fixed points};
\node at (2.5,0.3) {\footnotesize Projecting};
\draw [thick, ->] (2,0)--(3,0);
%Projection
\draw [thick] (3.5,-1.5)--(3.5,1.5)--(7.5,1.5)--(7.5,-1.5)--(3.5,-1.5);
\draw [thick, darkred] (5.5,0) circle (1cm);
\node at (5.5,-1.25) {\footnotesize\color{darkred}Entangling surface};
\node at (5.5,0) {\footnotesize$\mathcal I^+/\pi$};
\node at (6.8,1) {\footnotesize$\mathcal I^-/\pi$};
%CAPTION
\node[text width=16cm, text justified] at (1,-4){
\small {\hypertarget{Fig:12}\bf Fig.~12}:
\sffamily{The boundary manifold $\mathcal B$ is constructed by using the antipodal symmetry. The total \\ boundary its reduced to a single sphere after gluing on the fixed points. In the last step a \\ stereographic projection has been realized in order to see the entangling surface $S^1$. }}; 
\end{tikzpicture}
\end{center}
%%%%%%%%%%%%%%%%%%%%%%%%%%%
And now the $q-$th power of the reduced density matrix in terms of the CFT partition function can be computed on the branched cover $\mathcal B_q$ using Cardy-Calabrese formalisms previously mentioned
\begin{equation}\label{rhoZ}
{\Tr}\,\hat\rho_{\mathcal I^+}^{\, q} = 
\frac{\mathcal Z_{\rm CFT}[\mathcal B_q]}
{(\mathcal Z_{\rm CFT}[\mathcal B_1])^q}~,
\end{equation}
and consequently, using the isomorphism $\mathcal B\cong \mathcal I^+\cong S^3$ 
\begin{equation}\label{Sq}
\log {\Tr}\,\hat\rho_{\mathcal I^+}^{\,q}
= \log \mathcal Z_{\rm CFT}[\mathcal I^+_q]
- q\log\mathcal Z_{\rm CFT}[\mathcal I^+_1]~.
\end{equation}
Next, we observe that the replica $\mathbb Z_q$ symmetry 
extends from the branched cover boundary manifold into the bulk,  where it is 
captured by and implemented through the orbifold  
$\widehat{dS}_4=dS_4/\mathbb Z_q$. In the AdS/CFT approach to holographic entanglement entropy we need to assume the extension of the boundary replica symmetry into the bulk manifold $\mathcal M_q$ and then this allows to implement the $\mathbb Z_q$ action through the orbifold $\widehat{\mathcal M}_q \equiv \mathcal M_q/\mathbb Z_q$. In our case we do not need to construct the branched cover bulk manifold because the replica symmetry is already inherited by $\widehat{dS}_4$.

And now we use the locality of the gravity action $I[\mathcal M_q]  = q I[\widehat{\mathcal M}_q]$ allow us to write
\begin{align}
\mathcal Z_{CFT} \approx \exp\left\{-\frac{q}{2}I_E[\mathfrak R_N]\right\}~,
\end{align}
therefore
\begin{equation}\label{Sq}
\log {\Tr}\,\hat\rho_{\mathcal I^+}^{\,q}
= \log \mathcal Z_{\rm CFT}[\mathcal I^+_q]
- q\log\mathcal Z_{\rm CFT}[\mathcal I^+_1]~.
\end{equation}
Which by using \eqref{Ract} onto \eqref{SEds} we obtain the Gibbons-Hawking entropy
\begin{align}\label{bSE}
\boxed{\mathcal S_E = \frac{\pi\ell^2}{G_4}~.}
\end{align}
which would corresponds to the entanglement entropy between the future and past boundary CFTs. 
Notice that the Renyi entropy \eqref{renyi} for this background is $q$-independent and equals modular and entanglement entropy, in agreement with \cite{Dong:2018cuv}. 
\\

From the canonical thermodynamic relation $\beta F = \beta E - \mathcal S_{dS}$, where the Gibbs free energy would be defined as the quantum gravity partition function which we assume to be well approximated in the semiclassical limit by its on-shell Einstein action $\beta F = -\log \mathcal Z_{QG} \approx I_E$. If we do not consider conical deficit ($q\rightarrow 1$) the action recovers minus the entropy  it follows that $E = 0$. Now by considering the mechanism mentioned in this section, inertial observer to be dual to the thermofield double state implies that its density matrix is thermal. Furthermore, the entanglement entropy obeys the canonical relation \eqref{FSE}, which it is defined in terms of the boundary field theory partition function, with $\mathcal H_a$ the modular Hamiltonian. Now we have act with the replica symmetry which made inclusion of the Nambu-Goto co-dimension 2 boundary term which modifies the value of the free energy. And combining this, and using the result that entanglement entropy  corresponds to the Gibbons-Hawking entropy  we obtain that
\begin{align}
\boxed{\expval{\mathcal H_a} = \left( 1 - \frac{1}{q}\right)\frac{\ell}{G_4}~.}
\end{align}
This implies that using the defects we may be able to define the $dS$ energy in terms of the expectation value of the boundary modular Hamiltonian $\expval{\mathcal H_a} \equiv E$. Notice that in the tensionless limit we can not see this. 

%%%%%%%%%%%
\section{Entanglement between disconnected bulk regions}
As we have seen in the previous section, the extended coordinates \eqref{metric} made manifest the existence of a bulk antipodal symmetry which maps points from one Rindler wedge into the other. Observers into each Rindler wedge would be causally disconnected from each other, and we will see on this section that they can be described as to be entangled with no need of a holographic prescription. We now will study the bulk antipodal symmetry, for this we can notice that, after Wick rotation each Rindler wedge have the topology of an $S^4$ (see \eqref{gE}), we can compute the total volume of the Eculidean version of \eqref{metric} as
\begin{align}
\int_{\mathfrak R_N\cup\mathfrak R_S} d^4x \sqrt{g} = 2~\text{Vol}(S^4)~,
\end{align}
where each Rindler wedges contributes with a factor of Vol$(S^4)$. With this in mind, the total bulk topology 
\begin{equation}
\big(\mathfrak R_N \cup \mathfrak R_S\big)_E 
\cong S^4_{\mathfrak R_N} \cup S^4_{\mathfrak R_S} ~.
\end{equation}
Exactly as the previous section, the inverted antipodal map in the bulk is defined in terms of the embedding coordinates but now it maps between each Euclidean Rindler wedge 
\begin{align}\label{Pi}
\Pi \equiv S^4_{\mathfrak R_N} \rightarrow S^4_{\mathfrak R_S}~,
\end{align}
by using the parametric shift on the extended coordinates as 
\begin{align}
\theta \longmapsto \pi - \theta~,\qquad  \phi \longmapsto \pi - \phi
\end{align}
which in the embedding coordinates reads
\begin{align}
\Pi(X) = -X~, \qquad X \in \mathbb R^5~.
\end{align}
The existence of the map~\eqref{Pi} amounts to effectively describe
the bulk topology as a single 4-sphere modulo antipodal symmetry,
\emph{viz.} 
\begin{equation}\label{S4modPi}
\big(\mathfrak R_N \cup \mathfrak R_S\big)_E~\big/~ \Pi~
\cong~S^4~.
\end{equation}
Thus, we have the isomorphisms
\begin{equation}
\big(\mathfrak R_N \cup \mathfrak R_S\big)_E~\big/~ \Pi~
\cong \mathfrak R_N\cong\mathfrak R_S~,
\end{equation}
that provide three equivalent domains of integrations in terms
of which the gravitational partition function can be formulated.
\\

By now assuming the existence of a quantum gravity theory whose infrared limit reduces to Einstein gravity and the quantum partition function can be approximated by the Euclidean Einstein gravity action. Following the same recipe of the above section we compute the entanglement entropy but now with no necessity of holography, but just treating each Rindler wedge as two complementary entangled subsystems. 
\\

The entanglement entropy between northern and southern 
observers is given by
\begin{equation}\label{EEbulk}
\mathcal S_{\rm E}= 
- {\Tr}\,(\hat\rho_{\mathfrak R_S}\log\hat\rho_{\mathfrak R_S})~,
\end{equation}
where the reduced density matrix 
\begin{equation}
\hat\rho_{\mathfrak R_S}\equiv{\Tr}_{\mathfrak R_N}(\hat\rho_{\rm total})~,
\end{equation}
and $\hat\rho_{\rm total}$ is the density matrix associated to 
the total bulk system $\mathfrak R_N\cup\mathfrak R_S$. 
And as it said it before the entanglement entropy can be computed as 
\begin{equation}\label{Eq1}
\mathcal S_{\rm E}
= \lim_{q\to1}\frac{\log {\Tr}\,\hat\rho_{\mathfrak R_S}^{\,q}}{1-q}
= -\partial_q \log {\Tr}\,\hat\rho_{\mathfrak R_S}^{\,q}\big|_{q=1} ~.
\end{equation}
Here, by means of the replica method, we compute
\begin{equation} 
{\Tr}\,\hat\rho_{\mathfrak R_S}^{\,q} 
=\frac{\mathcal Z_{\rm QG} [S_q^4]}{(\mathcal Z_{\rm QG} [S^4])^q}~,
\end{equation}
Where the branched cover of the $S^4$ is denoted as $S_q^4$. Then using the entanglement entropy recipe we obtain
\begin{align}
\log {\Tr}\,\hat\rho_{\mathfrak R_S}^{\,q}
=\log \mathcal Z_{\rm QG}[S^4_q]- q\log\mathcal Z_{\rm QG}[S^4]  
\approx -q \Big(I_E[S^4/\mathbb Z_q]- I_E[S^4] \Big)~.
\end{align}
In the above, we have made use of the locality of the gravity
action to write $I_E[S^4_q]=qI_E[S^4/\mathbb Z_q]$. 
The manifold $S^4/\mathbb Z_q$ is by construction an Eucludian 
Rindler wedge plus the corresponding defect. Finally we obtain
\begin{align}
\log {\Tr}\,\hat\rho_{\mathfrak R_S}^{\,q}
=-2q \Big(1-\frac{1}{q}\Big)\frac{\pi\ell^2}{G_4}~.
\end{align} 
We conclude that the entanglement entropy~\eqref{Eq1} is given by
\begin{equation}\label{bulkE}
\boxed{\mathcal S_{\rm E} = \frac{2\pi\ell^2}{G_4} = 2\,\mathcal S_{dS}~.}
\end{equation}
Which is twice the Gibbons-Hawking entropy.
\subsection{Area law from bulk fixed points}\label{SEEtop}
As we have seen, the boundary entanglement entropy \eqref{bSE} is half of the bulk entanglement entropy \eqref{bulkE}, this extra term can be understood by looking at the fixed points of the $\mathbb Z_q$ action on the Euclidean bulk. 

In \eqref{gE} we see that the Euclidean bulk can be seen as warp product of two $S^2$ labeled by $L$ and $R$, then the $\mathbb Z_q$ actin can be performed in $S^2_L$ or $S_R^2$. By choosing the action into $S^2_R$, the set of fixed points associated to this $\mathcal F_R$  gives two copies of $S^2_L$, one for each pole of this 
\begin{align}
\mathcal F_R \equiv \{\vartheta = 0\}\cup\{\vartheta = \pi\} \cong S^2_L \cup S^2_L~.
\end{align}
And the same for the fixed points of acting of the left sphere. 
\begin{equation}
\mathcal F_L \equiv \{\theta=0\} \cup \{\theta=\pi\} \cong S^2_R\cup S^2_R~.
\end{equation}
Which are just the same because the warp factor does not play a r\^ole. Then the set of fixed points of the $\mathbb Z_q$ action on the bulk is always
\begin{align}
\mathcal F \cong S^2 \cup S^2~.
\end{align}
We can then rewrite the bulk entanglement entropy as the area law for the set of fixed points 
\begin{align}\label{Sfixed}
\boxed{\mathcal S_E = \frac{\text{Area}(\mathcal F)}{4 G_4}~.}
\end{align}
This is twice the area of the cosmological horizon. If we only consider one of the Rindler wedges, then \eqref{gE} only considers one of the $S^2$ terms, and the set of fixed points would correspond to a single two-sphere, in which case the area of the fixed points is exactly the area of the horizon Area$(\mathcal F) = \text{Area}(\mathcal H)$, and the Gibbons-Hawking formula is recovered. 

Finally, is worth to notice the lack of divergences in this entropies. This finitude of the entropy has been previously discussed in \cite{Solodukhin:2011gn} where the black hole entropy is matched with the entanglement entropy, and the divergence is given in the Newton's constant. Also for a different interpretation of the lack of the universal divergent term on the black hole entropy equals entanglement entropy scenario is given in \cite{McGough:2013gka}, where the Bekenstein-Hawking Entropy is interpreted as the  Topological Entanglement entropy. 

%% file: Chapters/Microscopic.tex
% Chapter 5

\chapter{Conformal description of the Cosmological Horizon} % Main chapter title

\label{Chp:Microscopic} % For referencing the chapter elsewhere, use \ref{Chapter1} 

%----------------------------------------------------------------------------------------

%----------------------------------------------------------------------------------------
Dimensional reduction to a two-dimensional QFT must flow to a CFT due to the C-theorem \cite{Zamolodchikov:1986gt}. In two dimensions it is known that certain CFT's posses the characteristics that the conformal symmetry is broken when the theory is quantized. This is known as conformal anomaly, and can be read it from the non-vanishing of the trace of the stress-tensor \cite{Alves_2004, Polchinski_1998}, they are related to topological invariants depending on the dimensions. For two-dimensional CFTs, the anomaly trace reads
\begin{align}\label{tranom}
\expval{T^\mu_{~~\mu}} = \frac{c}{12}\mathcal R~,
\end{align}
where $c$ is the central charge of the theory (see \autoref{CFT}). We can try to identify the two-dimensional branes $\Sigma_{N,S}$, induced by the conical deficit \eqref{deficit} that have been introduced in \autoref{dS}, with a conformal field theory by seeing how the localized stress-tensor on the branes \eqref{braneT}. Using $h_{ij}h^{ij} = 2$ the trace reads
\begin{align}\label{branetrace}
T^\mu_{~~\mu} = \frac{1}{2G_4}\left(1 - \frac{1}{q}\right)~.
\end{align}
Let us assume for a moment that this corresponds to a CFT with trace anomaly \eqref{tranom}, by this we can use \eqref{branetrace} and isolate the central charge, \textit{viz}
\begin{align}
\frac{6}{G_4}\left(1 - \frac{1}{q}\right) = c \mathcal R~, 
\end{align}
using $\mathcal R = 2/\ell^2$ we finally get
\begin{align}\label{Cq}
\boxed{c = \frac{3\ell^2}{G_4}\left(1 - \frac{1}{q}\right)}~.
\end{align}
We will show how to achieve this central charge from two different recipes. One corresponds to made use of the Kerr/CFT correspondence \cite{Guica:2008mu} (resumed in \autoref{KerrCFT} by extending the $dS_4$ geometry and analyze the symmetries near the horizon, as it is done in \cite{Lin:1999gf, Carlip:1999cy}. As the Killing horizon corresponds to null horizon, it is expected that the ASG shares some properties with the BMS group \cite{Bondi:1962px, Sachs:1962wk, Sachs:1962zza}, for a discussion on this see \cite{Koga:2001vq}. 
\\

We will also see that it is possible to compare the reduced action for gravity on the de Sitter deformed background and compare with the Liouville theory and compute the central charge by using this theory.
\\

Finally we made use of Cardy formula defined in \autoref{CFT} to obtain that the entropy corresponds to the relative entropy, which recovers the Gibbons-Hawkin entropy in the tensionless limit. 

%%%%%%%%%%%%%%%%%%%%%%%%%
\section{De Sitter entropy from asymptotic symetries}
In this section we will see how the orbifolded horizon of \autoref{Sec:conical}, allow us to mimic the NHEK metric \eqref{NHEK} by applying a boost into the deformed geometry. This based on that the isometry group of both geometries are the same and still relying on the study of the ASG \eqref{ASG} in the same way as the original Kerr/CFT correspondence. Also it can be checked that this corresponds to a exact solution of Einstein equation (after regularizing the conical deficit by the insertion of the pair of Nambu-Goto actions) and that the boost does not produce an angular momenta as a conserved quantity, and therefore does not modify the cosmological horizon. 

We show that the boost parameter allow us to make use of the Kerr/CFT correspondence for $dS$ vacuum and the microscopic entropy of the CFT living at the Cosmological Horizon gives the relative entropy of two states living on the Liouville CFT spectrum.%
\subsection{Symmetries and Boost}
As is showed in \autoref{SubSec:Deformation} the symmetries of the deformed $dS_4$ geometry the isometry group corresponds to $SL(2, \mathbb R)\times U(1)$ exactly as the NHEK geometry \eqref{NHEKerr} at each fixed $\theta$. This is a strong hit that we may made use of the Kerr/CFT correspondance which relies heavily in the study of the diffeomorphisms group. Also as is presented by Carlip in \cite{Carlip:1999cy}, the Cosmological Horizon can be described by a two-dimensional CFT, and then extended by Solodukhin in \cite{Solodukhin:1998tc} interpreting this CFT as the Liouville theory. 

As mentioned, the metric \eqref{metric} can be put in a NHEK form by performing a boost with an arbtirary function of $q$, such that in the tensionless limit this vanishes and the $dS_4$ original geometry is recovered
\begin{align}
\frac{\ell}{q}\phi \rightarrow \frac{\ell}{q}\phi + \mathcal B(q)t~.
\end{align}
In doing so, the line elements becomes
\begin{equation}
\label{NHEK}
ds^2= \cos^2\theta\left(-f(r)dt^2 + \frac{dr^2}{f(r)} \right)
+\ell^2 d \theta^2 +  \sin^2\theta\left(\frac{\ell}{q} d\phi + \mathcal B(q)\,dt \right)^2 ~,
\end{equation}
with $f(r) = 1-r^2/\ell^2$ corresponding the metric function and $\mathcal B(q)$ an arbitrary function of the deformation parameter $q$ the will be referred as the boost parameter. The condition that we must recover the metric in the tensionless limit implies that must that the boost parameter must vanishes in this limit, we choose 
\begin{align}
\mathcal B(q) = b\left(1-\frac{1}{q}\right)~,
\end{align}
with $b$ an arbitrary real number. Then the boosted metric $dS_4^{(\mathcal B, \, q)}$ recovers the original geometry as
\begin{align}
dS_4 = \lim_{q \rightarrow 1} dS^{(\mathcal B ,\,  q)}_4~.
\end{align}

The deformed $dS_4^{(\mathcal B , \, q)}$ is still an exact solution to Einstein equations with an arbitrary $\mathcal B$-parameter, after compensate with the fixed Nambu-Goto action for the defects (showed in \autoref{SubSec:Deformation}).
It can be checked using different methods \cite{Guo:2016aa, Sekiwa:2006aa, Abbott:1981ff} that this procedure does not bring in angular momentum as a global conserved charge. 

\subsection{Boundary conditions and asymptotic symmetries.} 
In general, one must check that the choice of boundary conditions is such that they are neither too weak --in order to avoid divergent generators, nor too strong to render the theory trivial.
\\

Our aim is to describe the CFT at the horizon $\mathcal{H}$ whose generators will have the same properties as the Kerr/CFT case in the limit $r \rightarrow \ell$. Therefore, we considering the same boundary condition as for the extreme black hole.\\
Due to the isometries of \eqref{NHEK} and form of the metric \eqref{NHEK}, we will impose boundary conditions in complete analogy to \cite{Guica:2008mu}. But as we will take the limit to the horizon (i.e., $r\rightarrow \ell$), then we must impose boundary condition in order os $r-\ell$. Considering small fluctuations
of the background metric $\delta g_{\mu\nu} = h_{\mu\nu}$,  we will use $\hat r = r - \ell$, the boundary conditions reads exactly like \eqref{BC}.
%\framebox{FIX, NOT FULLY UNDERSTOOD...}
%
The original non-trivial diffeomorphisms compatible with the asymptotic behaviour \cite{Guica:2008mu} can be translated to 
\begin{equation}
\label{diff}
\zeta_\epsilon = \epsilon (\phi) \partial_\phi - \hat r \epsilon' (\phi) \partial_r ~.
\end{equation} 
The parameter $\epsilon=\epsilon(\phi)$ is periodic with period of $2\pi q^{-1}$. It is thus convenient to introduce the mode
expansion $\epsilon_n(\phi)=-\exp\{-in\phi\}$,  which amounts to label
\begin{equation}
\label{basis}
\zeta_{(n)} = -\exp\{-in\phi\} (\partial_\phi+ i n \hat r \partial_r)~,
\end{equation}
where the zero mode $\zeta_{(0)}=-\partial_\phi$ is a $U(1)$ generator
and the (classical) Lie bracket of two vector fields satiesfies Witt algebra
\begin{equation}
i\,[\zeta_{(m)}, \zeta_{(n)} ] = (m-n)\, \zeta_{(m+n)} ~.
\end{equation}
This correspond to a single copy of the conformal group in the circle. 
\subsection{Asymptotic Charges}\label{SubSec:Charges}
A covariant definition of conserved quantities for General Relativity was given in \cite{Barnich:2001jy} in terms of a background metric $g$ and a perturbation $h$,
\begin{align}\label{fullcharge}
Q_{\zeta} [g] = \frac{1}{8\pi G}\int\limits_{\mathcal{H}} k_{\zeta}[h ,g].
\end{align}
where
\begin{align}\label{twoform}
k_\zeta [h, g] =  -\frac{1}{4} \epsilon_{\alpha\beta\mu\nu} \Big[ 
\zeta^\nu \nabla^\mu h - \zeta^\nu \nabla_\sigma h^{\mu\sigma} 
&+ \zeta_\sigma\nabla^\nu h^{\mu\sigma} +\frac{1}{2} h \nabla^\nu \zeta^\mu
-h^{\nu\sigma}\nabla_\sigma \zeta^\mu \nonumber \\ 
&+ \frac{1}{2} h^{\sigma\nu} ( \nabla^\mu\zeta_\sigma+ \nabla_\sigma\zeta^\mu)
\Big] dx^\alpha \wedge dx^\beta ~.
\end{align}
The indices are raised and lowered with the background metric. This is a two-differential form then it works for an arbitrary dimension, we will work only in the four-dimensional case.
%
%%%%%%%%%%%%%%%
\subsection{Algebra of asymptotic charges}
At this point, we can use the Barnich-Brandt-Compere formalism \cite{Barnich:2003xg, Barnich:2006av, Barnich:2001jy, Compere:2007az} in order to obtain the central extension of the Witt algebra generated by the diffeomorphisms \eqref{diff} of the form
\begin{align}\label{qmalgebra}
\left\{Q_{\zeta_{m}},Q_{\zeta_{n}}\right\} = -i(m-n)Q_{\zeta_{m+n}} + \frac{1}{8\pi G_{4}}\int\limits_{\mathcal{H}} k_{\zeta_{m}}[\mathcal{L}_{\zeta_{n}}g,g]~.
\end{align}
Here, $\mathcal{L}_{\zeta_{m}}g$ is the Lie derivative of the metric along $\zeta_{m}$ and $k_{\zeta_{m}}$ is the two form is defined in Eq. \eqref{twoform}

The Lie derivatives of the components of the metric \eqref{NHEK} is given by the set of relations 
\begin{align}\label{liederivatives}
\left(\mathcal{L}_{\zeta_{n}}g\right)_{tt} &= 2 i n ((r-\ell) /\ell^2)\exp\{-in\phi\}\cos^2\theta~, \nonumber \\ 
\left(\mathcal{L}_{\zeta_{n}}g\right)_{t\phi}&= 2in\mathcal B (\ell/ q) \exp\{-in\phi\}\sin^2\theta~, \nonumber \\ 
\left(\mathcal{L}_{\zeta_{n}}g\right)_{rr} &= 2i n (\ell^3/(\ell-r)(\ell + r)^2) \exp\{-in\phi\}\cos^2\theta~, \nonumber \\ 
\left(\mathcal{L}_{\zeta_{n}}g\right)_{r\phi} &= 2n^2(\ell^2/(\ell+r))\exp\{-in\phi\}\cos^2\theta~, \nonumber \\ 
\left(\mathcal{L}_{\zeta_{n}}g\right)_{\phi\phi} &= 2in (\ell^2/q^{2})\ell^2\exp\{-in\phi\}\sin^2\theta~.
\end{align}
As the computation is performed at the horizon $\mathcal{H}$, the integration is performed over the angles in the limit $r\rightarrow \ell$. We will use $b = 2/3$ in order to recover the results of \eqref{Cq}.  This leads to the central extension for the asymptotic algebra \eqref{qmalgebra}, in this case given by
\begin{align}\label{centralext}
\frac{1}{8\pi G_{4}}\int\limits_{\mathcal{H}} k_{\zeta_{m}}[\mathcal{L}_{\zeta_{n}}g,g] = -i\frac{\ell^2 }{4G_{4}}\left(1 -\frac{1}{q}\right) \, m^{3}\delta_{m+n,0}~.
\end{align}
In the quantum version, the extension from the Dirac brackets to commutators ($\{ \, , \, \}\rightarrow \frac{1}{i\hbar}[\, , \,]$) and the define the dimensionless quantum generators by
\begin{align}
\hbar L_{m} = Q_{m} + \frac{\ell^2 \mathcal B}{8G_4}\delta_{m,0}~,
\end{align}
which defines the asymptotic algebra as
\begin{align}\label{viralg}
[L_{m},L_{n}] = (m-n)L_{m+n} + \frac{c}{12}(m^3 - m)\delta_{m+n,0},
\end{align}
where the central charge is
\begin{align}
\boxed{c = \frac{3\ell^2}{\hbar G_{4}}\left(1 -\frac{1}{q}\right)}~.
\end{align}
The central charge computed by this Kerr/CFT machinary, depends explicitly on the boost parameter, which is the analogue of the angular momenta-dependence of the NHEK central charge. We can see that in the tensionless limit this value can not be computed in this way. We procedure now to see the temperature of the quantum states near the cosmological horizon, and use CFT tools to compute the entropy associated to it. 
\subsection{Bunch-Davies vacuum}
\label{BunchDavies}
%%%%%%%%%%%%%%%%%
On general grounds, when considering a quantum field
theory on a curved spacetime there is no unique definition
of vacuum, but only an observer-dependent notion of it.
The underlying reason for this ambiguity is simply the 
fact that time evolution is implemented by the Hamiltonian of 
the system acting on a given operator as time derivative. 
Its action thus depends on the choice of time. As a consequence, 
different choices of time yield to different definitions of energy, 
and therefore there is no unique way to define a vacuum state.

When considering the static
embedding of de Sitter space, such state is an entangled state between 
the two causal patches.  When tracing the density matrix of one of these 
patches with respect to the other one, the resulting reduced density matrix 
is thermal. Any geodesic observer in dS space hence detects a thermal 
bath of particles with a temperature
\begin{equation}\label{B0}
T_{dS}= \frac{1}{2\pi\ell}~.
\end{equation}

The Bunch-Davies vacuum \cite{Bunch_1978}, also refered as Euclidean vacuum, 
is defined by the thermal density matrix $\rho=\exp\{\beta \mathcal H\}$. 
Let us assume the existence of such state on the $dS^{(q)}_4$ background~\eqref{metric}, whose macroscopic thermal properties 
are thus encoding within the Boltzmann factor $\exp\{-\omega/T_{dS}\}$ 
(in natural units), where $\omega$ is the energy eigenstate and 
$T_{dS}$ is the dS temperature~\eqref{B0}. 
On the region near the boundary defects~\eqref{SigmaNS}, wherein the $dS^{(q,\mathcal B)}_4$ metric 
is well approximated by
\begin{equation}\label{B1}
ds^2\approx - \left(  1 - \frac{r^2}{\ell^2}\right) dt^2 
+ \frac{dr^2}{   1 - \frac{r^2}{\ell^2} }  + \cdots
\end{equation}
The near horizon geometry $r\rightarrow\ell$ is more 
conveniently expressed in term of the dimensionless 
Rindler coordinates $\eta = t\ell$ and $R^2 = 2(1- r/\ell)$,
in terms of which~\eqref{B1} transforms to
\begin{equation}\label{B2}
ds^2 \approx - R^2d\eta^2 + dR^2~,\quad  R\ll 1~.
\end{equation}

A quantum field on the background~\eqref{B1} close
to the boundary defects can be expanded in eigenstates with 
energy $\omega$. For the case of an scalar field we have
\begin{equation}\label{B3}
\Phi(t, r)= \sum_\omega \phi_\omega(r) \exp\{-i\omega t\}~,
\qquad \omega>0~.
\end{equation}
When going to the near horizon sector~\eqref{B2}, the mode
expansion~\eqref{B3} is given in terms of the dimensionless
Rindler coordinates  by
\begin{equation}\label{B4}
 \Phi(\eta, R)= \sum_N \phi_N(R) \exp\{-i N \eta\}~, 
 \qquad N\equiv \ell \omega~.
\end{equation}
In turns, the Boltzmann factor in terms of the near horizon 
time $\eta$ is
\begin{equation}\label{B5}
\exp\{-\omega/T_{dS}\}= \exp\{-N/T\}~,
\end{equation}
where $N$ is the number of states with temperature 
\begin{equation}\label{B6}
\boxed{T = \frac{1}{2\pi}}~.
\end{equation}
\subsection{Entropy}\label{KerrS}
All the above description corresponds to the conformal algebra for each of the branes. Now, we proceed to apply the thermodynamic Cardy's formula for a canonical ensemble, which relates the microscopic degrees of freedom of a unitary CFT to its central charge and temperature, for each of this branes and sum both of them to obtain the total entropy of the configuration showed in \autoref{Sec:conical}. 
\\

Using the Cardy formula \eqref{CanoCardyMovers} on the branes, which the near temperatures has been derived in \autoref{BunchDavies}, we obtain
\begin{align}\label{Sq}
\mathcal S^{(q,\mathcal B)}_{dS_4} = \frac{\pi \ell^2}{\hbar G_{4}} \mathcal T_q = \frac{\mathcal A_{\mathcal H}}{4\hbar G_4}\left(1-\frac{1}{q}\right)~,
\end{align}
The interpretation of this entropy will be done in \autoref{SubSec:Relative} down below.
%%%%%%%%%%%
%%%%%%%%%%%%%%%%%%%%%%%
\section{De Sitter entropy from Liouville theory}
Here we will made use of Liouville conformal field theory (summarized in \autoref{Chp:Liouville}) as an effective theory to describe the antipodal branes. It must be notice that the definitions giving on \autoref{Chp:Liouville} are the standard ones of the conformal field theory department, which definition of the stress-tensor may differ with the one defined on gravity. Therefore one must be careful with the constant at the moment of comparing, this is also recalled at the moment of computation through the chapter. 
\\

We recall that the (Euclidean version of the) full gravity action~\eqref{Itotal} 
consists on a bulk term plus a pair of co-dimension 2 Nambu-Goto terms
\begin{align}\label{IEtotal}
I^E_{\rm total}[\widehat{dS}_4]
&=-\frac{1}{16\pi G_4} \int_{\widehat{dS}_4}
d^4 x \sqrt{g}\Big( R- \frac{6}{\ell^2}\Big)
+\sum_{I=N, S} \mathcal T_q
\int_{\widehat\Sigma_I} d^2 y\sqrt{h}\nonumber\\
&=: I_{\rm bulk}[\widehat{dS}_4]
+I_{\rm NG}[\widehat\Sigma_N]
+I_{\rm NG}[\widehat\Sigma_S]~.
\end{align}

The bulk integral \eqref{IEtotal} can be dimensionally reduced down to two dimensions 
\begin{equation}
I_{\rm bulk}[\widehat{dS}_4] 
\overset{\rm dim\,red}\longrightarrow
I_{2d}[\widehat\Sigma_N\big] 
+ I_{2d}[\widehat\Sigma_S\big]~,
\end{equation}
as to define an effective action on each of the boundary 
defects, using $i = N,S$
\begin{equation}\label{effNS}
I_{\rm eff}[\widehat\Sigma_i] 
= I_{2d}[\widehat\Sigma_i] 
+ I_{\rm NG}[\widehat\Sigma_i] ~,
\end{equation}
then the on-shell total action is simply 
\begin{equation}\label{totaleff}
I^{E}_{\rm total}[\widehat{dS}_4]
\approx \sum\limits_i I_{\rm eff}[\widehat\Sigma_i] ~.
\end{equation}
The reduced Euclidean action $I_{2d}$ follows from integrating out 
the spindle coordinates $(\theta, \phi)$ and using the line 
element ~\eqref{metric} as
\begin{align}
\label{4d2d}
I_{\rm bulk}[\widehat{dS}_4]
={}&-\frac{1}{16\pi G_4} \int_{dS_4^{(q)}} d^4 x \sqrt{g} 
\Big( R- \frac{6}{\ell^2}\Big)~,\nonumber \\
%={}&\frac{\ell^2}{16\pi q G_4} \int
%d\theta\, d\phi \,d^2x \sqrt{\gamma}\, 
%\sin \theta \cos^2\theta \Big( R- \frac{6}{\ell^2}\Big)  
\approx{}&
-\frac{\ell^2}{8qG_4} \int_{\widehat\Sigma_N\cup\widehat\Sigma_S} d^2y\sqrt h\,\mathcal R~,
%=I_{2d}[\widehat\Sigma_N\big] + I_{2d}[\widehat\Sigma_S\big] ~,
\end{align}
where $\mathcal R=\mathcal R[h]$ is the 
two-dimensional Ricci scalar of the induced metric $h$ that covers the $dS_2$ space.
To compute this we made use of Einstein equations
$\ell^2 R_{mn}=3g_{mn}$ 
with $(m,n)$ angular coordinates, and also that on shell
$R - 6/\ell^2 = 3\mathcal R_{ij}$. Therefore, using~\eqref{4d2d}, we define%
\begin{equation}
I_{2d}[\widehat\Sigma_i]\equiv 
-\frac{\ell^2}{8 q G_4} 
\int_{\widehat\Sigma_i} d^2y\sqrt h\,\mathcal R~,
\end{equation}
and the total effective action~\eqref{effNS} reads
\begin{equation}\label{IeffN}
I_{\rm eff}[\widehat\Sigma_i] \approx 
- \frac{\ell^2}{8qG_4} 
\int_{\widehat\Sigma_i} d^2y\sqrt h\,\mathcal R
+\frac{1}{4 G_4} \Big(1-\frac{1}{q}\Big)
\int_{\widehat\Sigma_i} d^2y\sqrt h ~,
\end{equation}
which on-shell value is 
\begin{equation}\label{onshellN}
I_{\rm eff}[\widehat\Sigma_i]\approx
\Big(1-\frac{2}{q}\Big) \frac{A_{\widehat\Sigma_i}}{4G_4}
= \Big(1-\frac{2}{q}\Big) \frac{\pi\ell^2}{G_4}~,
\end{equation}
where $A_{\widehat\Sigma_i}$ denotes the area
of $\widehat\Sigma_i$, computed as
\begin{equation}
A_{\widehat\Sigma_i}
=\int d^2y\sqrt{h}
=\int_{0}^{\beta}dt_E\int_{-\ell}^{\ell}dr = 4\pi\ell^2
= A_{\mathcal H}~, 
\end{equation}
with $\beta=T^{-1}_{dS}=2\pi\ell$ is the inverse of the $dS$ temperature \eqref{TdS}

\subsection{Microscopic theory.}
Now we can observe that the reduced action on the branes \eqref{IeffN} can be cast onto the form of the Liouville theory (see \autoref{Chp:Liouville})  
\begin{equation}\label{IL}
I_{\rm L}[g, \Phi; \gamma]= -\frac{1}{2} \int_{M_2} d^2y \sqrt g 
\Big(g^{ij}\partial_i\Phi \partial_j\Phi + Q\mathcal R \Phi 
+4\pi \mu e^{2\gamma\Phi}\Big)~.
\end{equation}
Here, $(M_2, g)$ is a two-dimensional Euclidean manifold. 
And as is seen in \autoref{Chp:Liouville}, the action posses conformal invariance on the classical and quantum regime by imposing $Q =  \mathcal O(\gamma) + \mathcal O(\gamma^{-1})$, therefore, $Q$ has the same form on both limits.%
\\

This action differs 
from the standard one given in~\eqref{Lact} by 
an overall factor of $2\pi$: Such a normalization is 
needed in order to uniformize the definition of the 
stress-energy tensor while comparing~\eqref{IeffN}.
\\

Indeed, the on-shell action~\eqref{IeffN} corresponds precisely 
to the Liouville action~\eqref{IL} on a fixed background 
$(M_2, g)=(\widehat\Sigma_N, h)$, whe the 
Liouville field gets a \textit{vev} $\langle \Phi\rangle= \Phi_0$, i.e., $I_{\rm eff}[\widehat\Sigma_i]
\approx 
I_{\rm L}\big|_{\Phi_0}~$.
This on-shell correspondence holds provided
\begin{equation}\label{matching1}
\frac{\ell^2}{8qG_4}=\frac{Q\Phi_0}{2}~,\quad
\frac{1}{4G_4}\Big(1-\frac{1}{q}\Big)
=-2\pi\mu e^{2\gamma\Phi_0}~,
\end{equation}
as follows from simple inspection of~\eqref{IeffN}
and~\eqref{IL}. The first of these equations fix 
the expectation value of the scalar mode as
\begin{equation}\label{Phi0}
\Phi_0 = \frac{g^{-2}}{4qQ}~,
\end{equation}
where the dimensionless parameter 
$g^{-2}\equiv\ell^2/G_4\gg 1$. 
Since the background charge 
is $Q\gg 1$ in both semiclassical ($\gamma^2\ll1$) 
and quantum ($\gamma^2\gg1$) regimes, 
it follows that $\Phi_0\sim q^{-1}$.  

In addition, the Liouville equation of motion 
for a constant field $\Phi=\Phi_0$ reads
\begin{equation}\label{matching2}
Q\mathcal R + 8\pi \gamma \mu e^{2\gamma\Phi_0} 
= \frac{2Q}{\ell^2} + 8\pi \gamma \mu e^{2\gamma\Phi_0}=0~,
\end{equation}
for the curvature of the cosmological horizon.
Consistency of the second equation in~\eqref{matching1} 
with ~\eqref{matching2} yields
\begin{equation}\label{gq}
1+\frac{1}{\gamma^2} = 
\Big(1-\frac{1}{q}\Big)\frac{\ell^2}{2 G_4}~.
\end{equation}

As we will see below, the semiclassical 
limit of~\eqref{gq} provides a nontrivial link between the Liouville 
coupling $\gamma$ and the gravitational coupling $g^{-2}=\ell^2/G_4$. Also we have sen in \autoref{Chp:Liouville} that the central charge of the theory is related with the value of $Q$ and therefore $\gamma$. This central charge will be used latter in order to compute the entropy of the configuratio. \\ As it was said it in \autoref{Chp:Liouville}, the states \eqref{Lstates} must be analyzed in the limit of $\phi_0\rightarrow \infty$ (regardless the sign of $\mu$) , which in our case corresponds to the tensionless limit, therefore we can see that the tensionless limit provides the analysis in order to see the normalizable wave functions on the quantum Liouville theory, therefore even that we have choose a particular fixed value for the Liouville field, we still have the option to perform the quantum local analysis by using $q$ as the parameter. 
%~~~~~~~~~~~~~~~~~~~~~~~~~~~~~~~~~~
\subsubsection{Central charge and Cardy formula.}\label{CCandCardyS}
%~~~~~~~~~~~~~~~~~~~~~~~~~~~~~~~~~~
In the semiclassical regime $\gamma^2\ll1$, 
where thus $Q\sim\gamma^{-1}$, there exists 
a $\mathcal O(1/\gamma^2)$ contribution to the 
Liouville central charge\footnote{We recall again that all the definitions must be shifted with the ones given in \autoref{Chp:Liouville} because of the redefinition of the proportionally term in front of the action.}.
\begin{equation}
c = 1+6Q^2\approx \frac{6}{\gamma^2}~,
\end{equation}
whose value can be computed in terms of the gravity
couplings and the orbifold parameter $q$. From the 
semiclassical limit of~\eqref{gq}, we find 
\begin{equation}\label{cq}
\boxed{c_q= \Big(1-\frac{1}{q}\Big)\frac{3\ell^2}{G_4}}~.
\end{equation}

As it is noticed, the central charge depends on the $q$-parameter, it depends on how deformed is the dS spacetime, and it is not absolute as it is usual in the Liouville theory. This has been previously study in different contexts on AdS$_3$/CFT$_2$ correspondence \cite{Banados:1998wy}, and microscopic description of black hole horizons \cite{Solodukhin:1998tc, Carlip:1998wz, Carlip:2001wq}. 

We further observe that in order to ensure unitarity
of the theory $c_q > 0$ the orbifold 
parameter must be strictly greater than one.  
Hereafter we assume this to be case\footnote{We can see from \eqref{matching1} and \eqref{matching2} that if $q<1$, then the curvature on the branes would be negative. On the other hand, the unity lower bound on $q$ ensures 
the reality of the couplings $\gamma$ and $\ell$ in~\eqref{gq}. }.

Having obtained the central charge~\eqref{cq} and by virtue of
the thermal (canonical) Cardy formula~\cite{Cardy:1986ie, Bloete:1986qm} (see \autoref{KerrS})
\begin{equation}\label{CardyE1}
{\mathcal S}_q^{\,\rm Cardy} 
= \frac{\pi^2}{3} c_{q, L}\, T_L 
+ \frac{\pi^2}{3} c_{q, R}\, T_R~,
\end{equation}
a $q$-dependent entropy can be computed (as usual, 
$L$ and $R$ label left and right-movers central charge 
and temperature). For a non-chiral Liouville theory, 
we have
\begin{equation}\label{cT}
c_{q, N}=c_{q, S}=c_q ~, \qquad
T_L=T_R=\frac{1}{2\pi}~,
\end{equation}
where $T_{L,R}$ is the near horizon temperature~\eqref{B6} 
at the boundary defects (derived in \autoref{BunchDavies}). 
Using~\eqref{cq} and~\eqref{cT} in the Cardy formula~\eqref{CardyE1},
we find
\begin{equation}\label{CardyE2}
\mathcal S_q^{\,\rm Cardy}
= \Big(1-\frac{1}{q}\Big)\frac{\pi\ell^2}{G_4} = \frac{\mathcal A_{\mathcal H}}{4\hbar G_4}\left(1 - \frac{1}{q}\right) ~.
%=\Big(1-\frac{1}{q}\Big)\frac{\mathcal A_{\mathcal H}}{4G_4} ~.
\end{equation}
Where in the second quality we have recovered the value of $\hbar$, and we can notice that this coincides with \eqref{Sq} the one obtained in \autoref{KerrS}. The applicability of the Cardy formula is discussed in \cite{Hartman:2014oaa} where the extended Cardy regime corresponds to a thermal theory with large central charge as it is in our scenario. 

\section{Modular free energy and Cardy entropy}\label{SubSec:Relative}
As we have seen previously in \autoref{Sec:EE}, we can compute the modular entropy by taking minus the derivative respect to the temperature. 
In \cite{Baez2011RenyiEA} there has been proposed a relation between the free energy and the $q$-parameter of the Renyi entropy, where the Renyi parameter is interpreted as a dimensionless temperature of a Gibbs state \cite{Rangamani:2016dms}. 
We will show in \autoref{Chp:dSEntanglement} how the dS entropy can be interpreted as the entanglement entropy between the past and future CFT on $\mathcal I^\pm$ respectively\cite{Arias:2019aa}. This result came from the fact that the Renyi entropy in this setup is a $q$-independent quantity, and thereby equals entanglement entropy, and also modular entropy. 
Also, as we have used the canonical ensemble (see \autoref{Sec:Cardy}), and moreover, the energy of pure dS$_4$ spacetime is zero \cite{Corichi:2003kd, Gomberoff:2003ea, Ashtekar:2000sz}. Therefore, we follow to propose the interpretation of the product of the Cardy entropy \eqref{CardyE2} and a dimensionless temperature $T =1/q$ as the modular free energy.

Regarding the $q$-dependance of the Cardy entropy~\eqref{CardyE2}, we follow to use the relation between the Renyi entropy $\mathcal S_q$ and the Free energy $\mathcal F$ proposed by Baez \cite{Baez2011RenyiEA}, that states that Renyi entropy is a $q$-deformation of the usual definition of entropy, and is proportional to the free energy in the canonical ensemble, 
\begin{align}\label{BaezR}
\mathcal F  = -(T-T_0)\mathcal S_{T_0/T}~,
\end{align}
where $T_0$ is the temperature at thermal equilibrium, and $\mathcal S_{T_0/T}$ corresponds to the Renyi entropy being the Renyi parameter the ratio $T_0/T$. This was previously conjectured in \cite{Beck_1993} using the normalized temperature $T_0 = 1$. 

This relation can also been rewritten in terms of the thermal entropy $\mathcal S_{\rm th}$ as 
\begin{align} \mathcal S_q = \left(\frac{q}{q-1}\right)\frac{1}{T_0}\int_{T_0/q}^{T_0}\mathcal S_{\rm th}~dT ~. \end{align} 

Baez has proposed that the Renyi can be related with the Free energy as in \eqref{BaezR}, and through a $q^{-1}-$derivative\footnote{The concept of $q-$derivative appears firstly in the context of quantum groups, see for instance~\cite{Kac_2002}~.}
\begin{align}\label{Baezq}
\mathcal S_q = -\left(\frac{d\mathcal F}{dT}\right)_{q^{-1}} \equiv - \frac{\mathcal F(T/q) - \mathcal F(T)}{T/q - T} ~.
\end{align}
 Thereby, considering the normalized temperature, and the Renyi parameter as 
\begin{align}\label{qTemp}
q = \frac{T_0}{T} = \frac1T~,
\end{align}
the relation simply renders
\begin{align}
\mathcal F_q = \left(1 - \frac1q\right)\mathcal S_q~,
\end{align}
and both quantities are equal up to the fudge factor $(1 - q^{-1})$~. Considering a modular density matrix $\rho^q = e^{-qH}$, can be seen as a modular free energy with temperature $q = T^{-1}$ as we will show now. 

It has been proposed that the entanglement entropy $\mathcal S_{E}$ between the conformal boundaries of dS$_4$ equals the Gibbons--Hawking entropy of the Cosmological Horizon \cite{Arias:2019aa}, which has been previously proposed in black hole physics \cite{Solodukhin:2011gn}. In \cite{Arias:2019aa} it is noticed that the Renyi entropy was $q-$independent (thereby equals modular entropy) and matches exactly the dS$_4$ entropy, that is in agreement with the computation done using the dS/dS correspondence~\cite{Dong:2018cuv}
\begin{align}\label{SdSR}
 \mathcal S_q = \frac{\pi\ell^2}{G_4} = \mathcal S_{\rm dS}~.
\end{align}
Combining \eqref{SdSR} and \eqref{BaezR} we obtain
\begin{align}
\mathcal F_q = \left(1-\frac{1}{q}\right)\mathcal S_{dS}~,
\end{align}
that is exactly the Cardy entropy in the canonical ensemble. Thereby using the Baez formula \eqref{Baezq}, and the temperature $T = q^{-1}$ we obtain
\begin{align}
\mathcal S_q = -\left(\frac{\mathcal{S}_q^{\rm Cardy}}{dq^{-1}} \right)_{q^{-1}}  = \frac{\pi\ell^2}{G_4}~,
\end{align}
in agreement with the previous result of~\cite{Arias:2019aa}~. This proposal enhance the possibility to link the thermal Cardy entropy and the Renyi entropy (thereby with entanglement entropy). As the Renyi entropy is independent of the $q-$parameter this quantity equals the modular entropy, and remain fix in the tensionless limit $q\to1$ that recovers the original dS smooth geometry. 
This also can be seen it by considering the universal terms of the free energy in two-dimensional conformal field theories \cite{Affleck:1986bv, Bloete:1986qm}, where the free energy is
\begin{align}
\mathcal F = \frac{\pi }{6}\left(\frac{c_L}{\beta_L} + \frac{c_R}{\beta_R} \right) + \alpha\mu^2\beta~,
\end{align}
where $\alpha$ is some constant, $\beta$ is the inverse of the termperature, and $\mu^2$ is a term that appears in the confomal anomaly as
\begin{align}
\langle T \rangle = \frac{c}{12}(\mathcal R + \mu^2)~.
\end{align}
Using the stress tensor localized in the branes \eqref{Tij}, the two-dimensional Ricci scalar $\mathcal R = 2\ell^{-2}$, and the Liouville central charge \eqref{cq}, we conclude that 
\begin{align}
\mu^2 = 0~.
\end{align}
Hence, using the normalized temperature of the equilibrium $\beta = 1$, we conclude
\begin{align}
\mathcal F = \left(1-\frac1q\right)\frac{\pi\ell^2}{G} = \mathcal S^{\rm Cardy}_q~,
\end{align}
in agreement with the previous computation using the Baez formalism. We would like in the future to find a precise link in field theory between the Cardy entropy and the free energy that recovers the above presented result. 
The disagreement between the vacuum temperatures is due to the conventions on the definition of density matrix where the temperature is defined with an extra $2\pi$ coefficient in \autoref{CCandCardyS}.

Then Cardy entropy equals the free energy which corresponds to the modular free energy for temperature $1/q$, where $q$ is the same orbifold anisotropic parameter that will be used latter as the replica parameter in the computation of entanglement entropy. 

\section{dS$_3$ holography from $4$ dimensions}\label{largeq}
We have previously study the tensionless limit giving us the possibility to understand the dS entropy in four dimensions as entanglement entropy or modular entropy. 
We follow to take a different limit, in which the orbifold geometry recovers the dS$_3$ spacetime in global coordinates, and the original codimension-2 defects are realized as boundaries in one lower dimension. 
The procedure gives the correct central charge of the two-dimensional CFT at the euclidean boundary of global dS$_3$ \cite{Strominger:2001pn, Park:1998qk, Banados:1998tb, Nojiri:2001mf,Klemm:2001ea, Cacciatori:2001un, Klemm:2002ir}. Moreover, this approach is in agreement with \cite{Klemm:2002ab, Klemm:2002aa, Dyson:2002nt} where the dual CFT does not capture the thermal nature of dS. 

In this section we will take the large $q$-limit ($q \rightarrow \infty$) which corresponds to the zero radius limit of the spindle geometry, as it is shown in Figure~\hyperlink{Fig:12}12. 
\vspace{0.7cm}
\begin{center}
\begin{tikzpicture}[scale=0.7]
%2-sphere
\draw [thick] (-4,0) circle (2cm);
\node at (-5.5,2) {\footnotesize$S^2$};
\draw[thick, dashed] (-2,0) 
arc[start angle=0,end angle=180, 
x radius=2, y radius=0.5];
\draw[thick] (-2,0) 
arc[start angle=0,end angle=-180, 
x radius=2, y radius=0.5];
%Gamma
{\color{darkred}
\draw[thick, dashed] (-2.4,1.2) 
arc[start angle=0,end angle=180, 
x radius=1.6, y radius=0.2];
\draw[thick] (-2.4,1.2) 
arc[start angle=0,end angle=-180, 
x radius=1.6, y radius=0.2];
\node at (-2.1,1.5) {\footnotesize{$\Gamma_\theta$}};
}
%Orbifold arrow
{\color{darkblue}
\node at (-0.5,0.45) {\footnotesize{$S^2/\mathbb Z_q$}};
\draw [thick, ->] (-1.3,0) -- (0.35,0);
}
%Spindle
\fill[fill=yellow!10] (2,2.3)--(1,1.7)--(6,1.7)--(7,2.3);
\fill[fill=blue!10] (2,-1.7)--(1,-2.3)--(6,-2.3)--(7,-1.7);
\draw [thick] (2,2.3) --(7,2.3);
\draw [thick] (1,1.7) --(6,1.7);
\draw [thick] (1,1.7) --(2,2.3);
\draw [thick] (7,2.3) --(6,1.7);
\coordinate (N) at (4,2);\coordinate (S) at (4,-2);
\node [darkred] at (N) {\textbullet};
\node [darkred] at (S) {\textbullet};
\draw[thick] (N)to[out=-20,in=20](S);
\draw[thick] (N)to[out=-150,in=150](S);
%\draw [thick](3.05,0) ellipse (1.05cm and 0.3cm);
\draw[thick, dashed] (5.1,0) 
arc[start angle=0,end angle=180, 
x radius=1.05, y radius=0.3];
\draw[thick] (5.1,0) 
arc[start angle=0,end angle=-180, 
x radius=1.05, y radius=0.3];
%\node at (4.4,2.9) 
%{\footnotesize{$c_q(\Sigma_N)$}$= (1-\tfrac{1}{q})\tfrac{3\ell^2}{G_4}$};
%\node at (4.4,-2.9) 
%{\footnotesize{$c_q(\Sigma_S)$}$= (1-\tfrac{1}{q})\tfrac{3\ell^2}{G_4}$};
\draw [thick] (2,-1.7) --(7,-1.7);
\draw [thick] (1,-2.3) --(6,-2.3);
\draw [thick] (1,-2.3) --(2,-1.7);
\draw [thick] (7,-1.7) --(6,-2.3);
\draw [thick, ->] (4.05,0) -- (5.05,0);
\node at (6,0) {\footnotesize{$\ell_q=\frac{\ell}{q}$}};
\node at (7,1.7) {\footnotesize $\Sigma_N$};
\node at (1,-1.7) {\footnotesize $\Sigma_S$};
%\node at (6,0) {\footnotesize{$\ell_q=\frac{\ell}{q}$}};
%Largeqarrow
{\color{darkblue}
\node at (8.4,0.35) {\footnotesize{$q\to\infty$}};
\draw [thick, ->] (7.5,0) -- (9.3,0);
}
%ZEROREADIUSLIMIT
\fill[fill=yellow!10] (10,2.3)--(9,1.7)--(14,1.7)--(15,2.3);
%\node [darkred] at (10,2) {\textbullet};
\draw [thick] (10,2.3) --(15,2.3);
\draw [thick] (9,1.7) --(14,1.7);
\draw [thick] (9,1.7) --(10,2.3);
\draw [thick] (15,2.3) --(14,1.7);
\fill[fill=blue!10] (10,-1.7)--(9,-2.3)--(14,-2.3)--(15,-1.7);
%\node [darkred] at (10,-2) {\textbullet};
\draw [thick] (10,-1.7) --(15,-1.7);
\draw [thick] (9,-2.3) --(14,-2.3);
\draw [thick] (9,-2.3) --(10,-1.7);
\draw [thick] (15,-1.7) --(14,-2.3);
\draw [thick, dashed] (12,-2) --(12,2);
\node at (14,0) {\footnotesize{dS$_3$}};
\draw [thick, ->] (12,2) -- (12,1);
\node at (11.7,1.3) {\footnotesize{$z$}};
%\node at (12, 2.9) 
%{\footnotesize{$c_\infty(\Sigma_N)$}$=\tfrac{3\ell}{4G_3}$};
%\node at (12,-2.9) 
%{\footnotesize{$c_\infty(\Sigma_S)$}$=\frac{3\ell}{4G_3}$};
%CAPTION
\node[text width=16cm, text justified] at (5,-5) 
{\small {\hypertarget{Fig:12}\bf Fig.~12}: 
\sffamily{
The large $q$ limit of the spindle $S^2/\mathbb Z_q$. The two-dimensional geometry between the defects $\Sigma_N$ and $\Sigma_S$ fully shrinks to one dimension.
}};
\end{tikzpicture}
\end{center}
\vspace{0.5cm} 
In the global coordinates, the large $q$-limit takes the form
\begin{align}
\label{dS3embedding}
X_0 = \sqrt{\ell^2 - \xi^2}\,\cos\theta\,\sinh(t/\ell)~&,\quad
X_1 =\sqrt{\ell^2 - \xi^2}\,\cos\theta\,\cosh(t/\ell)~,  \\ \nonumber
X_2 = \xi\,\cos\theta ~&,\quad
X_3 =\ell \sin\theta~, \quad X_4 = 0~,
\end{align}
which parametrizes the dS$_3$ embedding on the $4$-dimensional hyperboloid with radius $\ell$. The geometry resembles to 
\begin{align}
\lim_{q\to\infty}ds^2 = dz^2 + \cos^2(z/\ell)h~,
\end{align}
where $h$ corresponds to the induced metrics on the defects \eqref{h}. The latter corresponds to the global foliation of dS$_3$ spacetime by mapping
\begin{align}
z \rightarrow iT~,\quad t \rightarrow i\tau~.
\end{align}
Then, the compact coordinate $z$ becomes the global time, and the induced metrics becomes the unit $2-$sphere, \textit{viz}
\begin{align}
\lim_{q\to\infty}ds^2 = -dT^2 + \cosh^2(T/\ell)d\Sigma_2^2~.
\end{align}
Is important to notice, that in this approach, the original codimensional-two defects $\Sigma_{N,S}$ are sent to $T \rightarrow \pm \infty$, that corresponds to the conformal boundaries of dS$_3$,
\begin{align}\label{dS3limit}
(\widehat{\rm dS}_4, \widehat g_4)
&\stackrel{\substack{q\to\infty\vspace{-1mm}\\{}}}{\longrightarrow}
({\rm dS}_3, g_3) \\ \nonumber
(\Sigma_S, \Sigma_N) &\longmapsto (\mathcal I^+, \mathcal I^-)~.
\end{align}
We previously shown that both defects admits a description in terms of Liouville theory, that now is mapped to the dual CFTs, which is in agreement with the literature on dS$_3$/CFT$_2$ \cite{Klemm:2002ab, Klemm:2002aa, Maldacena:1998ab}. 
Also in this limit, the dimensionless temperature vanishes as it is proposed in the previous section, also in agreement with \cite{Dyson:2002nt}. 
Now the total central charge of the composite boundaries $\mathcal I^+ \cup \mathcal I^-$ can be computed by using the large $q$-limit in the Liouville central charge \eqref{Cq}
\begin{align}
c = c(\mathcal I^+) + c(\mathcal I^-) = c_\infty(\Sigma_N) + c_\infty(\Sigma_S) = \frac{6\ell^2}{G_4}~.
\end{align}
Now we follow to express the four-dimensional Newton constant in terms of the three-dimensional Newtons constant
\begin{align}
G_4 = {\rm Vol}(S^1)G_3~,
\end{align}
where Vol($S^1$) is the average volume of a meridian $\Sigma_\theta$ located at a polar angle as it is shown in Figure~\hyperlink{Fig:12}12. This average is given by
\begin{align}
{\rm Vol}(S^1) = 4\ell~,
\end{align}
finally obtaining 
\begin{align}
G_4 = 4\ell G_3~.
\end{align}
Thereby, the central charge in the large $q-$limit corresponds 
\begin{align}
\boxed{c = \frac{3\ell}{2G_3}}~,
\end{align}
exactly as has been previously derived in the dS$_3$/CFT$_2$ correspondence \cite{Strominger:2001pn, Park:1998qk, Banados:1998tb, Nojiri:2001mf,Klemm:2001ea, Cacciatori:2001un, Klemm:2002ir}. 
Notice that in such limit, the temperature of the Liouville theory $q^{-1}$ vanishes, and the dual theory is a non-thermal Euclidean Liouville theory with fixed values, in agreement with \cite{Klemm:2002ir}.

%%%%%%%%%

%% file: Chapters/Conclusions.tex
% Chapter 1

\chapter{Conclusions} % Main chapter title
\label{Chp:Conclusions} % For referencing the chapter elsewhere, use \ref{Chapter1} 
In this work two different approaches to describe the Gibbons-Hawking entropy as entanglement entropy has been pointed out. Both of them relies on the use of an extended static patch for $dS$ spacetime. Onto this static patch an orbifold on the horizon has allow us to interpretate the Gibbons-Hawking entropy between two disconnected conformal field theory living at past or future infinity of the spacetime, and the entanglement between two disconnected bulk region. The first one is motivated by the existence of the dS/CFT correspondence, in which we recover the semiclassical Gibbons-Hawking entropy. In the second we have just made use of the properties of causality that posses the spacetime manifold, and assuming the existence some quantum gravity theory that can be semiclassically approximated by the on-shell Einstein gravity. In this last we have obtain a quarter of the area law \eqref{Sfixed} for the area of the set of fixed points of the $\mathbb Z_q$ action on the bulk geometry, therefore entanglement follows simply by topology. Which in the case of restricting ourself of a single Rindler wedge we recover the $dS_4$ entropy as entanglement entropy. Also, as it is noticed, the Renyi entropy equals the modular entropy and the entanglement entropy, as it is a $q-$independent quantity, in agreement with \cite{Dong:2018aa}, and thereby we found a connection between Cardy entropy and the modular free energy using the $q^{-1}$-derivative as it is proposed by Baez \cite{Baez2011RenyiEA}. We also backup the latter using the universal terms in the free energy for two-dimensional thermal conformal field theories \cite{Affleck:1986bv}.
\\
The bulk $\mathbb Z_q$ symmetry is interpreted as a massive observer in $dS$ space that back-reacts such that the $SO(3)$ symmetries of the horizon are broken just to a discrete $\mathbb Z_q$ symmetry that give rise to two codimension 2 defects that contain the worldline of the observers defined on the fixed points of an $S^2/\mathbb Z_q$ orbifold. Such geometry enhance the possibility to take different limits that realize holography in dS spacetime.
\\

Also this has suggest a possibility to define the four-dimensional $dS$ energy in terms of the expectation value of the modular Hamiltonian, which in the tensionless limit is zero. As a future direction on this subject would be to extended this notion of entanglement on the bulk, that is provided just by topology, to AdS$_4$ spacetime with no holographic framework. This by noticing that if we glue together two AdS spaces along their conformal boundary we could treat this as two different Rindler wedges and perform the same calculation that has been done in \autoref{SEEtop} but for a $\mathbb H_2/\mathbb Z_q$ orbifold. For more discussion on this and generalization to multiple boundaries manifold see \cite{Arias:2019aa}.
We would like to also obtain quantum corrections to the entanglement entropy using the formulation given by Faulkner, Lewkowycz, and Maldacena in \cite{Faulkner:2013ana}, and the study of inclusion of higher curvature corrections to gravity \cite{Hubeny:2019bje, Arenas-Henriquez:2017xnr, Arenas-Henriquez:2019rph, Dong:2013qoa, Giribet:2018hck, Harper:2018sdd,Rivera-Betancour:2018qva,Haehl:2017sot, Anastasiou:2017rjf}~.
\\

Another approach that we have use on this work, relies on two-dimensional CFT tools, in which a relation between the central charge of the theory an its entropy has been pointed out by Cardy. We have made use of the isometry group of the extended static patch, that allow us to mimic the NHEK \eqref{NHEK} geometry  and made use of the Kerr/CFT mechanism and analyze the asymptotic symmetry group obtaining a Virasoro algebra providing the central charge. Also for this conformal description, based on Solodukhin description of spherical killing horizons,we have proposed the Liouville CFT for a fixed value of the Liouville field to describe the defects. The resulting entropy via Cardy's formula in both cases give rises to the relative entropy that relates the amount of free energy that it cost to generate this defects. Therefore in the tensionless limit the quarter of the horizon's area entropy its recovered. 
\\

Also in the last section, we compute the modular free energy giving the Gibbons--Hawking entropy. Moreover, we take the large $q-$limit in order to realize dS$_3$ holography. In this approach, the defects are mapped to the conformal boundaries of dS$_3$, and is it in agreement with the literature where the conformal screens are described using Liouville theory. We are currently exploring the same approach in AdS spacetime and the higher dimensional realization of the BTZ black hole and AdS$_3$ holography. 
We summarize the result of the last section in the following diagram, where we want to emphasize the 
\bigskip
\begin{center}
\begin{tikzpicture}
\node at (0,0) {\framebox{dS$_4/\mathbb Z_q$}};
\node at (-5,0) {\framebox{{\footnotesize Global} dS$_3$}};
\node at (5,0) {\framebox{{\footnotesize Static} dS$_4$}};
\node at (-2.5,0.3) {\color{darkred}\footnotesize $q \to \infty$};
\draw [->] (-1.5,0)--(-3.5,0);
\node at (2.5,0.3) {\color{darkred}\footnotesize $q \to 1$};
\draw [->] (1.5,0)--(3.5,0);
\node at (0,-3){
\framebox{\footnotesize
\begin{tabular}{c} 
Liouville theory \\on $\Sigma_{N, S}$   
\end{tabular}}
};
\node at (-5,-3){
\framebox{\footnotesize
\begin{tabular}{c} 
Non-thermal \\ Liouville theory $\mathcal I^\pm$   
\end{tabular}}
};
\draw [-,thick] (-0.141,-2) arc (-180:0:2pt);
\draw[->,thick](0,-2)--(0,-0.5);
\node at (0,-1.3) {\scriptsize Defects~~(cod-2)};
\draw [-,thick] (-5.141,-2) arc (-180:0:2pt);
\draw[->,thick](-5,-2)--(-5,-0.5);
\node at (-5.25,-1.3) {\scriptsize Boundaries~~(cod-1)};
\node at (-2.5,-2.7) {\color{darkred}\footnotesize $q \to \infty$};
\draw [->] (-2,-3)--(-3,-3);
\end{tikzpicture}
\end{center}
\bigskip

Regarding future work, we are currently trying to formulate (higher spin) gravity theories as topological field theories of the AKSZ type~\cite{Alexandrov:1995kv} (following \cite{Boulanger:2011dd, Bonezzi:2016ttk, Bonezzi:2016ttk, Arias:2016ajh}) which are formulated in multi-boundary manifolds. In this shceme is possible to add extended objects of different codimensions, such that the Hilbert spaces are assigned to boundaries, exactly as we shown in the last section. In this moduli space, the dualities between different Hilbert spaces are associated to defects and boundaries involving a ladder of dualities between different codimensions, and different limits. Therefore, this thesis corresponds just to a particular realization of the latter.

%% file: Appendices/AppendixB.tex
% Appendix C

\chapter{Saddle Point Approximation} % Main appendix title
\label{App:B} % For referencing this appendix elsewhere, use \ref{AppendixA}
The saddle point approximation is just a mathematical trick that allow us to compute integrals that have the form
\begin{align}
I = \int_{-\infty}^\infty dx \exp\{-f(x)\}~,
\end{align}
we know that is have a point $f(x_0)$ that is very small minimum point, therefore $\exp\{-1\}$ would be way bigger than $\exp\{-f(x_0)\}$. Expanding around this point
\begin{align}
f(x) \sim f(x_0) + \frac{1}{2}(x-x_0)^2 f''(x_0) + \dots~,
\end{align}
where we have used the fact that we are expanding around a minimum, i.e., $f'(x_0) = 0$. Therefore the integral renders
\begin{align}
I \sim{}& \int_{-\infty}^\infty dx \exp\{-f(x_0) - \frac{1}{2}(x-x_0)^2 f''(x_0) + \dots\} \nonumber \\ \sim{}& \exp\{-f(x_0)\}\int_{-\infty}^\infty dx \exp\{-\frac{1}{2}(x-x_0)^2 f''(x_0)\}~,
\end{align}
and using Gaussian integration we obtain
\begin{align}
I \sim \exp\{-f(x_0)\}\sqrt{\frac{2\pi}{f''(x_0)}}~.
\end{align}
This is very used in physics to calculate this sort of integrals, they appear for example in statistical mechanics on the partition functions
\begin{align}
\mathcal Z = \int \mathcal D\phi_i \exp\{-\beta F[\phi_i]\}~,
\end{align}
expanding the free energy as above
\begin{align}
F[\phi_i] \sim F[\phi_{o,i}] + \frac{1}{2}\sum_{ij}(\phi_i - \phi_{o,i})(\phi_j - \phi_{o,j})\frac{\partial^2 F}{\partial\phi_i \partial\phi_j}~,
\end{align}
where $F[\phi_{0,i}]$ is the ground state contribution, and the other term gives a gaussian integral that can be solved in the limit $\beta\rightarrow \infty$. 

%% file: Appendices/AppendixC.tex
% Bppendix B

\chapter{Thermofield Double State} % MBin Bppendix title
\label{App:C} % For referencing this Bppendix elsewhere, use \ref{BppendixB}
As an easy example of an entangled state, we refer to the \emph{thermofield double state} \cite{ISRAEL1976107} which it normalized quantum state is defined in terms of the Energy $E$ as 
\begin{align}
\ket{\Psi}  = \frac{1}{\sqrt{\mathcal Z}}\sum_q \exp\left\{-\frac{\beta E_q}{2}\right\}\ket{q}_a\otimes\ket{q}_b~,
\end{align}
this will also be the proposed dual state of the entangled Rindler observed proposed in \autoref{Chp:dSEntanglement}. 
\\

The reduced density matrix of this system becomes simply the normalized Gibbs state for the modular Hamiltonian 
\begin{align}
\rho_a ={}& \frac{1}{\mathcal Z}\sum_q \exp\{-\beta E_q\}\ket{q}_{a~a}\bra{q}~, \nonumber \\ ={}& \frac{1}{\mathcal Z}\exp\{-\beta \mathcal H_a\}~.
\end{align}
This corresponds to \emph{purify} a thermal state on the subsystem $a$ in the total Hilbert space, just by copying vectors from $a$ to $b$, and this measure the entanglement of this purification of a thermal state.
\\

The entanglement entropy of this corresponds to 
\begin{align}\label{FSE}
\mathcal S_E ={}& -\Tr_a\left[ \rho_a(-\beta\mathcal H_a - \log \mathcal Z)\right] \\ ={}& \beta\left[ \expval{\mathcal H_a} - \mathcal F\right]~,
\end{align}
where $\mathcal F$ is the standard thermal free energy $\mathcal F = -\frac{1}{\beta}\log \mathcal Z$~.